\documentclass[longauth]{aa} % for the long lists of affiliations 

\usepackage{graphicx}
\usepackage{txfonts}
\usepackage{hyperref}
\usepackage{multirow}
\usepackage{marvosym}
\usepackage[euler]{textgreek}
\usepackage{xcolor}

\newcommand{\host}{WASP-189}
\newcommand{\planet}{WASP-189\,b}
\newcommand{\wasp}{WASP-South}
\newcommand{\trappist}{TRAPPIST-South}
\newcommand{\cheops}{CHEOPS}
\newcommand{\harps}{HARPS}
\newcommand{\coralie}{CORALIE}
\newcommand{\twomass}{2MASS}
\newcommand{\wise}{WISE}
\newcommand{\gaia}{Gaia}
\newcommand{\kepler}{Kepler}
\newcommand{\tess}{TESS}
\newcommand{\pytransit}{\texttt{pytransit}}
\newcommand{\batman}{\texttt{batman}}
\newcommand{\pipe}{PIPE}
\newcommand{\citepipe}{Brandeker et~al. (in~prep.)}

\begin{document} 

\title{The atmosphere and architecture of WASP-189\,b\\ probed by its CHEOPS phase curve}
\titlerunning{CHEOPS phase curve of WASP-189\,b}

\author{
        % Lead author
        A.~Deline\inst{\ref{Geneva},}\thanks{\email{adrien.deline@unige.ch}} \and
        % Major contributors
        M.~J.~Hooton\inst{\ref{Bern}} \and
        M.~Lendl\inst{\ref{Geneva}} \and
        B.~Morris\inst{\ref{Bern_CSH}} \and
        S.~Salmon\inst{\ref{Geneva}} \and
        % Science enablers
        G.~Olofsson\inst{\ref{Stockholm}} \and
        C.~Broeg\inst{\ref{Bern},\ref{Bern_CSH}} \and
        D.~Ehrenreich\inst{\ref{Geneva}} \and
        M.~Beck\inst{\ref{Geneva}} \and
        % Significant contributors
        A.~Brandeker\inst{\ref{Stockholm}} \and
        S.~Hoyer\inst{\ref{Marseille}} \and
        S.~Sulis\inst{\ref{Marseille}} \and
        V.~Van~Grootel\inst{\ref{Liege_STAR}} \and
        V.~Bourrier\inst{\ref{Geneva}} \and
        O.~Demangeon\inst{\ref{Porto},\ref{Porto_2}} \and
        B.-O.~Demory\inst{\ref{Bern_CSH}} \and
        K.~Heng\inst{\ref{Bern_CSH},\ref{Warwick},\ref{Munich}} \and
        H.~Parviainen\inst{\ref{Tenerife_IAC},\ref{Tenerife}} \and
        L.~M.~Serrano\inst{\ref{Torino}} \and
        V.~Singh\inst{\ref{Padova}} \and
        A.~Bonfanti\inst{\ref{Graz}} \and
        L.~Fossati\inst{\ref{Graz}} \and
        D.~Kitzmann\inst{\ref{Bern_CSH}} \and
        S.~G.~Sousa\inst{\ref{Porto}} \and
        T.~G.~Wilson\inst{\ref{StAndrews}} \and
        % Alphabetical author list
        Y.~Alibert\inst{\ref{Bern}} \and
        R.~Alonso\inst{\ref{Tenerife_IAC},\ref{Tenerife}} \and
        G.~Anglada\inst{\ref{Barcelona_ICE}} \and
        T.~B\'arczy\inst{\ref{Admatis}} \and
        D.~Barrado~Navascues\inst{\ref{Madrid}} \and
        S.~C.~C.~Barros\inst{\ref{Porto},\ref{Porto_2}} \and
        W.~Baumjohann\inst{\ref{Graz}} \and
        T.~Beck\inst{\ref{Bern}} \and
        A.~Bekkelien\inst{\ref{Geneva}} \and
        W.~Benz\inst{\ref{Bern},\ref{Bern_CSH}} \and
        N.~Billot\inst{\ref{Geneva}} \and
        X.~Bonfils\inst{\ref{Grenoble}} \and
        J.~Cabrera\inst{\ref{DLR_IPR}} \and
        S.~Charnoz\inst{\ref{Paris_IPGP}} \and
        A.~Collier~Cameron\inst{\ref{StAndrews}} \and
        C.~Corral~van~Damme\inst{\ref{ESA_ESTEC}} \and
        Sz.~Csizmadia\inst{\ref{DLR_IPR}} \and
        M.~B.~Davies\inst{\ref{Lund}} \and
        M.~Deleuil\inst{\ref{Marseille}} \and
        L.~Delrez\inst{\ref{Liege},\ref{Liege_STAR}} \and
        T.~de~Roche\inst{\ref{Bern}} \and
        A.~Erikson\inst{\ref{DLR_IPR}} \and
        A.~Fortier\inst{\ref{Bern},\ref{Bern_CSH}} \and
        M.~Fridlund\inst{\ref{Leiden},\ref{Onsala}} \and
        D.~Futyan\inst{\ref{Geneva}} \and
        D.~Gandolfi\inst{\ref{Torino}} \and
        M.~Gillon\inst{\ref{Liege}} \and
        M.~G\"udel\inst{\ref{Vienna}} \and
        P.~Gutermann\inst{\ref{Marseille},\ref{INSU}} \and
        J.~Hasiba\inst{\ref{Graz}} \and
        K.~G.~Isaak\inst{\ref{ESA_ESTEC}} \and
        L.~Kiss\inst{\ref{Konkoly}} \and
        J.~Laskar\inst{\ref{Paris_IMCCE}} \and
        A.~Lecavelier~des~Etangs\inst{\ref{Paris_IAP}} \and
        C.~Lovis\inst{\ref{Geneva}} \and
        D.~Magrin\inst{\ref{Padova}} \and
        P.~F.~L.~Maxted\inst{\ref{Keele}} \and
        M.~Munari\inst{\ref{Catania}} \and
        V.~Nascimbeni\inst{\ref{Padova}} \and
        R.~Ottensamer\inst{\ref{Vienna}} \and
        I.~Pagano\inst{\ref{Catania}} \and
        E.~Pall\'e\inst{\ref{Tenerife_IAC}} \and
        G.~Peter\inst{\ref{DLR_IOSS}} \and
        G.~Piotto\inst{\ref{Padova},\ref{Padova_University}} \and
        D.~Pollacco\inst{\ref{Warwick}} \and
        D.~Queloz\inst{\ref{Cambridge_Cavendish}, \ref{ETHZ}} \and
        R.~Ragazzoni\inst{\ref{Padova},\ref{Padova_University}} \and
        N.~Rando\inst{\ref{ESA_ESTEC}} \and
        H.~Rauer\inst{\ref{DLR_IPR},\ref{Berlin_TU},\ref{Berlin_FU}} \and
        I.~Ribas\inst{\ref{Barcelona_ICE},\ref{Barcelona_IEEC}} \and
        N.~C.~Santos\inst{\ref{Porto},\ref{Porto_2}} \and
        G.~Scandariato\inst{\ref{Catania}} \and
        D.~S\'egransan\inst{\ref{Geneva}} \and
        A.~E.~Simon\inst{\ref{Bern}} \and
        A.~M.~S.~Smith\inst{\ref{DLR_IPR}} \and
        M.~Steller\inst{\ref{Graz}} \and
        Gy.~M.~Szab\'o\inst{\ref{ELTE_Gothard},\ref{ELTE_MTA}} \and
        N.~Thomas\inst{\ref{Bern}} \and
        S.~Udry\inst{\ref{Geneva}} \and
        I.~Walter\inst{\ref{DLR_IOSS}} \and
        N.~Walton\inst{\ref{Cambridge_IoA}}
        }
\authorrunning{A.~Deline et al.}

\institute{
        Department of Astronomy, University of Geneva, Chemin Pegasi 51, 1290 Versoix, Switzerland\label{Geneva}
        \and
        Physikalisches Institut, University of Bern, Gesellsschaftstrasse 6, 3012 Bern, Switzerland\label{Bern}
        \and
        Center for Space and Habitability, University of Bern, Gesellsschaftstrasse 6, 3012 Bern, Switzerland\label{Bern_CSH}
        \and
        Department of Astronomy, Stockholm University, AlbaNova University Center, 10691 Stockholm, Sweden\label{Stockholm}
        \and
        Aix-Marseille Universit\'e, CNRS, CNES, Laboratoire d'Astrophysique de Marseille, 38 rue Fr\'ed\'eric Joliot-Curie, 13388 Marseille, France\label{Marseille}
        \and
        Space sciences, Technologies and Astrophysics Research (STAR) Institute, Universit\'e de Li\`ege, All\'ee du 6 Ao\^ut 19C, 4000 Li\`ege, Belgium\label{Liege_STAR}
        \and
        Instituto de Astrof\'isica e Ci\^encias do Espa\c co, Universidade do Porto, CAUP, Rua das Estrelas, 4150-762 Porto, Portugal\label{Porto}
        \and
        Departamento de F\'isica e Astronomia, Faculdade de Ci\^encias, Universidade do Porto, Rua do Campo Alegre 687, 4169-007 Porto, Portugal\label{Porto_2}
        \and
        Department of Physics, University of Warwick, Gibbet Hill Road, Coventry CV4 7AL, United Kingdom\label{Warwick}
        \and
        University Observatory Munich, Ludwig Maximilian University, Scheinerstra\ss e 1, Munich 81679, Germany\label{Munich}
        \and
        Instituto de Astrof\'isica de Canarias, 38200 La Laguna, Tenerife, Spain\label{Tenerife_IAC}
        \and
        Departamento de Astrof\'isica, Universidad de La Laguna, 38206 La Laguna, Tenerife, Spain\label{Tenerife}
        \and
        Dipartimento di Fisica, Universit\`a degli Studi di Torino, Via Pietro Giuria 1, 10125, Torino, Italy\label{Torino}
        \and
        INAF, Osservatorio Astronomico di Padova, Vicolo Osservatorio 5, 35122 Padova, Italy\label{Padova}
        \and
        Space Research Institute, Austrian Academy of Sciences, Schmiedlstra\ss e 6, 8042 Graz, Austria\label{Graz}
        \and
        Centre for Exoplanet Science, SUPA School of Physics and Astronomy, University of St Andrews, North Haugh, St Andrews KY16 9SS, United Kingdom\label{StAndrews}
        \and
        Institut de Ci\`encies de l'Espai (ICE, CSIC), Campus UAB, Carrer de Can Magrans, s/n, 08193 Barcelona, Spain\label{Barcelona_ICE}
        \and
        Admatis, Kand\'o K\'alm\'an \'ut 5, 3534 Miskolc, Hungary\label{Admatis}
        \and
        Departamento de Astrof\'isica, Centro de Astrobiolog\'ia (CSIC-INTA), ESAC campus, 28692 Villanueva de la Ca\~nada, Spain\label{Madrid}
        \and
        Universit\'e Grenoble Alpes, CNRS, Institut de Plan\'etologie et d'Astrophysique de Grenoble, 38000 Grenoble, France\label{Grenoble}
        \and
        Institute of Planetary Research, German Aerospace Center (DLR), Rutherfordstra\ss e 2, 12489 Berlin, Germany\label{DLR_IPR}
        \and
        Universit\'e de Paris, Institut de Physique du Globe de Paris, CNRS, 75005 Paris, France\label{Paris_IPGP}
        \and
        European Space Research and Technology Centre (ESTEC), European Space Agency (ESA), Keplerlaan 1, 2201-AZ Noordwijk, The Netherlands\label{ESA_ESTEC}
        \and
        Centre for Mathematical Sciences, Lund University, Box 118, 22100 Lund, Sweden\label{Lund}
        \and
        Astrobiology Research Unit, Universit\'e de Li\`ege, All\'ee du 6 Ao\^ut 19C, 4000 Li\`ege, Belgium\label{Liege}
        \and
        Leiden Observatory, University of Leiden, PO Box 9513, 2300 RA Leiden, The Netherlands\label{Leiden}
        \and
        Department of Space, Earth and Environment, Chalmers University of Technology, Onsala Space Observatory, 43992 Onsala, Sweden\label{Onsala}
        \and
        Department of Astrophysics, University of Vienna, T\"urkenschanzstra\ss e 17, 1180 Vienna, Austria\label{Vienna}
        \and
        Division Technique, Institut National des Sciences de l'Univers (INSU), CS 20330, 83507 La Seyne-sur-Mer, France\label{INSU}
        \and
        Konkoly Observatory, Research Centre for Astronomy and Earth Sciences, Konkoly-Thege Mikl\'os \'ut 15-17, 1121 Budapest, Hungary\label{Konkoly}
        \and
        IMCCE, UMR8028 CNRS, Observatoire de Paris, PSL Universit\'e, Sorbonne Universit\'e, 77 avenue Denfert-Rochereau, 75014 Paris, France\label{Paris_IMCCE}\\\\
        \and
        Institut d'astrophysique de Paris, UMR7095 CNRS, Universit\'e Pierre \& Marie Curie, 98 bis boulevard Arago, 75014 Paris, France\label{Paris_IAP}
        \and
        Astrophysics Group, Keele University, Staffordshire, ST5 5BG, United Kingdom\label{Keele}
        \and
        INAF, Osservatorio Astrofisico di Catania, Via Santa Sofia 78, 95123 Catania, Italy\label{Catania}
        \and
        Institute of Optical Sensor Systems, German Aerospace Center (DLR), Rutherfordstra\ss e 2, 12489 Berlin, Germany\label{DLR_IOSS}
        \and
        Dipartimento di Fisica e Astronomia ``Galileo Galilei'', Universit\`a degli Studi di Padova, Vicolo Osservatorio 3, 35122 Padova, Italy\label{Padova_University}
        \and
        Cavendish Laboratory, JJ Thomson Avenue, Cambridge CB3 0HE, United Kingdom\label{Cambridge_Cavendish}
        \and
        Department of Physics, ETH Z\"urich, Wolfgang-Pauli-Strasse 27, 8093 Z\"urich, Switzerland\label{ETHZ}
        \and
        Center for Astronomy and Astrophysics, Technical University Berlin, Hardenberstra\ss e 36, 10623 Berlin, Germany\label{Berlin_TU}
        \and
        Institut f\"ur Geologische Wissenschaften, Freie Universit\"at Berlin, 12249 Berlin, Germany\label{Berlin_FU}
        \and
        Institut d'Estudis Espacials de Catalunya (IEEC), 08034 Barcelona, Spain\label{Barcelona_IEEC}
        \and
        ELTE E\"otv\"os Lor\'and University, Gothard Astrophysical Observatory, Szent Imre herceg utca 112, 9700 Szombathely, Hungary\label{ELTE_Gothard}
        \and
        MTA-ELTE Exoplanet Research Group, Szent Imre herceg utca 112, 9700 Szombathely, Hungary\label{ELTE_MTA}
        \and
        Institute of Astronomy, University of Cambridge, Madingley Road, Cambridge, CB3 0HA, United Kingdom\label{Cambridge_IoA}
        }

\date{Received January 1, 2021; accepted January 1, 2021}

\abstract
        % context heading (optional, leave it empty if necessary)
        {Gas giants orbiting close to hot and massive early-type stars can reach dayside temperatures that are comparable to those of the coldest stars. These `ultra-hot Jupiters' have atmospheres made of ions and atomic species from molecular dissociation and feature strong day-to-night temperature gradients. Photometric observations at different orbital phases provide insights on the planet's atmospheric properties.}
        {We aim to analyse the photometric observations of \host{} acquired with the Characterising Exoplanet Satellite (\cheops{}) to derive constraints on the system architecture and the planetary atmosphere.}
        {We implemented a light-curve model suited for an asymmetric transit shape caused by the gravity-darkened photosphere of the fast-rotating host star. We also modelled the reflective and thermal components of the planetary flux, the effect of stellar oblateness and light-travel time on transit-eclipse timings, the stellar activity, and \cheops{} systematics.}
        {From the asymmetric transit, we measure the size of the ultra-hot Jupiter \planet{}, $R_p=1.600^{+0.017}_{-0.016}\,R_J$, with a precision of 1\%, and the true orbital obliquity of the planetary system, $\Psi_p=89.6\pm1.2\deg$ (polar orbit). We detect no significant hotspot offset from the phase curve and obtain an eclipse depth of $\delta_\text{ecl}=96.5^{+4.5}_{-5.0}\,\text{ppm}$, from which we derive an upper limit on the geometric albedo: $A_g<0.48$. We also find that the eclipse depth can only be explained by thermal emission alone in the case of extremely inefficient energy redistribution.
        Finally, we attribute the photometric variability to the stellar rotation, either through superficial inhomogeneities or resonance couplings between the convective core and the radiative envelope.}
        {Based on the derived system architecture, we predict the eclipse depth in the upcoming Transiting Exoplanet Survey Satellite (\tess{}) observations to be up to $\sim165\,\text{ppm}$. High-precision detection of the eclipse in both \cheops{} and \tess{} passbands might help disentangle reflective and thermal contributions.
        We also expect the right ascension of the ascending node of the orbit to precess due to the perturbations induced by the stellar quadrupole moment $J_2$ (oblateness).}


\maketitle

\section{Introduction} \label{sec:intro}
        
        Extra-solar planets exhibit a wide range of sizes, compositions, temperatures, and system architectures. Hot Jupiters are among the most extreme of these worlds, orbiting so close to their host star that they can reach equilibrium temperatures at their surfaces beyond 2000\,K. The proximity of the star also creates strong tidal forces causing the planet rotation and revolution periods to synchronise. Once tidally locked, the planet always has the same hemisphere facing the star, and this strongly impacts the atmospheric circulation. Effects of stellar irradiation are further enhanced when the host is an early-type F or A star, hotter and more massive than the Sun \citep[e.g. ][]{collier_cameron_2010_wasp33, gaudi_2017_kelt9b}. Close-in gas giants orbiting such stars, dubbed `ultra-hot Jupiters', have cloud-free daysides with temperatures commensurate with the surface of cool stars, where most molecules are thermally dissociated and atoms are ionised \citep{evans_2017_wasp121b, bell_2018_uhj, kitzmann_2018_kelt9b, parmentier_2018_wasp121b, lothringer_2018_uhj, fossati_2021_kelt-9b}. Partially ionised atmospheres inhibit atmospheric circulation from the dayside to the nightside of the planet (via Lorentz forces), resulting in strong temperature contrasts of about $1000$~K \citep{komacek_2016_hj_day-night_temp}. Colder nightside temperatures allow for various condensation processes to occur, as exemplified by the measurement of a different iron composition at the morning and evening twilights of the ultra-hot gas giant WASP-76b \citep{ehrenreich_2020_wasp76b-iron, kesseli_2021_wasp76b, wardenier_2021_wasp76b}. Due to their elevated temperatures and large day-to-night contrasts, ultra-hot gas giants are especially amenable to mapping their atmospheres with observations at various phase angles; that is, in transit (nightside), occultation (dayside), and in-between (phase curve). Key insights on ultra-hot gas giant atmospheres can thus be revealed by observing them at both infrared and optical wavelengths. In fact, these exoplanet daysides emit thermal radiation in the optical domain, giving rise to deep eclipses and large phase-curve amplitudes \citep[e.g. ][]{bourrier_2020_wasp-121b}.
        
        \cite{lendl_2020_wasp-189b} recently reported on occultations of the ultra-hot gas giant \planet{} \citep{anderson_2018_wasp-189b} observed with the Characterising Exoplanet Satellite \citep[\cheops{} --][]{benz_2021_cheops} in the visible wavelength range (330-1100\,nm). The occultation depth of $87.9\pm4.3\,\text{ppm}$ appears compatible with an unreflective atmosphere heated to $3425\pm27\,\text{K}$ when assuming inefficient heat redistribution \citep{lendl_2020_wasp-189b}. Using the ultra-high photometric precision of \cheops{}, \cite{lendl_2020_wasp-189b} also found the fast-rotating, gravity-darkened host star to cause an asymmetric transit light curve, allowing a direct inference on the true obliquity of the planet's orbital spin axis.
        
        Here, we report on the first full phase-curve observations of \planet, which were obtained with \cheops. We jointly analyse them with the occultations previously published by \cite{lendl_2020_wasp-189b}. Together, these observations cover two full planetary orbits, six eclipses and 3.5 transits of \planet. We complement these photometric observations with high-resolution spectroscopy to refine the stellar properties (Sect.~\ref{sec:star}). We describe the \cheops{} observations and their reduction in Sect.~\ref{sec:obs} and the light curve analysis in Sect.~\ref{sec:lc}. Finally, we discuss our results in Sect.~\ref{sec:results}.

\section{Host star \host} \label{sec:star}
        
        \subsection{An oblate, gravity-darkened fast rotator}
                
                The host star \host{}~(HD\,133112; HR\,5599) is a hot A4 star with an effective temperature of 8000\,K (see Table~\ref{tab:star}). Transits of a gas giant were reported by \cite{anderson_2018_wasp-189b} using \wasp{} \citep{pollacco_wasp, collier-cameron_wasp} and \trappist{} \citep{gillon_trappist,jehin_trappist}. Follow-up spectroscopy with \coralie{} \citep{queloz_coralie} and \harps{} \citep{mayor_harps} allowed \cite{anderson_2018_wasp-189b} to reveal the star as a fast rotator ($v_\star\sin i_\star\sim100\,\text{km}.\text{s}^{-1}$). Rapid rotation is common to early-type stars \citep[F, A, B, O -- e.g.][]{royer_2007_rotation_A-type_stars, dufton_2013_rotation_B-type_stars} that tend to be radially distorted by the centrifugal force resulting in oblate shapes. The surface gravity of an oblate star hence varies as a function of latitude causing a change of local temperature and brightness \citep{von-zeipel_rotating-stars}: the equator appears darker than the poles, a phenomenon known as gravity~darkening \citep[GD --][]{claret_gravity-darkening, espinosa-lara_gravity-darkening}. When a planet on a misaligned orbit transits such a star, the non-radially symmetric brightness distribution of the stellar disk will create an asymmetry in the photometric transit light curve. The asymmetry can be used to retrieve the absolute orientation of the system (stellar inclination and orbital obliquity) as modelled by \cite{barnes_gravity-darkening} and observed for several targets \citep{barnes_koi-13.01, szabo_2012_koi-13, ahlers_mascara4b_grav-dark, lendl_2020_wasp-189b}.
                
                The CHEOPS phase-curve observations reported in this work furthermore reveal a photometric variability attributed to \host \ (see Sect.~4.5 for details).

        \subsection{Refined stellar properties from spectroscopic measurements}
                
                To support our analysis of the \cheops{} observations of the \host{} system, we computed the stellar parameters listed in Table~\ref{tab:star} using the same methods as described in \cite{lendl_2020_wasp-189b} with the inclusion of a new validation procedure detailed in \cite{bonfanti_hd108236}.
                
                For consistency, we briefly summarise the methods used for the derivation of the stellar properties.
                The first method consists of using synthetic spectra to fit spectral lines observed with the \harps{} spectrograph \citep[][programmes 0100.C-0847 and 0103.C-0472, observed in 2018 and 2019, respectively]{mayor_harps} with synthetic spectra. The modelling of the stellar atmosphere and evolution then allows us to compute fundamental parameters. The first output of the analysis is the projected rotation speed of the star $v_\star\sin i_\star$. By assuming several conditions on iron atmospheric content (excitation equilibrium of FeI and FeII, ionisation equilibrium of Fe and minimum standard deviation of Fe abundance), we could compute, respectively, the effective temperature $T_\text{eff}$, the surface gravity $\log g$, and the microturbulent velocity $v_\text{mic}$. We also directly derived the iron abundance $\left[\text{Fe}/\text{H}\right]$ from this method.
                The second method we used is based on the infrared flux method \citep[IRFM --][]{blackwell_irfm}, which allows us to determine the angular diameter of \host{} from its relationship with the bolometric flux and the flux received on Earth. We used a Markov chain Monte Carlo (MCMC) formulation of the IRFM method as described in \cite{schanche_2020_irfm_mcmc}. It consists of building synthetic spectral energy distributions (SEDs) using stellar parameters from the previous step ($T_\text{eff}$, $\log g$ and $\left[\text{Fe}/\text{H}\right]$) and comparing them to photometry in various passbands: \gaia{} passbands $G$, $G_\text{BP}$, and $G_\text{RP}$ \citep{gaia_edr3}; \twomass{} passbands J, H, and K \citep{skrutskie_2mass}; and \wise{} passbands W1 and W2 \citep{wright_wise}. Once the angular diameter is estimated, we derive the stellar radius $R_\star$ with the \gaia{} EDR3 parallax after correction of the offset \citep{lindegren_gaia_edr3_parallax}.
                Finally, our third method is based on modelling the stellar evolution with two different codes: PARSEC \citep{marigo_parsec_2017} via the isochrone placement algorithm \citep{bonfanti_parsec_2015, bonfanti_parsec_2016}, and CL\'ES \citep{scuflaire_cles}. The input parameters are the previously determined effective temperature $T_\text{eff}$, iron abundance $\left[\text{Fe}/\text{H}\right]$ and stellar radius $R_\star$, from which the two models provide the stellar mass $M_\star$ and the system age $t_\star$. We combine the posterior probability distributions of the two results using the procedure detailed in \cite{bonfanti_hd108236}.
                All computed stellar parameters are listed in Table~\ref{tab:star}.
                
                \begin{table}
                        \caption{Properties of the star \host.}
                        \label{tab:star}
                        \centering
                        \resizebox{\columnwidth}{!}{
                                \begin{tabular}{lcr}
                                        \hline\hline
                                        Parameter & Value & Source \\
                                        \hline
                                        \multirow{5}{*}{Names \& aliases} & WASP-189 & \multirow{5}{*}{Simbad \tablefootmark{1}} \\
                                        & HD\,133112 & \\
                                        & HR\,5599 & \\
                                        & TIC\,157910432 & \\
                                        & Gaia~DR2~6339097679918871168 & \\
                                        \hline
                                        V-band magnitude & $6.618\pm0.011$ & Simbad \tablefootmark{1} \\
                                        Gaia G-band magnitude & $6.5537\pm0.0004$ & Gaia archive \tablefootmark{2} \\
                                        \hline
                                        $T_\text{eff}\ \left[K\right]$ & $8000\pm80$ & spectroscopy \\
                                        $M_\star\ \left[M_\odot\right]$ & $2.031^{+0.098}_{-0.096}$ & evolution model \\
                                        $R_\star\ \left[R_\odot\right]$ & $2.365\pm0.025$ & IRFM \\
                                        $\log g\ \left[\log_{10}\left(\text{cm}.\text{s}^{-2}\right)\right]$ & $3.9\pm0.2$ & spectroscopy \\
                                        $v_\text{mic}\ \left[\text{km}.\text{s}^{-1}\right]$ & $2.70\pm0.30$ & spectroscopy \\
                                        $\left[\text{Fe}/\text{H}\right]$ & $0.29\pm0.13$ & spectroscopy \\
                                        $t_\star\ \left[\text{Gyr}\right]$ & $0.73^{+0.19}_{-0.20}$ & evolution model \\
                                        $L_\star\ \left[L_\odot\right]$ & $20.64\pm0.93$ & $L_\star=4\pi R_\star^2 \sigma_\text{SB} T_\text{eff}^4$ \\
                                        $v_\star\sin i_\star\ \left[\text{km}.\text{s}^{-1}\right]$ & $93.1\pm1.7$ & spectroscopy \\
                                        \hline\hline
                                \end{tabular}
                        }
                        \tablefoot{
                                The methods used to obtain the stellar parameters are described in the text. The stellar luminosity $L_\star$ is computed following the Stefan–Boltzmann law and using the Stefan–Boltzmann constant $\sigma_\text{SB}$.
                                \tablefoottext{1}{SIMBAD astronomical database from the Centre de Donn\'ees astronomiques de Strasbourg\footnote{\url{http://simbad.u-strasbg.fr/simbad/}}.}
                                \tablefoottext{2}{Archive of the Gaia mission of the European Space Agency\footnote{\url{https://gea.esac.esa.int/archive/}}}
                        }
                \end{table}

\section{CHEOPS observations and data reduction} \label{sec:obs}
        
        \subsection{Eclipse and phase-curve observations} \label{ssec:obs}
        
        \cheops{} made several photometric observations of the WASP-189 system in the visible wavelength range (330-1100\,nm -- see Fig.~\ref{fig:star_sed}), covering four eclipses and two full orbits of WASP-189\,b. The four eclipses and two transits out of the first phase-curve observation were analysed in \cite{lendl_2020_wasp-189b}. In this work, we present the joint analysis of the data previously published together with both full phase curves.
        
        As \cheops{} orbits the Earth on a low-altitude Sun-synchronous trajectory, the target star is periodically occulted by our planet, causing interruptions in the photometric sequence, referred to as Earth occultations and visible in the light curve as gaps. In addition, the spacecraft regularly crosses the South Atlantic anomaly (SAA) where the Earth's magnetic field concentrates high-energy particles that strongly deteriorate the quality of the images and make them scientifically useless. In order to save the downlink bandwidth, observations acquired during SAA crossings are not transferred to the ground. Therefore, \cheops'{} photometric light curves feature gaps on a nearly periodic basis due to these two expected phenomena with typical durations per \cheops{} orbit ranging from 25 to 40\,min for the Earth occultation and from 0 to 18\,min for the SAA.
        
        Each of the \cheops{} observations can be referred to using a file key that is a unique identifier in the mission database. The observations log is shown in Table~\ref{tab:obs} and lists the file keys of the data sets analysed in this work together with additional information.
        
        The earliest data acquired with \cheops{} are the eclipse light curves. The images were taken at the very beginning of the mission, even before the start of the nominal science mission, as part of the Early Science Programme (ESP), which is dedicated to demonstrating the photometric capabilities of \cheops{}. For this reason, the eclipse data sets were obtained in different conditions. The first difference is the location of the target point spread function (PSF) on the instrument detector. The \cheops{} detector is a charge-coupled device (CCD) with an image section of 1024$\times$1024 pixels from which a 200$\times$200-pixel sub-array is extracted around the target PSF that is used to compute the photometry. The plate scale of the instrument is 1\,arcsecond per pixel, and the PSF, defocused by design, spreads across a radius of about 16\,pixels. The first eclipse observation was taken with the PSF in its initial default position at the center of the detector, $\left(x, y\right)=\left(512, 512\right)$. To minimise the noise induced by newly appearing hot pixels, the PSF location was changed to the lower right part of the detector, $\left(x, y\right)=\left(664, 296\right)$, for the other three eclipse observations, and again to the upper left part of the CCD, $\left(x, y\right)=\left(263, 842\right)$, for the phase-curve observations.
        
        The other change that occurred at the beginning of the \cheops{} mission and affected the eclipse data sets is related to the temperature of the detector. With the aim of reducing the impact of dark current and hot pixels on the photometry, the CCD temperature was lowered during some ESP observations including the four \planet{} eclipses (see Table~\ref{tab:obs}). The nominal science CCD temperature was then set to $-45^\circ\text{C,}$ and the \host{} phase curves were obtained in these conditions.
        Finally, the capability of the \cheops{} telescope to reject stray light varies with several parameters such as the angular distance between the line of sight and the Earth limb, the illumination of the Earth limb, the airglow of the Earth's atmosphere, and the brightness and location of the Moon. The main effect of stray light is an increase of the background level of images that is expected to be corrected during the data processing. We note that the second and third eclipse observations of \planet{} have some images with very high background levels compared to the other eclipse observations. For reasons that are not yet well understood, correlations between photometric flux and background level start to appear for extreme values of the background, and the two mentioned eclipse observations are affected by this effect. Discarding frames with extremely high background levels is a straightforward solution to this problem as discussed in Section~\ref{ssec:data_filtering}.
        
\begin{figure*}
                \centering
                \includegraphics[width=.97\hsize,trim={1cm 0.5cm 2cm 1.5cm},clip]{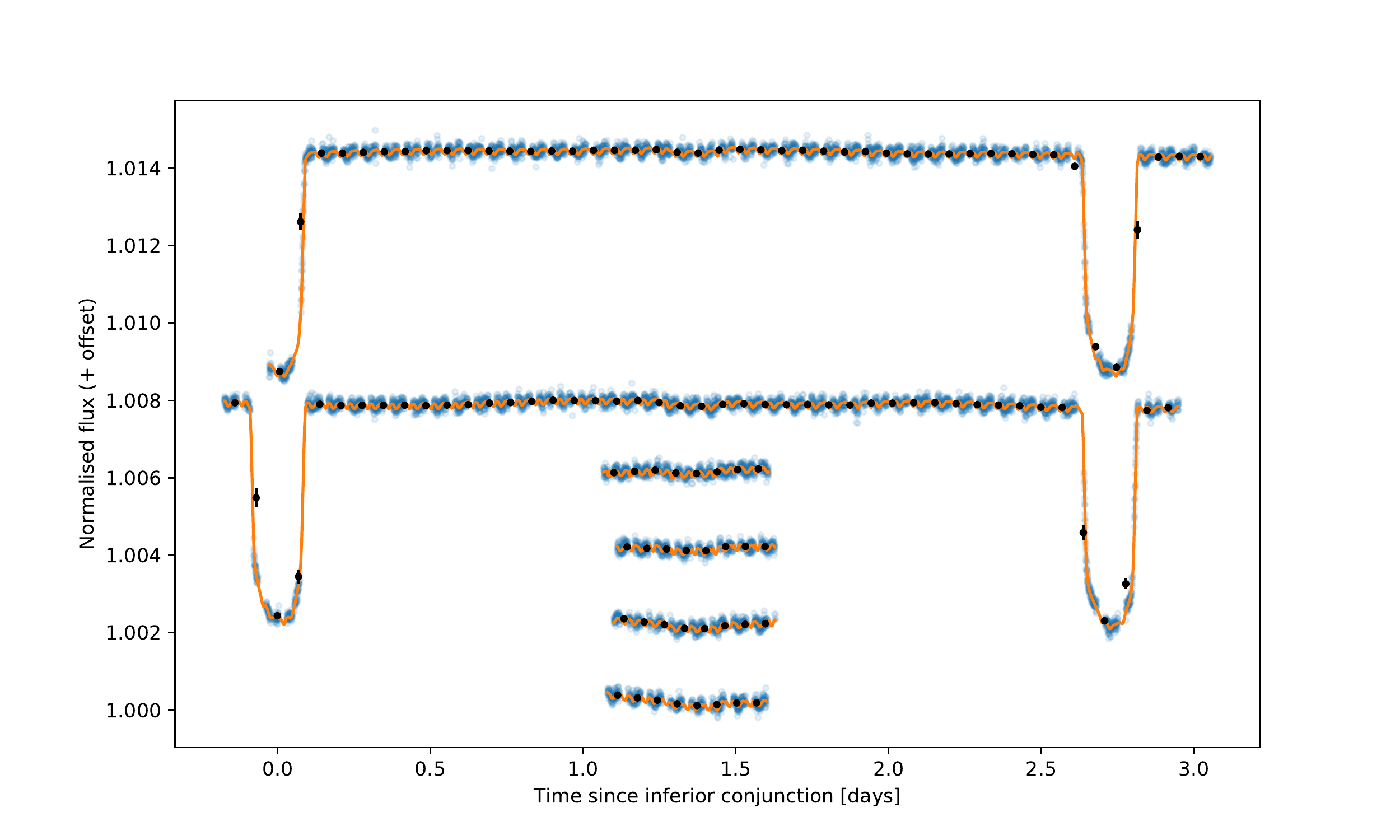}
                \caption{Raw light curves of the four eclipses and the two phase curves measured by \cheops. The represented data were extracted by aperture photometry by the mission data reduction pipeline for a circular aperture radius of 25~pixels. Each data set is shifted upward with respect to the one observed before for visualisation purposes. The raw data points are shown in blue and binned once per \cheops{} orbit (black). The orange solid line is the best-fit model obtained with the modelling described in Section~\ref{sec:lc}. The six observations are represented as a function of time since inferior conjunction according to the best-fit parameter values.}
                \label{fig:raw_lc}
        \end{figure*}

        Apart from the differences described previously, the data sets analysed in this work were obtained in the same conditions.
        The cadence of the observations was one 200$\times$200-pixel sub-array every 4.8\,s (upper limit imposed by the saturation of the CCD) but, in order to be able to transfer data to the Earth with the available downlink bandwidth, images were stacked together on board by groups of seven. This resulted in an effective integration time of 33.6\,s ($=7\times4.8\,\text{s}$) per downloaded frame. The image read-out is performed simultaneously to the next exposure, leading to an effective data cadence equal to the integration time (duty cycle of 100\%). In addition, \cheops{} also provides smaller circular images, called imagettes, with a radius of 25~pixels and at a higher cadence (one imagette every 4.8\,s). These imagettes do not allow us to perform precise aperture photometry and were originally meant for visual inspection and cosmic ray correction, but they can prove to be useful for PSF photometry, as described in Section~\ref{sssec:psf_phot}.
        The read-out frequency at which the CCD pixels were read is 100\,kHz for all \host{} observations.
        The duration of the observations was about 12.5\,hours for each of the eclipses (about 3 eclipse durations for the out-of-eclipse baseline) and about 75\,hours for each of the phase curves (encapsulating a full orbit plus an additional transit).
        The raw light curves extracted by aperture photometry are shown on Figure~\ref{fig:raw_lc}.
        
        \begin{table*}
                \caption{List of the \cheops{} observations.}
                \label{tab:obs}
                \centering
                \resizebox{\textwidth}{!}{
                        \begin{tabular}{cccccccc}
                                \hline\hline
                                \multirow{2}{*}{File key \tablefootmark{1}} & \multirow{2}{*}{UTC start \tablefootmark{2}} & \multirow{2}{*}{UTC end \tablefootmark{2}} & \multirow{2}{*}{Type} & \multirow{2}{*}{$N_\text{frames}$} & Efficiency \tablefootmark{3} & PSF location & $T_\text{CCD}$ \\[.4ex]
                                & & & & & $\left[\%\right]$ & $\left(x, y\right)$ & $\left[^\circ\text{C}\right]$ \\
                                \hline
                                \texttt{CH\_PR100041\_TG000201\_V0200} & 2020-03-19 02:50 & 2020-03-19 15:20 & Eclipse & 736 & 57.8 & $\left(512, 512\right)$ & $-40$ \\
                                \texttt{CH\_PR100041\_TG000202\_V0200} & 2020-03-27 07:27 & 2020-03-27 20:07 & Eclipse & 804 & 61.5 & $\left(664, 296\right)$ & $-45$ \\
                                \texttt{CH\_PR100041\_TG000203\_V0200} & 2020-03-30 01:05 & 2020-03-30 13:29 & Eclipse & 838 & 66.0 & $\left(664, 296\right)$ & $-45$ \\
                                \texttt{CH\_PR100041\_TG000204\_V0200} & 2020-04-07 04:08 & 2020-04-07 17:05 & Eclipse & 943 & 70.6 & $\left(664, 296\right)$ & $-50$ \\
                                \texttt{CH\_PR100036\_TG000701\_V0200} & 2020-06-15 18:07 & 2020-06-18 21:07 & Phase curve & 4597 & 59.5 & $\left(263, 842\right)$ & $-45$ \\
                                \texttt{CH\_PR100036\_TG000702\_V0200} & 2020-06-21 08:25 & 2020-06-24 10:23 & Phase curve & 4393 & 57.7 & $\left(263, 842\right)$ & $-45$ \\
                                \hline\hline
                        \end{tabular}
                }
                \tablefoot{
                        The time spent accumulating photons for each frame (a single frame is here the result of 7 stacked images) is referred to as the integration time and is the same for all observations: $t_\text{int}=33.6\,\text{s}$. We note that the image read-out is performed in parallel of the next exposure leading to an effective data cadence equal to the integration time (duty cycle of 100\%).
                        \tablefoottext{1}{Each file key refers to a unique observation in the \cheops{} database.}
                        \tablefoottext{2}{UTC start and end are the starting and ending dates of the observation in UTC.}
                        \tablefoottext{3}{The efficiency represents the ratio between the observation time without interruptions (due to Earth occultation or SAA crossings) and the total observation duration.}
                }
        \end{table*}
        
        \subsection{Data reduction} \label{ssec:data_red}
                
                \subsubsection{Aperture photometry} \label{sssec:aper_phot}
                        
                        All \cheops{} observations were processed with the latest version of the Data Reduction Pipeline (DRP), tagged version \texttt{13.1.0}. We based the analysis presented in this work on these output photometric light curves that were produced following a procedure described in detail by \cite{hoyer_cheops-drp} and summarised below.
                        
                        The first step of the data processing is referred to as calibration and aims to correct the images for the effects related to the detector and the optics. It consists of subtracting the bias offset of the CCD, converting the digital counts back to electrons (gain), correcting for the non-linear response of the read-out electronics, subtracting the contribution of the dark current, and, finally, correcting for the photo-response non-uniformity (flat field). The reference files on which the calibration correction is based have been produced during the on-ground calibration campaign performed on the \cheops{} payload before launch \citep{chazelas_2019_cheops_calibration, deline_2020_cheops_calibration}.
                        
                        The second step is called the correction step and removes the spurious effects such as hits on the detector by high-energy particles (cosmic rays), background level caused by stray light, and smearing trails of bright close-by stars induced by the CCD read-out.
                        
                        The \cheops{} DRP applies an aperture photometry method on the corrected images that consists of determining the location of the PSF centre (using an iterative centroiding technique) and counting the flux in electrons that falls within a given radius from this centre. The DRP provides photometric light curves for four different aperture radii: a default radius of 25\,pixels, a small radius of 22.5\,pixels, a large radius of 30\,pixels, and an optimal radius that depends on the target observed\footnote{The plate scale of \cheops{} is 1\,arcsecond per pixel.}. The optimal aperture is designed to maximise the signal-to-noise ratio of the photometry and to minimise the contamination from nearby field stars based on simulations generated by the DRP using the Gaia catalogue \citep{gaia_catalogue} and a CHEOPS PSF template.
                        
                        The photometric contamination from nearby stars as estimated by the DRP in the simulations has two components: the direct contamination and the smearing trails. The direct contamination is caused by nearby stars with PSF falling entirely or partly inside the photometric aperture. The associated signal is a positive offset with usually small variations that are in phase with the rolling of \cheops{} around the line of sight. If the offset is significant, it must be accounted for as it dilutes the flux from the target and might lead to underestimate the transit depth for instance. The smearing trails are vertical features above and below the nearby star PSF on the CCD. As the field of view rotates, the nearby stars move on the CCD and the trails enter the aperture periodically, contaminating the photometry. In most cases, both contributions can be corrected as a function of the roll angle of the spacecraft (provided in the light curve products of the DRP).
                        In the case of \host{} observations, the brightest nearby stars have magnitude differences larger than 7.8 in the Gaia G band. They are expected to induce a direct contamination in the aperture smaller than 0.076\% with variations of the order of 0.001\%, and a signal due to smearing trails not greater than 0.021\%. We thus consider that the photometric dilution has a negligible effect on relative photometric features (e.g. transits or eclipses). The associated variations, however, are comparable to the estimated photometric precision of the data (signal-to-noise ratio of the order of unity) and are corrected in the analysis as a function of the roll angle of the spacecraft.
                        
                        During the full reduction process, the DRP runs diagnosis to determine if the data are valid according to several criteria: the location of \cheops{} with respect to the SAA, the values of various thermal sensors, the angular separations between the line of sight and three bodies (the Earth, the Moon, and the Sun), and the number of cosmic ray hits on the detector. If one of these criteria is out of range, the image is flagged accordingly. The DRP team recommends discarding all flagged images from photometric analyses.
                        
                        Figure~\ref{fig:raw_lc} shows the raw light curves obtained for the default aperture (radius of 25\,pixels). For each observation, we estimate the photometric precision by computing the median absolute deviation (MAD) of the flux jumps $df = f_{i+1}-f_i$ to be robust to outliers and to remove correlated signals (e.g. transit, eclipse, stellar activity). The root-mean-square (rms) noise level $\tilde{\sigma}$ of the observation is then derived with the following formula: $\tilde{\sigma}=\text{MAD}/2/\text{erf}^{-1}\!\left(0.5\right)$, with $\text{erf}^{-1}$ being the inverse error function. This technique allows us to find results consistent with the ones computed from best-fit residuals (see Sect.~\ref{ssec:results_perf}). We find that the photometric precision of all observations are very similar, with values at the raw cadence (33.6\,s) for the four eclipses and the two phase curves of 81.3\,ppm, 89.8\,ppm, 90.5\,ppm, 93.9\,ppm, 92.6\,ppm, and 93.3\,ppm, respectively. When applying the same method to several one-hour-long windows, we find median noise levels per one-hour bins of 11.3\,ppm, 11.5\,ppm, 10.8\,ppm, 11.2\,ppm, 12.1\,ppm, and 12.1\,ppm, respectively.

                \subsubsection{PSF photometry} \label{sssec:psf_phot}
                        
                        In addition to the aperture photometry provided by the \cheops{} DRP, we cross-checked our analysis and the results against another photometric extraction technique, PSF photometry. This method consists of fitting the images with a two-dimensional template PSF in order to determine the amplitude of the signal and derive the flux.
                        
                        We used a software package called \pipe, developed specifically for \cheops{}. The code is available on GitHub\footnote{\href{https://github.com/alphapsa/PIPE}{https://github.com/alphapsa/PIPE}} and will also be extensively described in \citepipe.
                        
                        \pipe{} first derives a PSF template library from the observed sub-arrays and imagettes, using a principal component analysis (PCA). The first five principal components (PCs) together with a constant background are then used to best fit the PSF of each image using a least-squares minimisation. The number of PCs to use is a trade-off between following systematic PSF changes and overfitting the noise, but the derived photometry changes only marginally. The principal advantage is instead that the PC coefficients can be used to track PSF changes and correct for the so-called ramp effect (Sect.~\ref{ssec:systematics}).

                        Apart from serving as an independent extraction method, \pipe{} has the advantage of being more robust against cosmic ray hits and telegraphic pixels, since these can be identified and flagged as statistically inconsistent with the PSF and thereby masked. Another advantage with PSF fitting is that accurate photometry can be extracted at faster cadence from the smaller imagettes. Since a large annulus around the target to measure the background is not required, it is simultaneously fitted with the PSF.
                        
                        \pipe{} provides the photometric light curves together with the computed PSF coordinates, the background level and the relative weights $U_X$ of the first five principal components of the PSF PCA. Similarly to the DRP, images can be flagged for several reasons, such as a centroid far away from the median centroid, high level of bad pixels, or poor PSF fit.
                        
                        Following the method described in the previous section, we obtain an estimate of the photometric precision that is very similar to the aperture photometry, albeit slightly better. At the raw cadence (33.6\,s), one obtains 76.9\,ppm, 84.4\,ppm, 86.8\,ppm, 88.6\,ppm, 86.6\,ppm, and 86.6\,ppm for the four eclipses and the two phase curves, respectively. When considering one-hour-long windows, one reaches 10.8\,ppm, 11.0\,ppm, 11.1\,ppm, 10.6\,ppm, 11.5\,ppm, and 11.2\,ppm, respectively.

\section{Light-curve analysis} \label{sec:lc}

        \subsection{Outlier removal} \label{ssec:data_filtering}
                
                Prior to the analysis of the light curves, we perform a filtering of the data by first discarding all points flagged during the photometric extraction process, either aperture photometry from the DRP or PSF photometry from \pipe. The flagged data points represent 458\,/\,12770 images (3.59\%) for the aperture photometry and 493\,/\,12770 (3.86\%) and 2351\,/\,89390 (2.63\%) for PSF photometry on the sub-arrays and imagettes, respectively.
                
\begin{figure}
                        \centering
                        \includegraphics[width=.97\hsize,trim={0cm 0.5cm 0cm 0cm},clip]{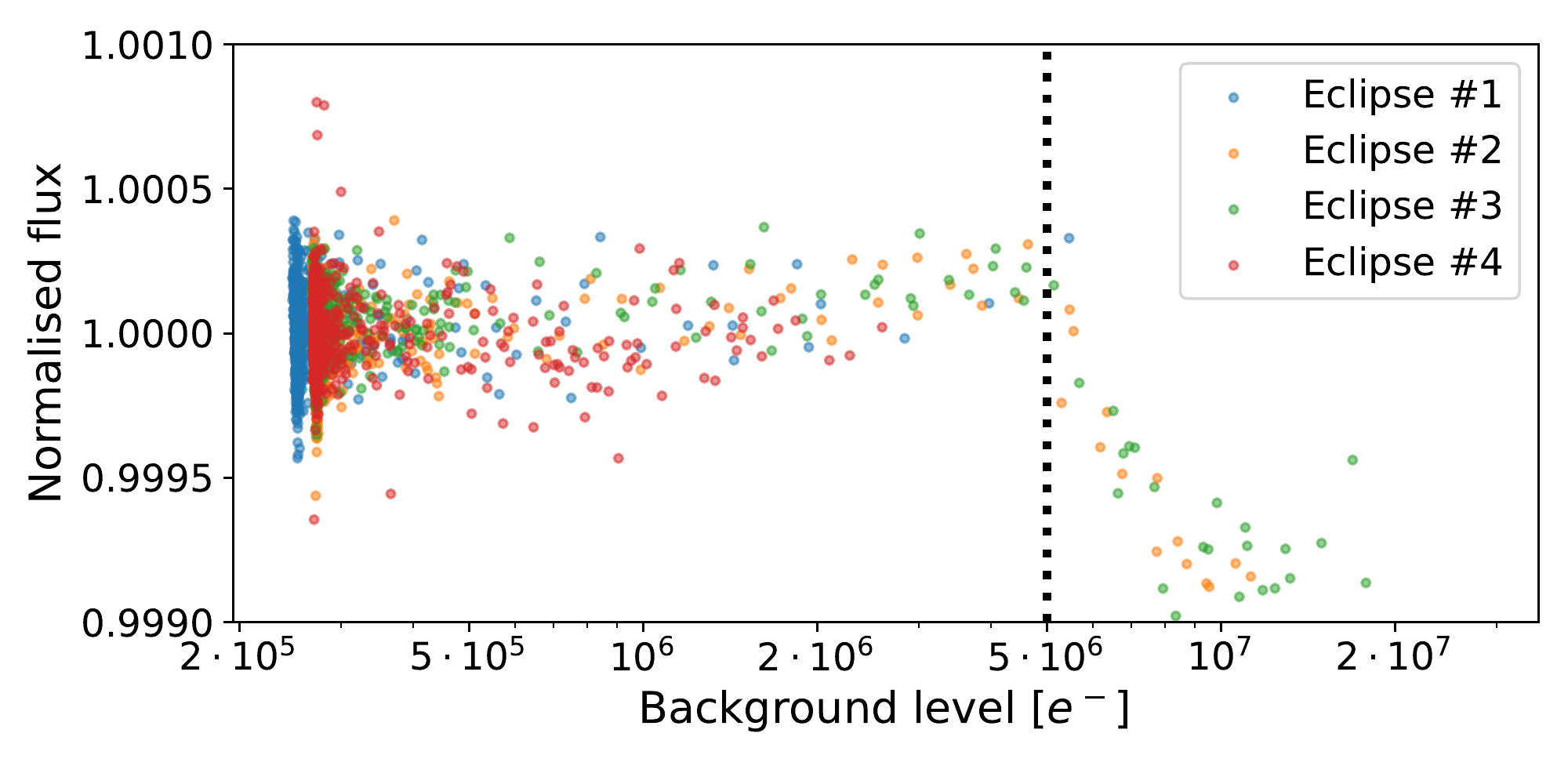}
                        \caption{Flux extracted with aperture photometry (DRP, aperture of 25\, pixels) as a function of the image background levels for the four eclipse observations of \planet{}. This figure highlights the unexpected anti-correlation occurring above a given background value. The vertical dotted black line shows the threshold used to discard data points.}
                        \label{fig:flux_vs_bkg}
                \end{figure}

                In addition, we apply two other criteria to identify data points as outliers and discard them.
                The first criterion is related to the level of background in the images. The change of background level is mostly due to variations in the amount of stray light entering the telescope along the orbit of \cheops, with the main sources of stray light being the Earth and the Moon. All observations are affected by a periodic increase of background level before and after Earth occultations. As mentioned in Section~\ref{ssec:obs}, the second and third eclipse observations of \host{} show much higher background levels (up to 10 times the nominal values) when compared to the other eclipse and phase-curve time series. While inspecting the housekeeping parameters for possible correlations, we noticed that these extreme background values resulted in an unexpected effect on photometry. As shown in Figure~\ref{fig:flux_vs_bkg}, the measured flux becomes suddenly strongly anti-correlated with the background level above a given background value. This effect is present for all photometric apertures (DRP) as well as in the PSF photometry of sub-arrays (\pipe). However, it is not visible in the PSF photometry of imagettes, possibly due to the higher photometric noise of the data linked to the shorter cadence. We checked the moon separation for these observations and it appeared to be greater than $140\deg$, and similar to the one of the first phase curve. The fourth eclipse had the smallest moon separation of the whole data set ($\sim35\deg$) but shows no extreme background levels. As this correlation is not yet understood and being investigated, we decided not to correct for it, but to clip out data points with background levels above a threshold that varies with the extraction technique (see Table~\ref{tab:bkg_thrsh}). This step results in discarding 45\,/\,3321 points (1.36\%) for the two smallest apertures (22.5 and the default 25 pixels), 65\,/\,3321 points (1.96\%) for the two largest apertures (30 and the optimal 40 pixels), and 52\,/\,3310 points (1.57\%) for the PSF photometry on sub-arrays.
                
                The second criterion is a sigma-clipping approach that differs slightly between eclipses and phase curves. Both methods rely on an estimate $\tilde{\sigma}$ of the noise level using a method robust against outliers that fits a Gaussian distribution to the histogram of the data and computes $\tilde{\sigma}$ from the Gaussian width.
                For the eclipses, we normalise each individual observation by its median. We then perform a least-squares fit of the data with a linear trend for each observation and a common eclipse model from \batman{} \citep{kreidberg_batman}. Points beyond $4\,\tilde{\sigma}$ from the median value of the residuals are flagged as outliers.
                For the phase curves, both time series are normalised by their global median value. The model used to de-trend the data is a combination of a symmetric transit model (no GD) from \batman{} and a single slope. For the DRP and \pipe{} sub-array photometry, data points are discarded when off by more than $6\,\tilde{\sigma}$ from the residual median. The sigma-clipping limit is reduced to $5\,\tilde{\sigma}$ for the \pipe{} light curves obtained from imagettes.
                The number of outliers flagged in the aperture photometry is 33\,/\,12267 points (0.27\%), 39\,/\,12267 points (0.32\%), 43\,/\,12247 points (0.35\%), and 55\,/\,12247 points (0.45\%) for the 22.5-, 25- (default), 30-, and 40-pixel (optimal) aperture radii, respectively. For the PSF photometry, we identify 10\,/\,12225 points (0.08\%) as outliers for sub-arrays and 15\,/\,87039 points (0.02\%) for the imagettes.

        \subsection{Flux normalisation} \label{ssec:flux_norm}
                
                The analysis of the photometric time series starts with the normalisation of the measured flux before modelling the systematics and astrophysical signals. As expected, the changes of observation conditions that occurred in-between eclipse observations (see Section~\ref{ssec:obs}) modified the absolute stellar flux measured with \cheops{}. The location of the PSF and the temperature of the detector affect the flux repartition on the CCD (PSF shape) and the conversion factor from photo-electrons to digital units (gain of the read-out electronics), which in turn have an impact on the measured photometry. In order to account for this effect, we decided to include in our model an individual normalisation factor for each of the six eclipse and phase-curve observations that is applied before any of the data analysis steps described hereafter.

        \subsection{Systematic noise} \label{ssec:systematics}
                
                The first systematic noise to be mentioned is specific to \cheops{} and is related to the rotation of the spacecraft around the Earth. In order to guarantee the thermal stability of the detector, the passively cooling radiator of the instrument must never face the Sun nor the Earth, which implies a continuous rolling of the spacecraft around its pointing direction with one full rotation per orbit. This nadir-locked orientation implies a rotation of the field of view around the target PSF, at a rate not necessarily constant, which causes nearby stars and stray light sources to move around and induce a periodic photometric noise. We implement a modelling of this systematic effect using the first terms of the following Fourier series:
                \begin{equation} \label{eq:roll_model}
                        \mathcal{S}_\text{roll}\!\left(\theta_\text{roll}\right) = \sum_{i=1}^{N}{a_i\cos\!\left(i\,\theta_\text{roll}\right) + b_i\sin\!\left(i\,\theta_\text{roll}\right)}
                ,\end{equation}
                 where $\theta_\text{roll}$ is the roll angle of \cheops{} provided in the data, and $a_i$ and $b_i$ are free parameters. We chose to limit the model to the first five terms of the series ($N=5$ in Eq.~\ref{eq:roll_model}) as a trade-off between the quality of the fit, the number of parameters and to avoid overfitting the data. In addition, we implemented one model for the four eclipse observations and another for the phase-curve observations, or in other words, one set of coefficients $\left(a_i, b_i\right)_\text{ecl}$ for the eclipses and one set $\left(a_i, b_i\right)_\text{PC}$ for the phase curves. This decision is motivated by the fact that there is a gap of nearly 70\,days between the eclipse observations and the phase-curve observations. Over this two-month time span, the Earth moves significantly around the Sun, modifying its position with respect to \host{} in the field of view of \cheops. As a consequence, the photometric systematics associated with the roll angle are different and could not be modelled by the same set of coefficients.
                
                Another well-known systematic effect in the \cheops{} data sets is the so-called ramp effect. The source of this ramp is a thermo-mechanical distortion of the telescope tube that induces variations in distance and orientation of various optical components, including in particular the secondary mirror. This distortion is hard to predict as it depends on the orientation of the spacecraft with respect to the Sun during the observation preceding the one being analysed. The ramp has been extensively studied, and it has been found that (a)~its duration can last up to more than ten\,hours with a behaviour similar to that of a thermal relaxation (exponential decay), (b)~its photometric effect is correlated with the shape of the PSF, and (c)~its photometric effect is correlated to the outputs from several thermal sensors, and in particular one referred to as \emph{thermFront\_2}. The study also concluded on a possible modelling of the effect as a linear relationship between the flux and the \emph{thermFront\_2} temperature \citep{maxted_2021_pycheops}. Based on this conclusion, we included a modelling of the ramp effect as a linear relationship between the flux and the telescope temperature as follows:
                \begin{equation} \label{eq:therm_model}
                        \mathcal{S}_\text{therm}\!\left(T_\text{thermFront\_2}\right) = c_\text{therm}\,\Delta T_\text{thermFront\_2}
                ,\end{equation}
                where $T_\text{thermFront\_2}$ and $\Delta T_\text{thermFront\_2}$ are the \emph{thermFront\_2} temperature and its deviation from its median value respectively, and $c_\text{therm}$ is a free parameter defining the strength of the correlation.
                
                To account for other long-term systematic effects, we included a trend model for each observation. This model includes a linear slope with time for each data set, plus an additional quadratic trend for each of the eclipse observations, as shown in the following equation:
                \begin{equation} \label{eq:trend_model}
                        \mathcal{S}_\text{trend}\!\left(t\right) = c_2\,\left(t-t_0\right)^2+c_1\,\left(t-t_0\right)
                ,\end{equation}
                where $t$ is the time, $t_0$ is the mid-time of each observation, $c_2$ and $c_1$ are the quadratic and linear trend coefficients for each observation ($c_2=0$ for the phase curves). The nature of the corrected trends is not clearly determined and could be due to imperfect instrumental long-term stability, stellar activity or both. However, in light of the modelling of stellar activity in the phase curves discussed in Section~\ref{ssec:model_star}, we strongly suspect that most of the trends in the eclipse observations are of stellar origin, but this could not be assessed given the short duration of each observation.
                
                Based on the available housekeeping data, we performed an extensive study on possible correlations between the photometry and all the other parameters. We found no correlation other than the ones mentioned before that required modelling and correction.

        \subsection{Planetary model} \label{ssec:model_planet}
                
                The modelling of the photometric signal from \planet{} along its orbit is decomposed in three contributions that are described below. The first and second contributions are the transit model, for when the planet passes in front of the gravity-darkened star, and the eclipse model, for when the planet is hidden by the star. The third contribution is the phase-curve model that describes the flux received by the observer from the planetary surface as a function of its position around the star.
                
                \subsubsection{Transit model with stellar gravity darkening} \label{sssec:model_transit}
                        
                        The fast-rotating nature of \host{} causes the centrifugal force to have a non-negligible effect with respect to the surface gravity. As a consequence, the effective surface gravity at the stellar equator is smaller than the one at the poles and the star becomes oblate. Based on the von Zeipel theorem \citep{von-zeipel_rotating-stars}, one can show that the radiative flux at a given latitude on the rotating star is proportional to the local effective surface gravity \citep[e.g.][]{maeder_stars_2009}, and this implies the following relation:
                        \begin{equation}
                                \label{eq:temp-grav-beta}
                                T\!\left(\vartheta\right)=T_{\rm pole}\left(\frac{g_{\rm eff}\!\left(\vartheta\right)}{g_{\rm eff,\,pole}}\right)^\beta
                        ,\end{equation}
                        where $T\!\left(\vartheta\right)$ and $T_{\rm pole}$ are, respectively, the temperatures at a given colatitude $\vartheta$ and at the poles ($\vartheta=0$), $g_{\rm eff}\!\left(\vartheta\right)$ and $g_{\rm eff,\,pole}$ are, respectively, the effective surface gravities at the colatitude $\vartheta$ and at the poles, and $\beta$ is the GD exponent. The value of $\beta$ is 0.25 for a purely radiative envelope but can deviate from the theory as measured for the star Altair \citep[$\beta=0.190\pm0.012$, ][]{monnier_altair_2007}. The previous equation shows that, as the stellar rotation reduces the local effective gravity, the equator gets cooler and thus appears darker than the poles. The GD leaves a peculiar photometric signature when a planet transits in front of its star and hides regions with varying brightness, leading to asymmetric transit light curves when the orbit is misaligned.
                        
                        As shown by \cite{lendl_2020_wasp-189b}, the transit light curve of \planet{} shows such GD features and one must account for this effect in the modelling of the data. In our analysis, we make use of \pytransit{}\footnote{\url{https://github.com/hpparvi/PyTransit}} \citep{parviainen_pytransit}, version 2.5.13, which provides gravity-darkened transit models implemented based on \cite{barnes_gravity-darkening}\footnote{We note that there is a typo in Eq.~14 of \cite{barnes_gravity-darkening}, and the terms $(1-f^2)$ should be replaced by $(1-f)^2$.}. The base assumption of various equations of the code is that the gravitational potential follows a Roche model, which is equivalent to assuming that only the outer layers of the star are distorted by rotation, meaning that the inner layers are spherical, hence producing the same gravitational potential as if the whole mass were concentrated at the centre of the star. \pytransit{} represents the star as a discretised oblate sphere and computes the transit luminosity dip with a discretised planetary disc crossing and partially hiding the stellar object projected onto the plane of the sky. The effective surface gravity $g_{\rm eff}$ at each point on the stellar surface is evaluated from the Newtonian gravity force and the centrifugal force:
                        \begin{equation}
                                \label{eq:eff-grav}
                                \overrightarrow{g_{\rm eff}}\!\left(\vartheta\right)=-\frac{G\,M_\star}{r_\vartheta^2}\,\overrightarrow{u_r}+\left(\frac{2\pi}{P_\star}\right)^2R_\star\sin\!\left(\vartheta\right)\,\overrightarrow{u_{\rm x}}
                        ,\end{equation}
                        where $G$ is the universal gravitational constant, $M_\star$ is the stellar mass, $r_\vartheta$ is the distance from the stellar centre of the point considered, $P_\star$ is the rotation period of the star, $R_\star$ is the stellar radius, and $\vartheta$ is the colatitude of the point. The unit vectors $\overrightarrow{u_r}$ and $\overrightarrow{u_{\rm x}}$ point outwards, in the opposite direction to the stellar centre and perpendicularly to the spin axis of the star, respectively. The local stellar radius $r_\vartheta$ equals $R_\star$ at the equator and decreases down to $R_{\rm pole}=R_\star\left(1-f_\star\right)$ at the poles ($\vartheta=0$), where $f_\star$ is the stellar oblateness that can be expressed as a function of the stellar parameters:
                        \begin{equation}
                                f_\star=1-\frac{R_{\rm pole}}{R_\star}=\frac{2 \pi^2 R_\star^3}{2 \pi^2 R_\star^3 + G\,M_\star P_\star^2}=\frac{3 \pi}{2\,G\,\rho_\star P_\star^2}
                        \end{equation}
                        
                        The temperature map of the star is computed from Eqs.~\ref{eq:temp-grav-beta} and \ref{eq:eff-grav} for every discretised surface point. The conversion from temperature to measured flux requires two additional elements, which are the flux emission spectrum $\mathcal{S}\!\left(\lambda, T\right)$ for a given temperature $T$ and the instrument sensitivity or passband $\mathcal{T}_{\rm inst}\!\left(\lambda\right)$. \pytransit{} provides the option to use synthetic spectra from the PHOENIX library \citep{husser_phoenix_2013}, which are more suited than black-body laws to approximate the emission spectra of hot stars such as \host{}. We compute the CHEOPS passband by combining the optical throughput of the telescope and the quantum efficiency of the detector that are both available as reference files in the CHEOPS mission archive\footnote{\url{https://cheops-archive.astro.unige.ch/archive_browser/}}. The measured flux from a given point can then be computed after including the limb-darkening effect at the considered location:
                        \begin{equation}
                                \mathcal{F}\!\left(\vartheta, \mu\right) = \int_{\lambda=0}^{+\infty}\!\mathcal{S}\!\left(\lambda, T\!\left(\vartheta\right)\right) \mathcal{T}_{\rm inst}\!\left(\lambda\right) d\lambda \,\times\,\mathcal{I}\!\left(\mu\right)
                        ,\end{equation}
                        where $\mathcal{F}\!\left(\vartheta, \mu\right)$ is the local flux, $\lambda$ is the wavelength, $\vartheta$ is the colatitude of the point, and $\mu=\sqrt{1-x^2}$ with $x$ the normalised radial coordinate of the point. The term $\mathcal{I}\!\left(\mu\right)$ represents the local attenuation due to the limb darkening and is implemented in \pytransit{} with the quadratic law $\mathcal{I}\!\left(\mu\right)=1-u_1\left(1-\mu\right)-u_2\left(1-\mu\right)^2$, where $u_1$ and $u_2$ are the two limb-darkening parameters.
                        
\begin{figure}
                                \centering
                                \includegraphics[width=.95\hsize,trim={1cm 0.3cm 0.2cm 1cm},clip]{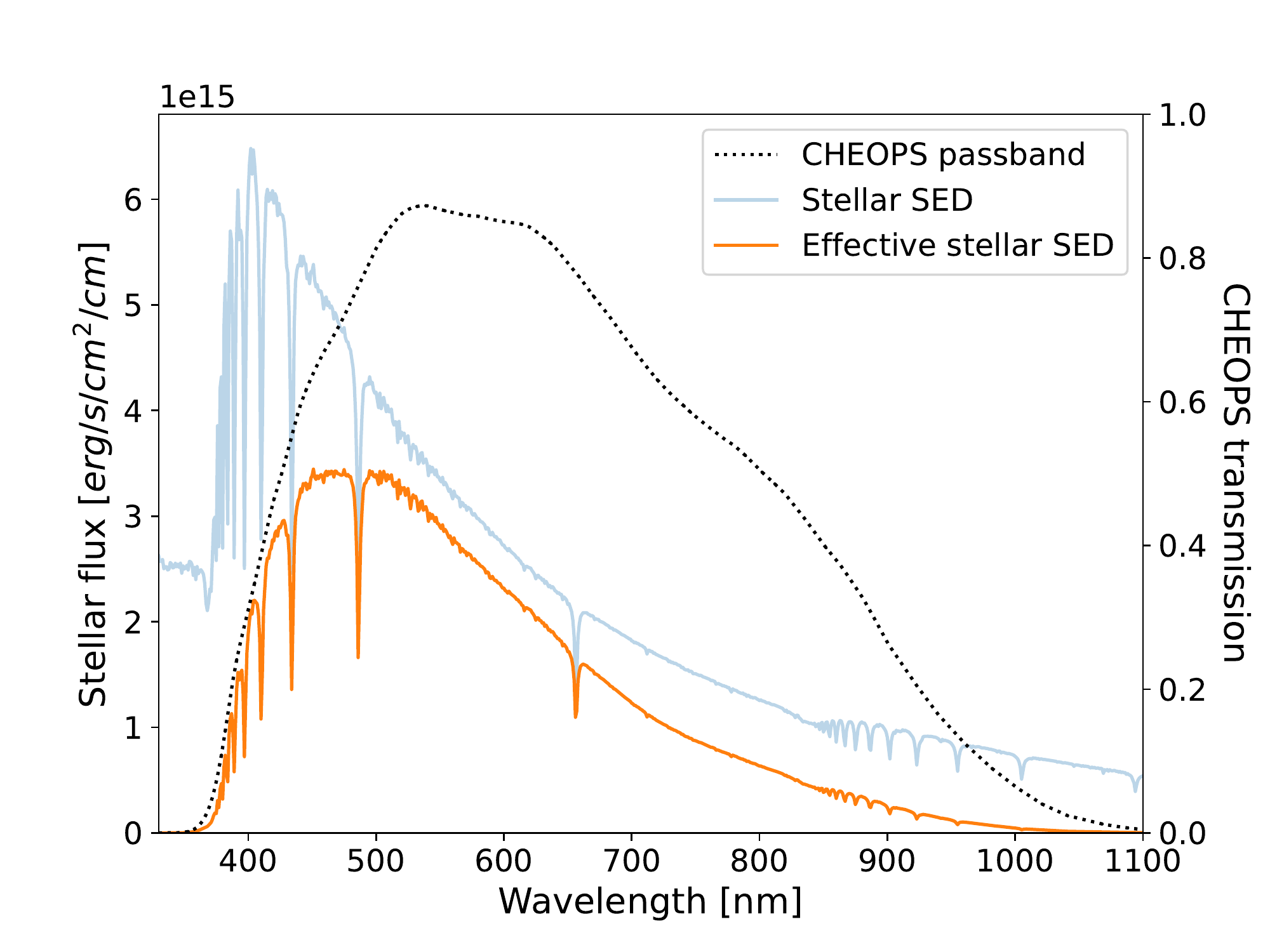}
                                \caption{Synthetic SED from the PHOENIX library \citep{husser_phoenix_2013} of a star similar to \host{}. The stellar properties are an effective temperature $T_\text{eff}=8000\,K$, a surface gravity $\log g=4.0,$ and a solar metallicity. The stellar SED in blue is represented over the \cheops{} passband (330-1100 nm) shown as a black dotted line. The effective SED as seen by the \cheops{} instrument is the solid orange line, for which one can see that most of the energy lies within the 400-800\,nm range and the contribution beyond 1000\,nm is marginal.}
                                \label{fig:star_sed}
                        \end{figure}

                        The implementation of \pytransit{} limits the wavelength range of the PHOENIX spectra $\mathcal{S}\!\left(\lambda, T\!\left(\vartheta\right)\right)$ from 300 to 1000\,nm, while the CHEOPS passband $\mathcal{T}_{\rm inst}\!\left(\lambda\right)$ covers the wavelength range from 330 to 1100\,nm. As shown in Fig.~\ref{fig:star_sed}, the combination of low stellar flux emission and poor CHEOPS sensitivity in the 1000-1100\,nm range makes the contribution of this part of the spectrum in the integrated stellar flux small ($<0.2\%$). Using \pytransit{} and a black-body approximation for \host, we estimated the effect of not accounting for wavelengths longer than 1000\,nm to have an impact on the transit depth smaller than 0.2\,ppm, which we considered to be negligible.
                        
                        The \pytransit{} model we implemented has the following parametrisation: the time of inferior conjunction $T_0$; the period of the planetary orbit $P$; the planet-to-star radii ratio $k=R_p / R_\star$; the normalised semi-major axis of the planetary orbit $a / R_\star$; the inclination of the planetary orbit $i_p$; the eccentricity $e$ and the argument of periastron $\omega$ combined into two parameters $e\cos\!\left(\omega\right)$ and $e\sin\!\left(\omega\right)$; the two coefficients $u_1$ and $u_2$ of the quadratic limb-darkening law $\mathcal{I}\!\left(\mu\right)$; the projected stellar rotation speed $v_\star\sin\!\left(i_\star\right)$; the temperature of the stellar poles $T_{\rm pole}$; the stellar inclination $i_\star$; the projected orbital obliquity $\lambda_p$; the GD exponent $\beta$; the stellar equatorial radius $R_\star$; and the stellar mass $M_\star$.
                        One must note that the time of inferior conjunction $T_0$ might differ from the mid-transit time (mid-time between first and fourth contacts) in the case of oblate stars. For configurations where the orbit is misaligned but not perpendicular to the stellar equatorial plane ($\lambda_p\neq90\deg$) and the impact parameter is not zero ($i_p\neq90\deg$), the oblateness of the star will cause a shift in mid-transit time with respect to $T_0$ due to a late ingress or an early egress depending on the system orientation.
                        The parameter $R_\star$ is the stellar radius at the equator and is the one used to normalise the planet radius $R_p$ and the semi-major axis $a$ and to compute two key parameters of the GD effect: the rotation period and the density of the star.
                        The orientation of the planetary system is fully described by the three angles $i_p$, $i_\star,$ and $\lambda_p$, which need to be unambiguously defined in the case of GD. The convention on angular geometry used in this work is detailed in Fig.~\ref{fig:gd_geometry}. The stellar inclination is allowed to vary from $0\deg$ (north pole on) to $180\deg$ (south pole on). The orbital inclination $i_p$ is also defined on the interval $\left[0, 180\right]\deg$ with $i_p=90\deg$ corresponding to a transit through the center of the stellar oblate disc. The projected orbital obliquity $\lambda_p$ varies in the $\left[-180, 180\right]\deg$ range and affects the orientation of the planet transit path around the centre of the stellar oblate disc.

                \subsubsection{Eclipse model with an oblate star} \label{sssec:model_eclipse}
                
                        Similarly to transits (see Section~\ref{sssec:model_transit}), the shape of the light curve during the occultation of the planet by its host will be affected by the oblateness of the star, causing a shift of the mid-eclipse time with respect to the time of superior conjunction and a change of shape of the ingress and egress. We note that this effect will be even more pronounced for grazing eclipses (and transits), but this does not concern the case of \planet{}.
                        
                        For our analysis, we implemented an eclipse model based on the gravity-darkened transit model from \pytransit{} to be consistent with the oblateness inferred by a given set of parameters. The model is the same as the transit model described in the previous section, with a few changes in the parameter values, as detailed below.
                        The value of the time of inferior conjunction $T_0$ is replaced by the time of superior conjunction, which is half an orbital period after $T_0$ for circular orbits and can be computed from Kepler's equation, relating the mean and eccentric anomalies, for eccentric orbits.
                        The values of the limb-darkening coefficients $u_1$ and $u_2$ and the GD exponent $\beta$ are set to zero to generate a uniform oblate stellar disc.
                        The value of the argument of periastron $\omega$, the inclination of the planetary orbit $i$ and the projected orbital obliquity $\lambda_p$ are modified as if the system were being observed from the other side. This transformation leads to the following sets of parameters: $\omega_\text{ecl}=\omega+\pi$, $i_{p,\,\text{ecl}}=\pi-i_p$, and $\lambda_{p,\,\text{ecl}}=-\lambda_p$.
                        
\begin{figure}
                                \centering
                                \includegraphics[width=.55\hsize,trim={0.2cm 0.3cm 0.2cm 0.1cm},clip]{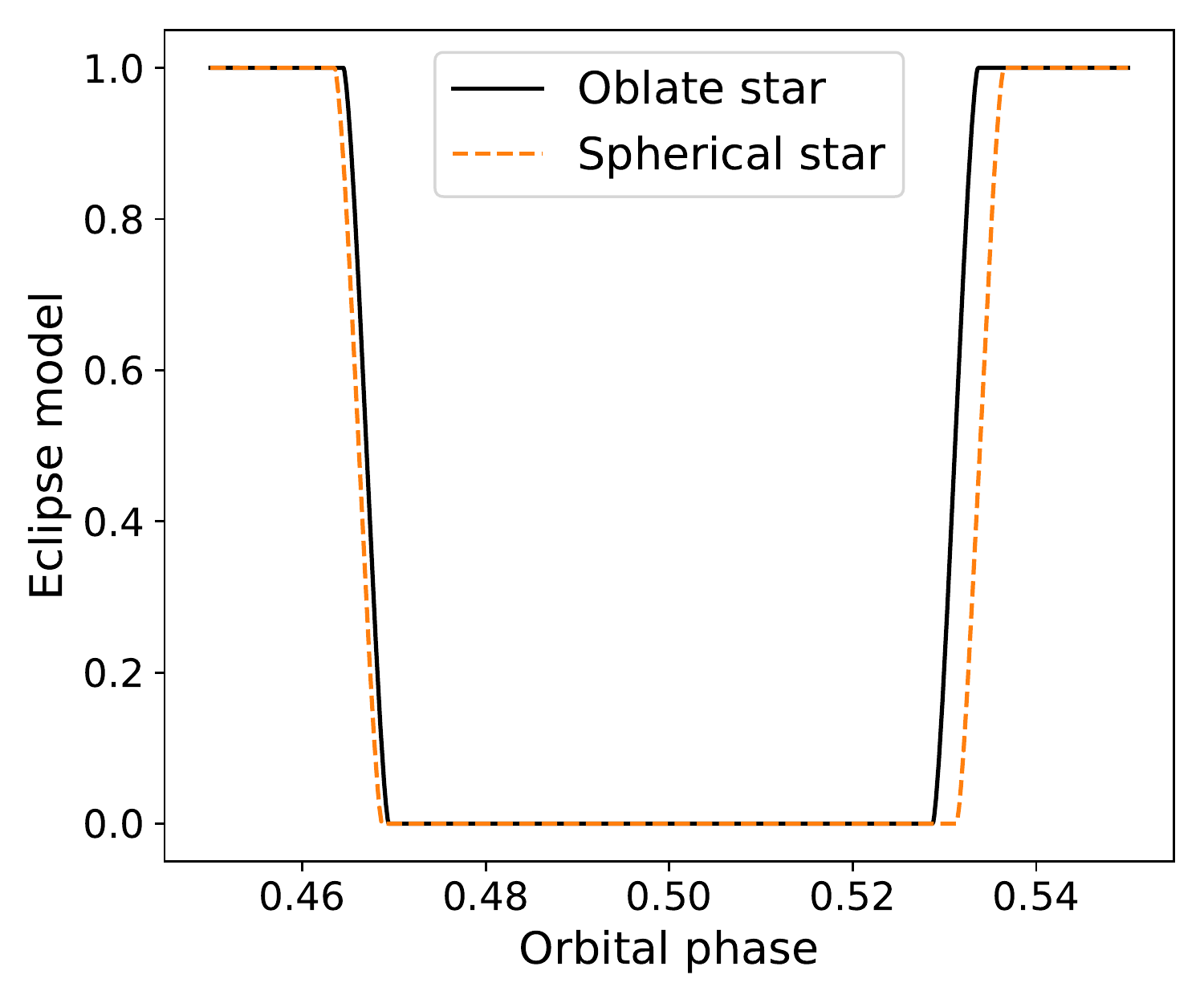}
                                \includegraphics[width=.4\hsize,trim={0.2cm -1.15cm 0.4cm 0.2cm},clip]{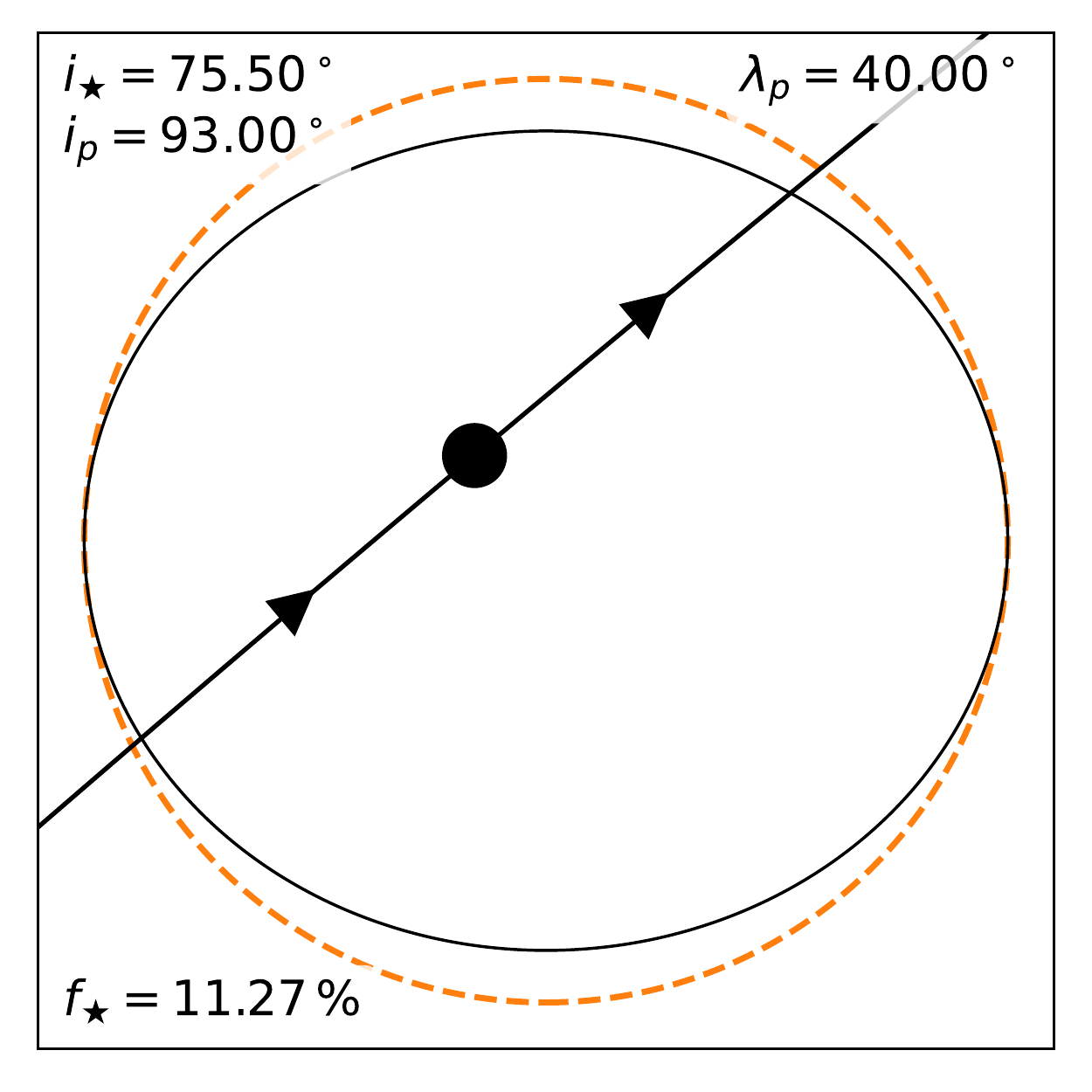}
                                \caption{Eclipse models with and without accounting for the oblateness of the host star. The parameters used in this example are the ones reported in \cite{lendl_2020_wasp-189b} except for the stellar rotation speed (increased by a factor of 2 to enhance the effect of rotation), the projected orbital obliquity (set to $40\deg$), and the orbital inclination (set to $93\deg$). The left panel shows the light curves normalised by the out-of-eclipse flux for the same system orientation with an oblate stellar disc (solid black line) and a spherical star (orange dashed line). On the right we show a projected view of the system during the eclipse event. The black ellipse and the orange dashed circle show the limb of the star in both cases with the same colour code. The path of the planet is represented by the straight black line and arrows, with the planet being to scale (small black disc).}
                                \label{fig:gd_eclipse_model}
                        \end{figure}

                        The generated light curve corresponds to a transit in front of a uniform oblate stellar disc and has to be normalised to obtain the eclipse model for a fast-rotating star. The normalisation is performed so that the flux when the planet is fully occulted is 0, and the out-of-eclipse flux is equal to 1 (see Fig.~\ref{fig:gd_eclipse_model}). The normalised eclipse light curve is then multiplied by the phase-curve signal computed in the next section, which defines the eclipse depth.
                        
                \subsubsection{Phase-curve signal} \label{sssec:model_pc}
                        
                        The phase-curve model used in this work is analytical and assumes the planet is a perfect Lambertian reflector (isotropic scattering) with a given geometric albedo $A_g$, and the thermal emission is approximated by a sinusoidal function of the planetary phase.
                        To describe the orbital phase of the planet, we defined the phase angle of the planet $\alpha=\arccos\!\left[-\sin\!\left(\omega+\nu\right)\sin\!\left(i_p\right)\right]$, where $\omega$ is the argument of periastron, $\nu$ is the true anomaly, and $i_p$ is the orbital inclination. The values of the phase angle range between 0 at superior conjunction and $\pi$ at inferior conjunction.
                        The reflected flux component due to the isotropic reflection of the stellar light off the planetary atmosphere can be written analytically \citep{sobolev_scattering_1975, charbonneau_tau_bootis_1999}:
                        \begin{equation}
                                \label{eq:pc_lamb_ref}
                                \frac{F_\text{refl}}{F_\star} = A_g\,\left(\frac{R_p}{a}\frac{1+e\cos\!\left(\nu\right)}{1-e^2}\right)^2\,\frac{\sin\!\left(\alpha\right)+\left(\pi-\alpha\right)\cos\!\left(\alpha\right)}{\pi}
                        ,\end{equation}
                        where $F_\star$ is the stellar flux, $A_g$ is the geometric albedo, $R_p$ is the planetary radius, $a$ is the semi-major axis, $e$ is the eccentricity, $\nu$ is the true anomaly, and $\alpha$ is the phase angle.
                        The thermal emission flux is approximated by the following function of the phase angle:
                        \begin{equation}
                                F_\text{therm} = \left(F_\text{day}-F_\text{night}\right)\frac{1+\cos\!\left(\alpha_\text{therm}\right)}{2}+F_\text{night}
                        ,\end{equation}
                        where $F_\text{day}$ and $F_\text{night}$ are the planetary fluxes of the dayside and the nightside, respectively, and
                        \begin{equation}
                                \alpha_\text{therm}=\arccos\!\left[-\sin\!\left(\omega+\nu-\phi_\text{therm}\right)\sin\!\left(i_p\right)\right]
                        ,\end{equation}
                        with $\phi_\text{therm}$ being the phase shift of the thermal emission accounting for hotspot offset.
                        In total, our phase-curve model makes use of four parameters in addition to the ones already provided to the transit model described in Section~\ref{sssec:model_transit}: the geometric albedo $A_g$, the planet dayside and nightside fluxes, and the hotspot offset $\phi_\text{therm}$.
                        
                        In the framework of this analysis, we also used another phase-curve model with a more complex implementation. The reflective component was more generic and allowed a divergence from a Lambertian profile as described in \cite{heng_reflected_light_2021}. The thermal emission of the planet was computed from 2D temperature maps and integrated in the \cheops{} passband as detailed in \cite{morris_2021_temp_maps}. This approach involved more free parameters and provided inconclusive results: the reflective component was consistent with a Lambertian profile and the thermal map was not constrained mostly due to the strong degeneracy between reflected light and thermal flux. We thus opted for the model with a Lambertian reflector and a sinusoidal thermal phase curve.
                        
                        In addition to the reflective and thermal flux of the phase-curve model, we implemented the possibility to fit for the ellipsoidal variations \citep{mazeh_2008_tidal_interaction} and the Doppler beaming \citep{maxted_2000_kpd_1930+2752}, both approximated by sinusoidal functions:
                        \begin{align}
                                &F_\text{ell} = 2\,A_\text{ell}\,\cos^2\!\left(\omega+\nu\right)\,\sin\!\left(i_p\right) ,\\
                                &F_\text{beam} = A_\text{beam}\,\cos\!\left(\omega+\nu+\pi\right)\,\sin\!\left(i_p\right),
                        \end{align}
                        where $\nu$ is the true anomaly, $\omega$ is the argument of periastron, $i_p$ is the orbital inclination, and $A_\text{ell}$ and $A_\text{beam}$ are the semi-amplitudes of the ellipsoidal variations and the Doppler beaming, respectively.
                        
                        The combined model of the light curve including the transit and eclipse models can be expressed as follows:
                        \begin{equation}
                                F_p = F_\text{tra} + \left(F_\text{refl}+F_\text{therm}\right) \times F_\text{ecl} + F_\text{ell} + F_\text{beam}
                        ,\end{equation}
                        where $F_\text{tra}$ is the gravity-darkened transit light curve, $F_\text{refl}$ and $F_\text{therm}$ are the reflected light and thermal emission from the planet, $F_\text{ecl}$ is the normalised eclipse model, and $F_\text{ell}$ and $F_\text{beam}$ represent the contributions from ellipsoidal variations and Doppler beaming.

                \subsubsection{Light-travel time} \label{sssec:model_ltt}
                        
                        The light-travel time (LTT) across the planetary system is accounted for in the model used in this work. The observation times are converted into reference times by correcting for the light-travel time along the projected distance between the current planet position and its position at inferior conjunction. This choice of reference frame allows us to synchronise the time of inferior conjunction $T_0$ in both time frames. For eccentric orbits, such a correction is slow as it has to be solved numerically, but fortunately it simplifies into an analytical formula for circular orbits:                         
                        \begin{equation}
                                \label{eq:ltt}
                                t_\text{ref} = t_\text{obs} - \frac{a}{c}\,\left[1-\cos\!\left(2\pi\frac{t_\text{obs}-T_0}{P}\right)\right]\,\sin\!\left(i_p\right),
                        \end{equation}
                        where $t_\text{ref}$ is the time corrected for LTT, $t_\text{obs}$ is the observation time, $a$ is the semi-major axis, $c$ is the speed of light, $T_0$ is the time of inferior conjunction, and $P$ and $i_p$ are the orbital period and inclination. In this work, we always use Eq.~\ref{eq:ltt} for LTT correction, even in the cases of non-zero eccentricity, as the orbit of \planet{} is expected to be close to circular ($e\sim0$). This approximation avoids the use of slow numerical implementation. The expected amplitude of the correction is of the order of 50\,seconds, which is the LTT between superior and inferior conjunctions.
                        In practice, the transit, eclipse, and phase-curve models (Sections~\ref{sssec:model_transit}, \ref{sssec:model_eclipse}, and \ref{sssec:model_pc}) of data points observed at times $t_\text{obs}$ are computed using the corresponding LTT-corrected times $t_\text{ref}$.

        \subsection{Stellar variability} \label{ssec:model_star}
                
\begin{figure}
                        \centering
                        \includegraphics[width=1\hsize,trim={0.2cm 0cm 2cm 0.2cm},clip]{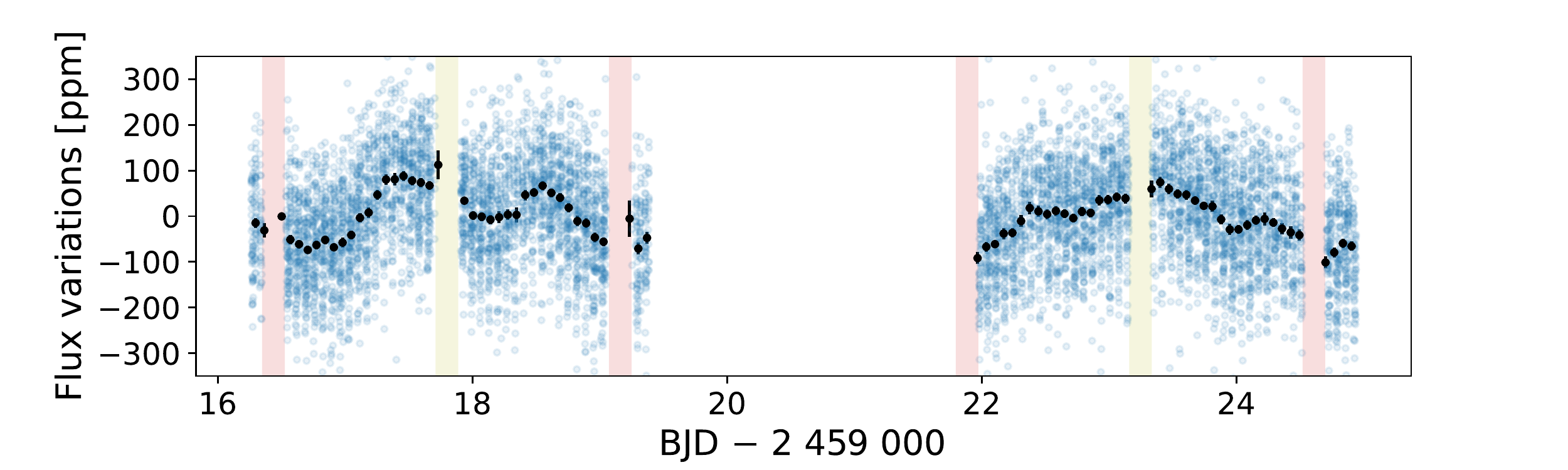}
                        \includegraphics[width=1\hsize,trim={0.2cm 0cm 2cm 1.5cm},clip]{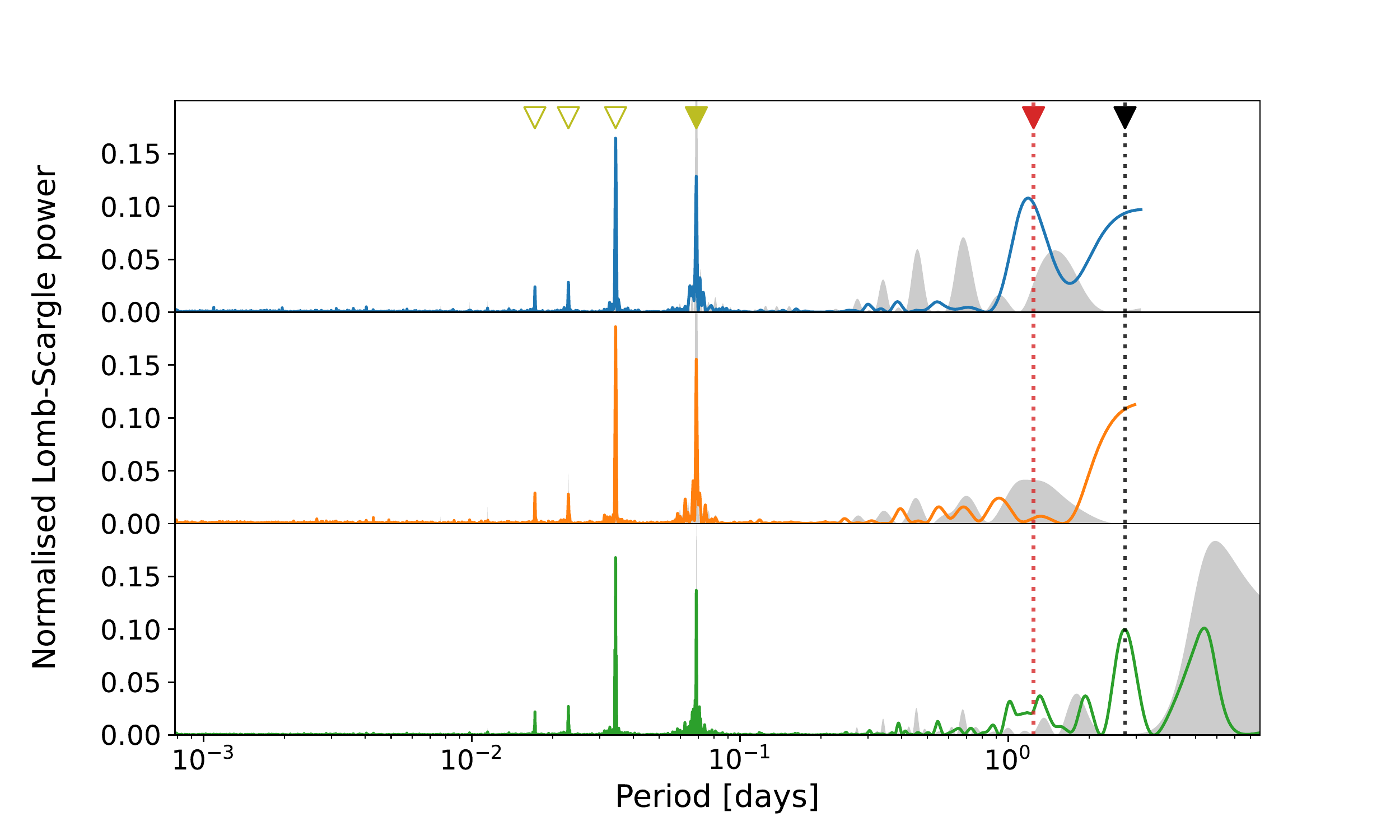}
                        \caption{Photometric variability observed in the two \planet{} phase-curve time series.
                        \emph{Top.} Raw flux (blue points) variations in ppm around the median value after removing outliers, de-trending for linear slope and hiding in-transit (light red shaded area) and in-eclipse (light yellow shaded area) data. Black points are raw flux variations binned once per \cheops{} orbit (98.90\,min).
                        \emph{Bottom.} Lomb-Scargle periodograms of the raw flux variations shown above. The top panel represents the power spectrum of the first phase curve observed for BJD\,<\,2\,459\,020, while the mid panel corresponds to the second phase-curve observation (BJD\,>\,2\,459\,021). The bottom panel is the periodogram of the combined phase-curve time series. The period range covered by each periodogram goes from the Nyquist period (twice the sampling time) to the duration of the time series, and thus encompasses all detectable periodicities. The coloured triangles in the top panel mark the periods of interest. The four leftmost ones (yellow) represent the fundamental (filled triangle) and harmonic (empty triangles) frequencies of the \cheops{} orbital period. The rightmost triangle (black) and the associated vertical dashed line correspond to the orbital period of \planet{}. The red triangle and dashed line mark the stellar rotation period of 1.24\,days computed from \cite{lendl_2020_wasp-189b}. The grey shaded area corresponds to the spectral window of the observation (spectral power induced by the sampling and the gaps of the data).}
                        \label{fig:pc_raw_variability}
                \end{figure}

                In addition to the instrumental systematics and the planet-related signal, the photometric time series feature another flux variability that we studied carefully before including a model for it.
                Fig.~\ref{fig:pc_raw_variability} shows the raw flux variations after the removal of outliers (see Section~\ref{ssec:data_filtering}) and the de-trending of a linear slope, where the variability is visible on top of the phase-curve signal from the planet. The first striking aspect is the absence of variability in the second phase curve while its detection is unambiguous in the first one, as revealed by the Lomb-Scargle periodograms \citep{lomb_1976, scargle_1982} visible in the same figure. In addition, the maximum of the peak is located at a period of 1.19\,days with a full width at half maximum of 0.43\,days, which is consistent with the stellar rotation period of 1.24\,days computed based on \cite{lendl_2020_wasp-189b}. We computed the spectral window of the observations to see how the sampling and gaps in the data impact the periodogram. The spectral window shows peaks corresponding to the orbital period of CHEOPS and its harmonics. This means these peaks are induced jointly by the photometric modulation with the roll angle of the spacecraft and the observation gaps due to Earth occultation and SAA crossings. Other peaks at longer periods are visible in the spectral window of the observations but none matches the signal at 1.19\,days (see Fig.~\ref{fig:pc_raw_variability}).
                
                The spectral type and temperature of \host{} places it at the limit between stars with and without chromospheres and coronae \citep{fossati_2018_xuv_A-stars}. Therefore, the nature of the stellar outer envelope is not well defined and might be either convective or radiative, which in turn implies the presence or absence of stellar spots, respectively. In the former case and the presence of spots, the stellar rotation is expected to create a photometric signature that could resemble the one observed in this time series. On the other hand, if the outer envelope is radiative, the lack of strong superficial magnetic activity leads to an absence of stellar spots and another mechanism is necessary to explain photometric variability in phase with the stellar rotation. Recent asteroseismologic studies \citep{balona_hot_stars_spots_2019, trust_A-star_rotation_2020} based on long-term observations with the \kepler{} space telescope \citep{koch_kepler, borucki_kepler} and the Transiting Exoplanet Survey Satellite \citep[\tess{} --][]{ricker_tess} argue in favour of the presence of inhomogeneities of unknown origin at the surface of hot stars that could explain photometric variability matching the stellar rotation. As shown in the bottom panel of Fig.~2 of \cite{trust_A-star_rotation_2020}, the photometric variability of the A-type star KIC\,6192566 measured by \kepler{} is remarkably similar to the one observed for \host, with a strong modulation matching the stellar rotation period that nearly disappears on a timescale of a few days, before reappearing. \cite{lee_2020_rotating_cores_pulsations} recently provided another possible explanation for such a photometric variability. They show that, when the convective core of an early-type star rotates slightly faster than its radiative envelope, it can excite non-radial pulsations (gravity modes) through resonance couplings. The frequency of the oscillation would then be at the rotating speed of the core, which is almost identical to the one of the outer layers.
                
                The properties of \host, namely its surface gravity $\log g$ and its effective temperature $T_\text{eff}$, places it in the instability strip where about 50\% of the stars are expected to be hybrid $\delta$\,Scuti\,-\,$\gamma$\,Doradus pulsators \citep{uytterhoeven_2011_A-F_star_variability}. $\delta$\,Scuti stars exhibit pulsations with periods typically ranging from dozens of minutes to several hours, while the $\gamma$\,Doradus variability is on longer timescales ranging from half a day to a few days. Therefore, $\gamma$\,Doradus pulsation modes could also explain the photometric variability seen in the \cheops{} phase curves. However, the surface gravity and effective temperature of the star are compatible with those of hybrid pulsators presenting both types of pulsations, albeit more in favour of $\delta$\,Scuti modes in this case. Based on the Lomb-Scargle periodograms shown in Fig.~\ref{fig:pc_raw_variability}, peaks with periods shorter than six\,hours are exact harmonics of the \cheops{} orbital period and cannot be attributed to $\delta$\,Scuti pulsations. This leaves only the mode at 1.19\,days to be compatible with $\gamma$\,Doradus pulsations, but it seems unlikely that only a single frequency would be visible. For completeness, we ran a non-adiabatic oscillation computation of a stellar model of \host{} to determine which types of modes are expected to be excited, as well as their oscillation frequencies. We used the non-adiabatic code MAD \citep{dupret_2001_mad, dupret_2005_mad}, with the inclusion of a time-dependent treatment for the interaction of convection with pulsation. The results show $\delta$\,Scuti oscillation modes are unstable for this stellar model at frequencies between 225 and 500\,\textmu Hz, or periods between 30 and 75\,min, but none in the $\gamma$\,Doradus regime that could explain the peak observed in the data.
                
                Based on this study on the possible origin of the photometric variability observed in the \cheops{} phase curves of \planet{}, we attributed the signal to the stellar rotation and decided to model it with a Gaussian process (GP) to account for its imperfect periodic nature. We implemented the GP model based on the code \texttt{celerite} \citep{foreman_celerite_2017} and, more specifically, using the kernel corresponding to a stochastically driven damped harmonic oscillator (SHO kernel). This kernel is efficient at modelling quasi-periodic oscillations and is defined by three hyper-parameters $S_0$, $Q$ and $P_0$ that drive the amplitude, the damping and the period of the oscillations, respectively. $Q$ is called the quality factor and has to be greater than 0.5 for oscillations to occur. In this case, the damped period of the oscillations is given by $P_\text{damped}=2Q/\!\sqrt{4Q^2-1}\,P_0$, which can be used to estimate the period really fitted by the GP model.
                
                The GP is fitted simultaneously with all other components of the model described previously. However, a Gaussian process is not a parametric function as its output also depends on the fitted data. Therefore, at each iteration the GP model is applied to the data after normalisation and subtraction of all parametric components of the model. The likelihood of the global model to represent the data is computed during this last step and involves the inversion of the covariance matrix generated from the kernel and the error bars of the data points. During the likelihood computation, an additional parameter $\sigma_w$ acting as a jitter term is added quadratically to the residual error bars (diagonal of the covariance matrix) in order to account for possible underestimation of uncertainties in the data.

        \subsection{Prior probabilities of the model parameters} \label{sssec:priors}
                
                To fit our model to the data, we compute the probability of the full model (normalisation factor, systematic noise, planet-related signal and stellar variability) considering the measured data set. The probability is computed by combining the likelihood of all points to be represented by the model for a given set of parameter values, and the prior probabilities placed on the model parameters. We sample the posterior probability using a MCMC algorithm based on the code \text{emcee} described in \cite{foreman_emcee}.
                
                We describe here the choice of prior probabilities made for the models described in Sections~\ref{ssec:flux_norm}, \ref{ssec:systematics}, \ref{ssec:model_planet}, and \ref{ssec:model_star}. Most of the parameters have broad uniform priors around the expected values with the only aim of reducing the size of the explorable parameter space and facilitate convergence of the fit. Choices of parameter prior probabilites motivated by other factors are detailed below. All prior probabilities are listed in Table~\ref{tab:result_parameters_lamb} alongside the best-fit parameter values.

                \subsubsection{Eccentricity and argument of periastron}
                        
                        In the final analyses presented in this work, the eccentricity $e$ used for the computation of the planet-related model is fixed to zero. This is justified by the fact that the orbits of ultra-hot Jupiters are expected to reach long-term stability on circular configuration. To support this choice of setting the eccentricity to zero, we ran a fit with free eccentricity values. We obtained from the results an upper limit at $3\sigma$ (99.93\% confidence) on the eccentricity of 0.027. The largest values of $e$ were reached for specific orbital configurations with the line of apsides along the line of sight ($\omega=90\pm180\deg$). When discarding these cases, the upper limit at $3\sigma$ on the eccentricity dropped to 0.005.

                \subsubsection{Limb-darkening coefficients}
                        
                        The values of the limb-darkening coefficients $u_1$ and $u_2$ are constrained by boundaries ensuring physically meaningful values of the intensity on the stellar disc, which correspond to a positive intensity everywhere on the stellar disc and a monotonically decreasing intensity towards the stellar limb. These conditions translate into three inequalities in the case of the quadratic limb-darkening law as detailed in \cite{kipping_ld}: $u_1+u_2 \leq 1$, $u_1 \geq 0$ and $u_1+2u_2 \geq 0$.
                        In addition to these boundaries, we included a prior probability on the coefficients, which corresponds to the likelihood of the quadratic limb-darkening law $\mathcal{I}\!\left(\mu\right)$ for a given pair $\left(u_1, u_2\right)$ to represent a theoretical stellar intensity profile. This intensity profile is generated using the code \texttt{LDCU}\footnote{\url{https://github.com/delinea/LDCU}} , which computes limb-darkening profiles and coefficients from two libraries of synthetic stellar spectra, ATLAS \citep{kurucz_atlas_1979} and PHOENIX \citep{husser_phoenix_2013}, following the method detailed by \cite{espinoza_ld}. In order to account for uncertainties $\sigma_i$ on the stellar parameters, several profiles are generated by drawing stellar parameter set from normal distributions $\mathcal{N}\!\left(\mu_i, \sigma_i^2\right)$. Each profile is interpolated with a cubic spline over 100 evenly-spaced $\mu$ points as done by \cite{claret_bloemen_gd-ld_coeffs} to uniformly spread the weighting of the profile across the stellar disc. The likelihood $\mathcal{L}_\text{LD}$ is computed as a function of the $\chi^2$, which is the sum of the quadratic distance of the theoretical profile points to the evaluated profile $\mathcal{I}\!\left(\mu\right)$, and the minimum $\chi^2$ value corresponding to the best possible fit of the theoretical profiles: $\ln\!\left(\mathcal{L}_\text{LD}\right)=\left(1-\chi^2/\chi^2_\text{min}\right)/2$. This value is added to the global log-likelihood value and acts as a penalty for coefficients that would generate a limb-darkening profile completely different from the best-fit to the synthetic atmospheric models. The inclusion of the limb-darkening likelihood in the MCMC runs was tested and validated as it resulted in a faster convergence of the exploration and a negligible effect on the final parameter values.

                \subsubsection{Gravity-darkening parameters}
                        
                        Prior probabilities are set on the parameters used to constrain the GD effects in the transit and eclipse models. The projected stellar rotation speed $v_\star\sin\!\left(i_\star\right)$, the temperature of the stellar poles $T_{\rm pole}$, the stellar equatorial radius $R_\star,$ and the stellar mass $M_\star$ have normal priors corresponding to the values and uncertainties listed in Table~\ref{tab:star}, with the temperature of the poles equal to the effective temperature.
                        We let the stellar inclination $i_\star$ vary uniformly in the range $\left[0-180\right]\deg$.
                        We place a normal prior on the projected orbital obliquity $\lambda_p$ based on the value derived from Doppler tomography by \cite{anderson_2018_wasp-189b} ($\lambda_p=89.3\pm1.4\deg$).
                        Finally, we fix the GD exponent $\beta$ to the value provided in Table~2 of \cite{claret_gd_2016} based on the stellar effective temperature, which gives $\beta = \beta_1/4 = 0.22$.
                        
                        We note that, when allowed to vary freely, the parameters $\beta$ and $\lambda_p$ converge to values that are unphysical and inconsistent with Doppler tomography, respectively. This degeneracy justifies the use of prior probabilities from Doppler tomography as detailed in Sect.~5 of \cite{hooton_mascara-1b}.

                \subsubsection{Phase-curve parameters}
                        
                        In the phase-curve model, we let the geometric albedo $A_g$ vary freely within the theoretical Lambertian range $\left[0, 2/3\right]$ and the hotspot offset $\phi_\text{therm}$ within the range $\left[-60, 60\right]\,\deg$. We constrain the dayside flux $F_\text{day}$ to positive values. We first ran an analysis allowing non-zero nightside flux that yields $F_\text{night}=9.1_{-6.8}^{+14.3}\,\text{ppm}$. We therefore fix its value to zero.
                
                \subsubsection{Ellipsoidal variations and Doppler beaming}
                        
                        The amplitudes of the ellipsoidal variations and the Doppler beaming are both fixed to zero.
                        When fitted for, the amplitude of the ellipsoidal variations are degenerate with the GP modelling the stellar variability, especially in the second phase curve where the variability is nearly absent, and we obtain $A_\text{ell}=25_{-16}^{+21}$\,ppm. Following the formula from \cite{esteves_optical_pc_2013} and using values from \cite{lendl_2020_wasp-189b}, we can estimate the order of magnitude of the semi-amplitude of the ellipsoidal variations:
                        \begin{equation}
                                A_\text{ell}\approx\alpha_\text{ell}\frac{M_p\sin\!\left(i_p\right)}{M_\star}\left(\frac{R_\star}{a}\right)^3\approx10\,\text{ppm}
                        ,\end{equation}
                        where $\alpha_\text{ell}$ is a coefficient of the order of unity and $M_p$ is the mass of \planet{}. The value we obtain by fitting \cheops{} data is much larger than this estimation, and it can be explained by the fact that large amplitudes $A_\text{ell}$ create a signal in the second phase curve that mimics the variability modelled by the GP in the first phase curve. The global model has a higher likelihood given the ability of the GP to correct for this induced signal, even though its amplitude is not realistic. We thus decide to fix $A_\text{ell}$ to zero to prevent this effect from biasing the results of the final analyses.
                        The amplitude of the Doppler beaming is consistent with zero when left free, so we remove a degree of freedom by fixing it to zero as well. We validate this choice by computing the expected amplitude following \cite{esteves_optical_pc_2013}: $A_\text{beam}\approx\alpha_\text{beam}\,K/c\approx1\,\text{ppm,}$ where $\alpha_\text{beam}$ is of the order of unity, $K$ is the radial-velocity semi-amplitude, and $c$ is the speed of light.
                        
                        In addition, we performed model comparison by including Ellipsoidal variations, Doppler beaming, or both phase-curve components. We compared the different best-fit models using the Bayesian information criterion (BIC) and the Akaike information criterion (AIC). The model that minimises both the BIC and AIC is the one without Ellipsoidal variations or Doppler beaming. We report here the BIC and AIC differences with the best model: $\Delta\text{BIC}=20.5$ and $\Delta\text{AIC}=13.1$ for Ellipsoidal variations only, $\Delta\text{BIC}=9.4$ and $\Delta\text{AIC}=2.0$ for Doppler beaming only, and $\Delta\text{BIC}=18.0$ and $\Delta\text{AIC}=3.2$ when including both. These numbers further validate the choice to discard the contribution of Ellipsoidal variations and Doppler beaming from our final model.

                \subsubsection{Gaussian process}
                
                        The last set of prior probabilities are placed on the Gaussian process hyper-parameters. The amplitude and damping hyper-parameters $S_0$ and $Q$ are sampled logarithmically and allowed to vary in an interval aimed uniquely to reduce the size of the parameter space and improve convergence speed of the MCMC run. The undamped period of the oscillations $P_0$ is constrained by a normal prior determined from the oscillation peak identified in the Lomb-Scargle periodogram (see Fig.~\ref{fig:pc_raw_variability}), which corresponds to a distribution of $1.2\pm0.2$\,days.

\section{Results and discussion} \label{sec:results}
        
        In this section, we present the results we obtained from the analyses of the data sets. We performed several fits on all available light curves obtained with either aperture or PSF photometry. This represented a total of six light curves, with four aperture sizes provided by the DRP and two \pipe{} time series extracted from sub-array images and at higher cadence from imagettes. The results obtained from all light curves lead to consistent values for every planetary parameter. The best precision was reached in the cases of aperture photometry with the so-called default aperture size (aperture radius of 25\,pixels) and PSF photometry on the sub-array images. Unfortunately, the gain in cadence provided by the imagettes did not provide additional constraints on the parameter values due to a too significant loss of precision.
        Even though the outcomes of the default aperture photometry and PSF photometry on sub-arrays were extremely similar, we noticed a slight improvement with the latter with some of the system properties better constrained. Therefore, in this work we present the best outcome of our analysis obtained with the MCMC analysis of the light curves extracted with \pipe{} from the 200$\times$200-pixel subarray images.
        The MCMC sampling of the posterior distribution was performed using 128 chains with burn-in phases longer than 20\,000\,steps to ensure convergence of the algorithm, and sampling phases of 16\,384\,steps.
        
        \subsection{Planetary parameters} \label{ssec:results_params}
        
                The values of the main parameters of our best fit are listed in Table~\ref{tab:result_parameters_lamb}. They are related to the architecture of the planetary system, the stellar activity or the ramp effect. The other parameters used to normalise and de-trend the light curve against time and roll angle have their best-fit values listed in Table~\ref{tab:result_other_parameters}.
                
\begin{figure*}
	                \centering
                        \includegraphics[width=.38\hsize,trim={0.4cm 0.4cm 0.38cm 0.35cm},clip]{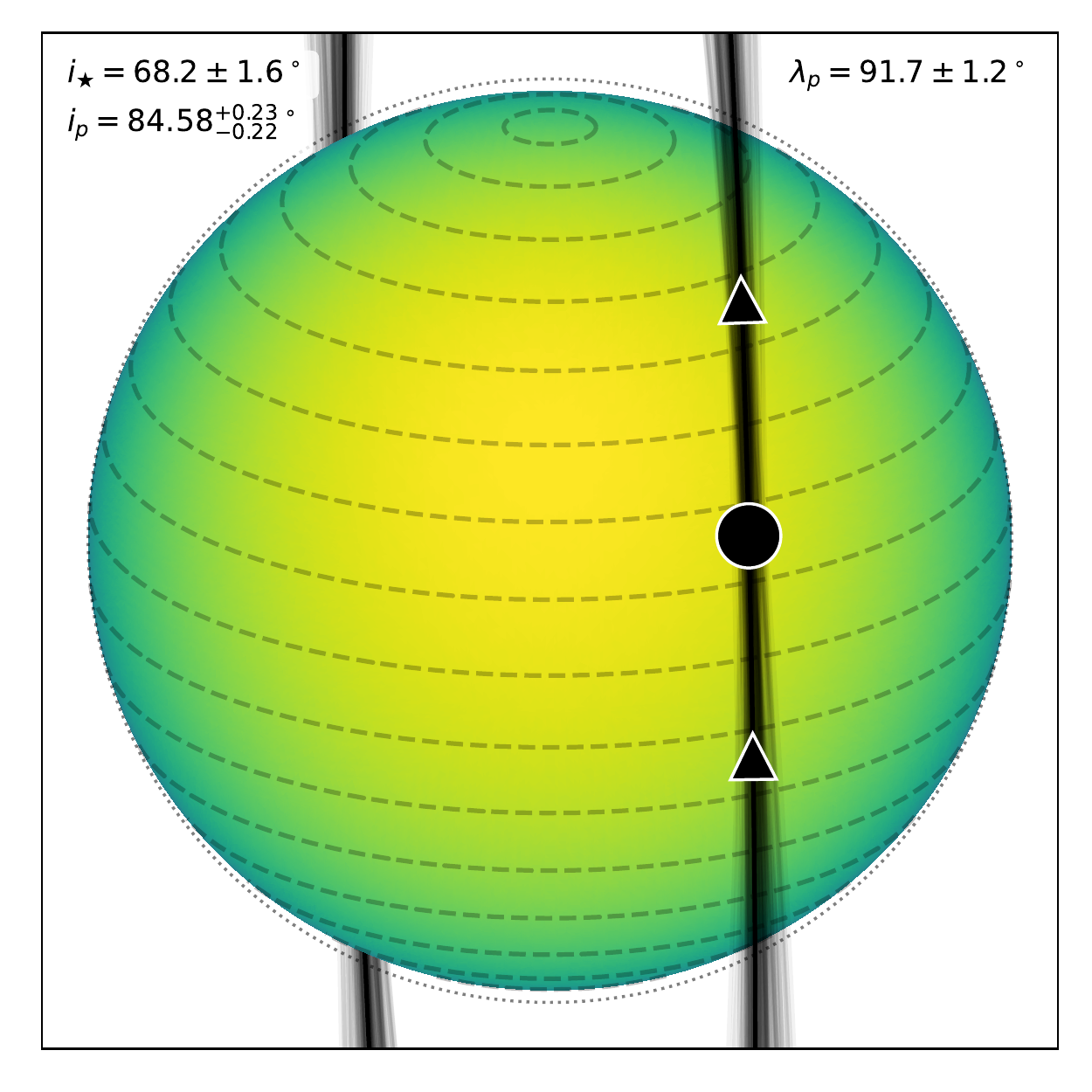}
                        \qquad
                        \includegraphics[width=.38\hsize,trim={0.4cm 0.4cm 0.38cm 0.35cm},clip]{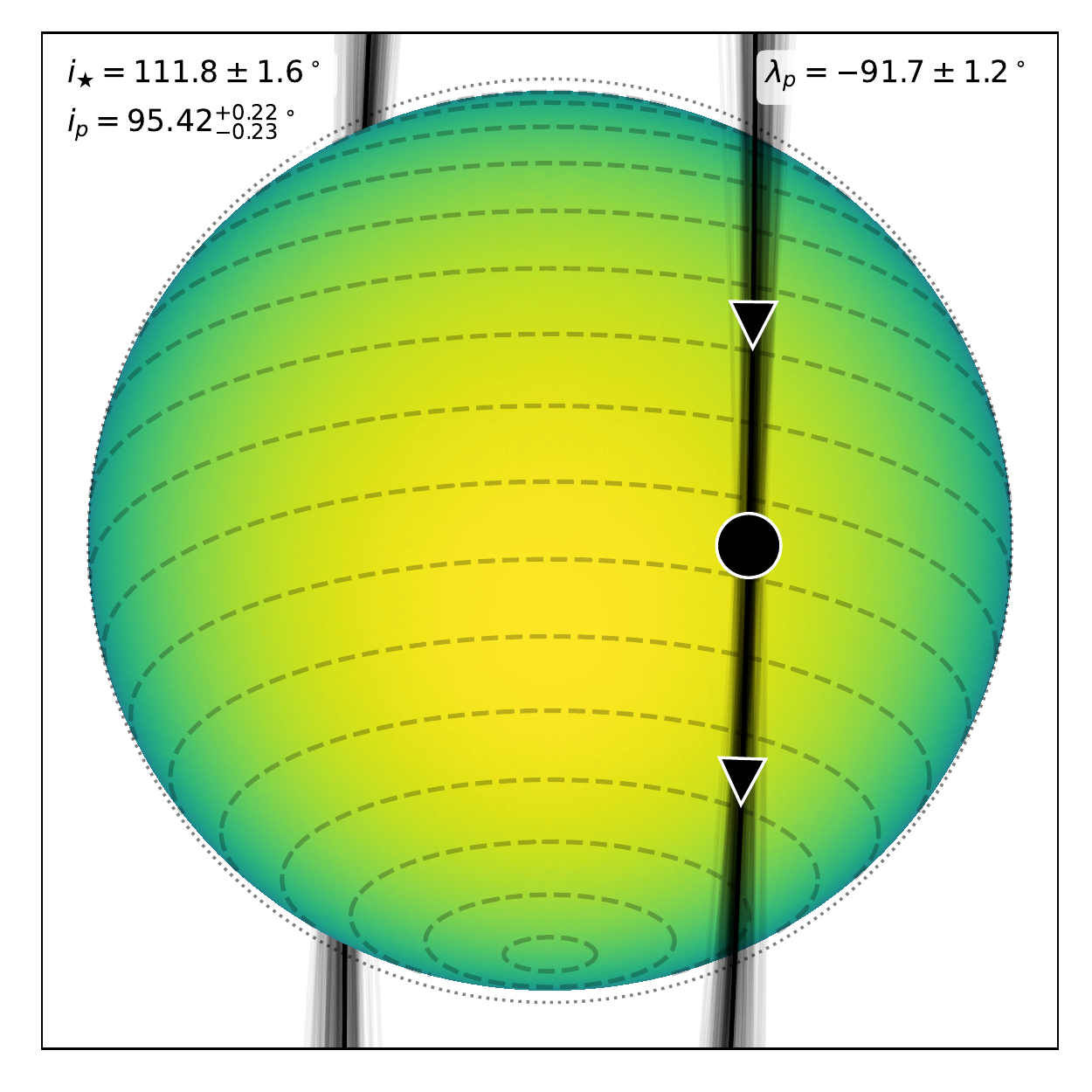}
                        \caption{Representation of the two possible orientations of the \host{} planetary system. The star is in its most probable state with the grey dashed curves marking constant stellar latitudes and with the north pole upward. In this representation, the stellar rotation causes the left and right hemispheres to appear blueshifted and redshifted, respectively. The stellar oblateness $f_\star=2.88^{+0.15}_{-0.12}\,\%$ is highlighted by the grey dotted circle. The colour scale represents the stellar flux as seen by \cheops{}, including the limb darkening, the gravity darkening and the instrument passband. The black disc shows the planet to scale, and the black shaded area is made of several orbits drawn from the posterior probability distribution and highlights the uncertainty we obtain on the orbital orientation of \planet. The black arrows indicate the direction in which the planet crosses the stellar disc. The values of the stellar inclination $i_\star$, the orbital inclination $i_p$ and the projected orbital obliquity $\lambda_p$ are displayed.}
                        \label{fig:system_view}
                \end{figure*}

                The system orientation is illustrated in Fig.~\ref{fig:system_view}, where the star is represented at its most probable inclination, and several planetary orbits sampled from the posterior distribution are shown.
                
                \begin{table*}
                        \caption{Planetary parameters obtained with the model assuming a Lambertian reflector and approximating the thermal emission as a sinusoidal function of the phase angle.}
                        \label{tab:result_parameters_lamb}
                        \centering
                        \begin{tabular}{llccr}
                                \hline\hline
                                Fitted parameter & Symbol & Value & Prior & Units\\
                                \hline
                                Time of inferior conjunction & $T_0$ & $16.434866\pm0.000060$ & $\mathcal{U}\!\left(16.4333, 16.4363\right)$ & $\text{BJD}-2\,459\,000$\\
                                Orbital period & $P$ & ${2.724035}_{-0.000023}^{+0.000022}$         & $\mathcal{U}\!\left(2.7213, 2.7268\right)$ & days \\
                                Planet-to-star radii ratio & $k=R_p/R_\star$ & ${0.06958}\pm{0.00016}$ & $\mathcal{U}\!\left(0.069, 0.072\right)$ & - \\
                                Normalised semi-major axis & $a/R_\star$ & ${4.587}_{-0.034}^{+0.037}$ & $\mathcal{U}\!\left(4.4, 4.9\right)$ & - \\
                                Orbital inclination & $i_p$ & ${84.58}_{-0.22}^{+0.23}$ & $\mathcal{U}\!\left(82.48, 85.59\right)$ & deg \\
                                \multirow{2}{*}{Eccentricity / argument of periastron} & $e \cos\!\left(\omega\right)$ & 0 & fixed & - \\
                                & $e \sin\!\left(\omega\right)$ & 0 & fixed & - \\
                                Geometric albedo\tablefootmark{$\dagger$} & $A_g$\tablefootmark{$\dagger$} & ${0.26}_{-0.17}^{+0.12}$\tablefootmark{$\dagger$} & $\mathcal{U}\!\left(0, 2/3\right)$ & - \\
                                Dayside thermal flux\tablefootmark{$\dagger$} & $F_\text{day}/F_\star$\tablefootmark{$\dagger$} & ${42}_{-30}^{+38}$\tablefootmark{$\dagger$} & $\mathcal{U}\!\left(0, +\infty\right)$ & ppm \\
                                Nightside thermal flux & $F_\text{night}/F_\star$ & $0$ & fixed & ppm \\
                                Hotspot offset\tablefootmark{$\ddagger$} & $\phi_\text{therm}$\tablefootmark{$\ddagger$} & ${-7}\pm{17}$\tablefootmark{$\ddagger$} & $\mathcal{U}\!\left(-60, 60\right)$ & deg \\
                                \multirow{2}{*}{Limb-darkening coefficients} & $u_1$ & ${0.414}_{-0.022}^{+0.024}$ & - & - \\
                                & $u_2$ & ${0.155}_{-0.034}^{+0.032}$ & - & - \\
                                Projected rotation speed & $v_\star\sin\!\left(i_\star\right)$        & ${92.5}_{-1.7}^{+1.8}$ & $\mathcal{N}\!\left(93.1, 1.7\right)$ & $\text{km}/\text{s}$ \\
                                Temperature of the stellar poles & $T_{\rm pole}$ & ${7967}_{-67}^{+69}$ & $\mathcal{N}\!\left(8000, 80\right)$ & K \\
                                Stellar inclination & $i_\star$ & ${68.2}\pm{1.6}$ & $\mathcal{U}\!\left(0, 180\right)$ & deg \\
                                Projected orbital obliquity & $\lambda_p$ & ${91.7}\pm{1.2}$ & $\mathcal{N}\!\left(89.3, 1.4\right)$ & deg \\
                                Gravity-darkening exponent & $\beta$ & 0.22 & fixed & - \\
                                Stellar radius & $R_\star$ & ${2.363}_{-0.024}^{+0.025}$ & $\mathcal{N}\!\left(2.365, 0.025\right)$ & $R_\odot$ \\
                                Stellar mass & $M_\star$ & ${2.073}_{-0.098}^{+0.090}$ & $\mathcal{N}\!\left(2.031, 0.098\right)$ & $M_\odot$ \\
                                Ellipsoidal variations amplitude & $F_\text{ell}/F_\star$ & 0 & fixed & ppm \\
                                Doppler beaming amplitude & $F_\text{beam}/F_\star$ & 0 & fixed & ppm \\
                                Ramp effect coefficient &$c_\text{therm}$ & ${-52}_{-33}^{+34}$ & $\mathcal{U}\!\left(-1000, 1000\right)$ & $\text{ppm}/\text{K}$ \\
                                GP amplitude & $\ln\!\left(S_0\right)$ & ${-22.21}_{-0.69}^{+0.84}$ & $\mathcal{U}\!\left(-25, -14\right)$ & - \\
                                GP quality factor & $\ln\!\left(Q\right)$ & ${0.85}_{-0.83}^{+0.89}$ & $\mathcal{U}\!\left(-4, 12\right)$ & - \\
                                GP undamped period & $P_0$ & ${1.27}_{-0.10}^{+0.12}$ & $\mathcal{N}\!\left(1.2, 0.2\right)$ & days \\
                                Jitter white noise & $\sigma_w$ & ${17.0}_{-3.4}^{+2.8}$ & $\mathcal{U}\!\left(0, +\infty\right)$ & ppm \\
                                \hline
                                Derived parameter & & \\
                                \hline
                                Optimal time of inferior conjunction & $T_{0,\,\text{opt}}$ & ${21.882937}\pm{0.000048}$ && $\text{BJD}-2\,459\,000$ \\
                                Planetary radius & $R_p$ & ${1.600}^{+0.017}_{-0.016}$ && $R_J$ \\
                                Semi-major axis & $a$ & ${0.05040}^{+0.00064}_{-0.00060}$ && AU \\
                                Impact parameter & $b$ & ${0.433}^{+0.014}_{-0.015}$ && - \\
                                Eccentricity & $e$ & 0  (fixed) && - \\
                                Eclipse depth & $\delta_\text{ecl}$ & ${96.5}^{+4.5}_{-5.0}$ && ppm \\
                                Stellar rotation period & $P_\star$ & ${1.198}^{+0.026}_{-0.025}$ && days \\
                                Stellar oblateness & $f_\star$ & ${2.88}^{+0.15}_{-0.12}$ && $\%$ \\
                                Stellar density & $\rho_\star$ & ${0.1617}^{+0.0088}_{-0.0089}$ && $\rho_\odot$ \\
                                True orbital obliquity & $\Psi_p$ & ${89.6}\pm{+1.2}$ && deg \\
                                GP damped period & $P_\text{damped}$ & ${1.32}^{+0.22}_{-0.11}$ && days \\
                                \hline\hline
                        \end{tabular}
                        \tablefoot{
                                The upper part of the table lists the fitted parameters with their corresponding prior probabilities. Uniform prior probabilities are represented with $\mathcal{U}\!\left(x_\text{min}, x_\text{max}\right)$, where $x_\text{min}$ and $x_\text{max}$ are the minimum and maximum allowed values, respectively. Normal (Gaussian) prior probabilities are written as $\mathcal{N}\!\left(\mu, \sigma\right)$, where $\mu$ and $\sigma$ are the mean and standard deviation of the normal distribution, respectively. The lower part of the table shows the values of parameters derived from the sampled parameter space.
                                \tablefoottext{$\dagger$}{The values of $A_g$ and $F_\text{day}$ both define the eclipse depth and are thus degenerate. This induces a strong linear correlation between the two parameters and the values reported in the table are not representative of any convergence. The eclipse depth obtained from $A_g$ and $F_\text{day}$ is a well-defined quantity and should be the one to refer to.}
                                \tablefoottext{$\ddagger$}{The hotspot offset is unconstrained for small values of $F_\text{day}$ where $A_g$ solely contributes to the phase curve amplitude. The value of $\phi_\text{therm}$ reported in the table is computed for values of $A_g<0.05$ to ensure reliable estimates of uncertainties.}
                        }
                \end{table*}
                
                All the fitted planetary parameters are consistent with values obtained in \cite{lendl_2020_wasp-189b}. We note that the quantities normalised by the stellar radius ($k$ and $a/R_\star$) are different, however. This is actually due to the fact that the stellar radius used for normalisation in \cite{lendl_2020_wasp-189b} is the polar radius, while it is the equatorial radius in our approach. Comparing absolute values of the planetary radius $R_p$ and the semi-major axis $a$ show consistencies within $0.1\,\sigma$.
                
                We note that the time of inferior conjunction $T_0$ and the orbital period $P$ are fully consistent with the values reported in \cite{anderson_2018_wasp-189b}. We obtain an exquisite precision on both parameters with fewer than two\,seconds for the period and an optimum $T_{0,\,\text{opt}}$ at about four\,seconds.
                
                From the fitted parameters, we derive a series of other useful parameter values, including the eclipse depth that is the amplitude of the phase curve (reflected and thermal) at superior conjunction ($\omega+\nu=270\deg$).
                It is important to mention that the amplitudes of the reflected and thermal parts of the phase curve are degenerate, especially where the hotspot offset is zero and the full amplitude is the sum of both contributions. This is the case for \planet,{} and the degeneracy is strong, with either a high geometric albedo $A_g$ and a low dayside thermal flux $F_\text{day}$ with unconstrained $\phi_\text{therm}$ values, or a more constrained offset for low $A_g$ (see Fig.~\ref{fig:corner_plot}). In order to provide reliable uncertainties on $\phi_\text{therm}$, the value reported in Table~\ref{tab:result_parameters_lamb} is computed for $A_g<0.05$, corresponding to a fully thermal phase curve. We find no indication for a hotspot offset from the sub-stellar point with the phase curve peaking at occultation.
                
                The true orbital obliquity $\Psi_p$ describes the relative orientation of the planetary orbit with respect to the spin axis of the star, and we compute it with the following angular relationship (see Fig.~\ref{fig:gd_geometry}):
                \begin{equation}
                        \Psi_p=\arccos\!\left[\cos\!\left(i_p\right)\cos\!\left(i_\star\right)+\cos\!\left(\lambda_p\right)\sin\!\left(i_p\right)\sin\!\left(i_\star\right)\right]
                ,\end{equation}
                where $i_p$ is the orbital inclination, $i_\star$ is the stellar inclination and $\lambda_p$ is the projected orbital obliquity. From the values of our MCMC runs, we obtain $\Psi_p=89.6\pm1.2\deg,$ which is fully consistent with a polar orbit.
                
                Despite the convergence of the fit towards a well-defined system orientation, there remains a degeneracy inherent to gravity-darkened transit photometry. The following four sets of angular parameters will produce the same photometric signal:
                \begin{enumerate}
                        \item $\left(i_\star,\ i_p,\ \lambda_p\right),$
                        \item $\left(i_\star,\ 180^\circ-i_p,\ 180^\circ-\lambda_p\right),$
                        \item $\left(180^\circ-i_\star,\ 180^\circ-i_p,\ -\lambda_p\right),$
                        \item $\left(180^\circ-i_\star,\ i_p,\ 180^\circ+\lambda_p\right).$
                \end{enumerate}
                From the observer's point of view, the difference between configurations~1 and~2 (and between~3 and~4) is that one is the symmetric of the other with respect to the `vertical' plane containing the line of sight and the spin axis of the star. Similarly, configuration~1 (respectively~2) is symmetrical to~3 (respectively~4) with respect to the `horizontal' plane containing the line of sight and perpendicular to the vertical plane mentioned previously. Spectroscopic observations such as Doppler tomography can rule out two out of four configurations by measuring if the transiting planet is above the blueshifted or the redshifted hemisphere of the rotating star. In the case of \planet{}, \cite{anderson_2018_wasp-189b} show by Doppler tomography that the polar orbit only goes over the redshifted hemisphere, which leaves only two possible scenarios, which are represented in Fig.~\ref{fig:system_view}. The second scenario (right-most panel of Fig.~\ref{fig:system_view}), where $i_\star > 90\deg$ and the south stellar pole is visible, cannot be ruled out based on the data available, and it gives the same true orbital obliquity.
                
                We constrained the stellar oblateness $f_\star$ of \host{} and were able to determine that its polar radius is about 2.9\% smaller than its equatorial radius due to its fast rotation. This corresponds to a temperature difference of about 200\,K between the poles ($8000\,\text{K}$ by construction) and the equator ($\sim7800\,\text{K}$).

        \subsection{Phase-curve constraints on the planetary atmosphere} \label{ssec:results_pc}
                
\begin{figure*}
                        \centering
                        \includegraphics[width=\hsize,trim={0.8cm 0.6cm 2.5cm 1.8cm},clip]{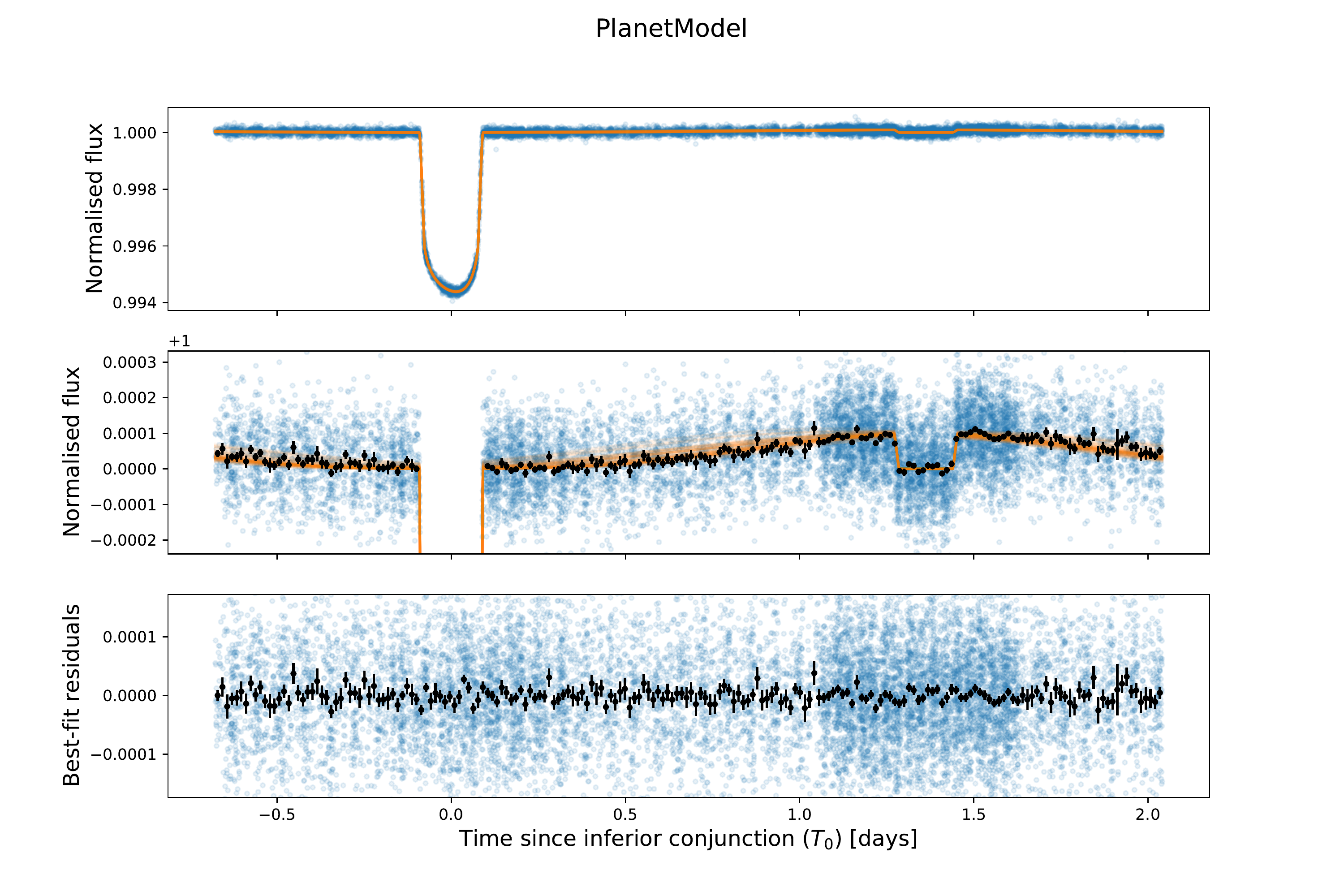}
                        \caption{Phase-folded phase curve of \planet{} after correction of instrumental systematics and stellar activity. This data set is obtained from a model assuming the planet is a Lambertian reflector and approximating the thermal emission with a sinusoid function. The data points de-trended from all systematics and stellar variability are shown in blue in the two upper panels. The mid-panel shows a zoomed-in view of the upper one to better visualise the phase-curve signal. The lower panel shows the residual after subtraction of the best-fit planet model. The black points represent the binned data with 100 bins per orbital period, which corresponds to a bin duration of about 40\,min. The faded orange lines are 100 samples drawn from the posterior probability obtained from the MCMC run, and thus highlight the uncertainty we obtain on the best-fit model.}
                        \label{fig:results_pc}
                \end{figure*}

\begin{figure}
                        \centering
                        \includegraphics[width=.9\hsize,trim={0.35cm 0.3cm 0cm 0.3cm},clip]{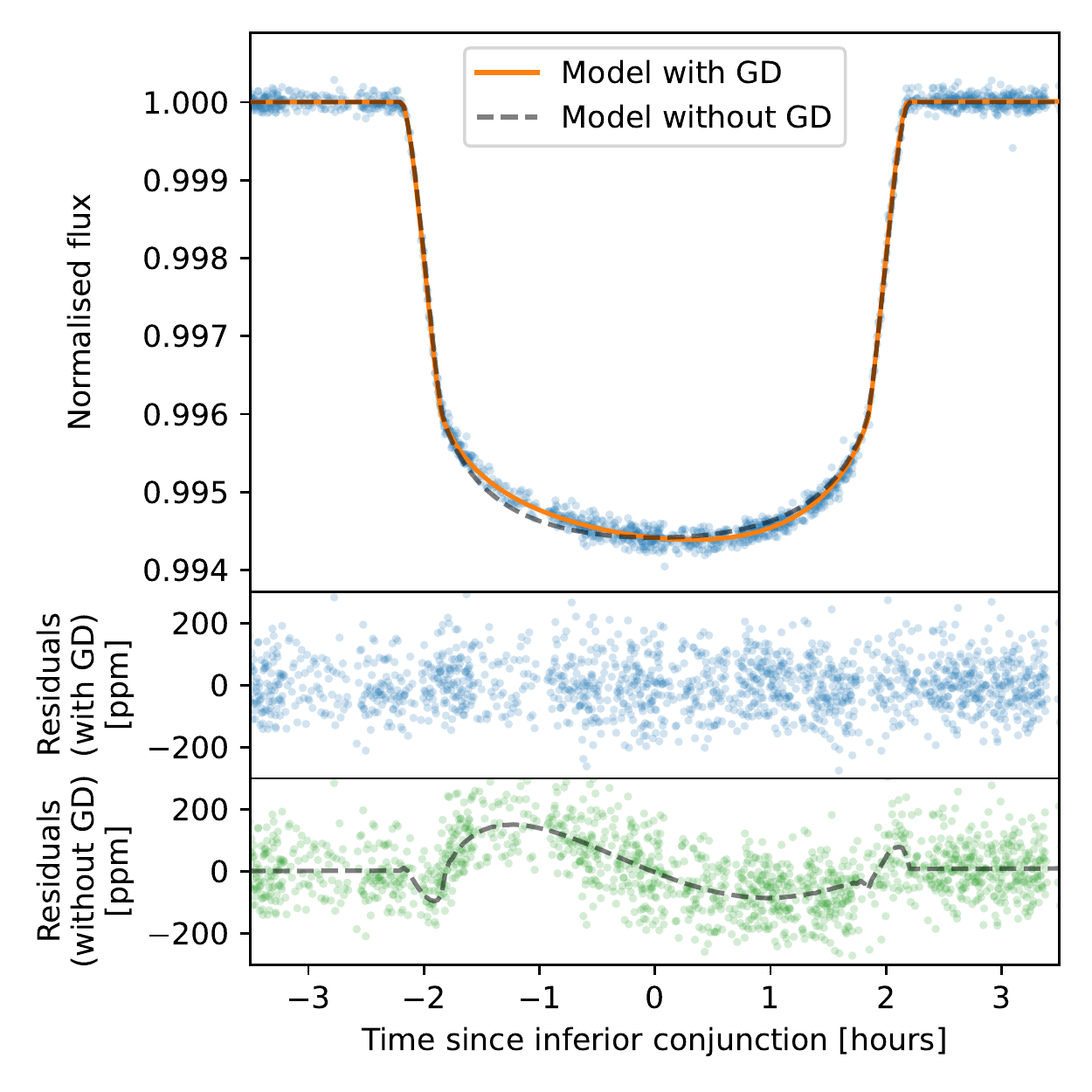}
                        \caption{Phase-folded transit light curve of \planet{} after correction of instrumental systematics and stellar activity. The top panel shows the normalised flux and two best-fit transit models with and without GD. The mid-panel (blue-point residuals) shows the best-fit residuals for the model accounting for GD. The bottom panel displays the residuals after removing a transit light curve without accounting for the gravity-darkened stellar photosphere. The black dashed line is the deviation of the model without GD from the model with GD.}
                        \label{fig:gd_transit}
                \end{figure}

                The light curve obtained after removing systematic noise and stellar activity assuming a Lambertian reflector and a sinusoidal thermal emission is shown in Fig.~\ref{fig:results_pc}. The phase curve highlights the exquisite photometric precision of \cheops{} while the planetary eclipse and the asymmetric transit are visible. Fig.~\ref{fig:gd_transit} shows the corrected phase-folded data around the mid-transit time and clearly reveals the effect of GD on the transit shape. The uncertainty on transit and eclipse parameters is small compared to the uncertainty found on the phase-curve signal. This is due to the photometric variability attributed to the stellar activity that is not perfectly periodic and thus hard to disentangle unambiguously from the planetary signal.
                
                As described in Section~\ref{ssec:results_params}, we computed the hotspot offset $\phi_\text{therm}$ reported in Table~\ref{tab:result_parameters_lamb} for parameter sets probing a phase-curve model dominated by thermal emission ($A_g<0.05$). We derived a hotspot offset consistent with zero ($\phi_\text{therm}=-7\pm17\,\deg$). The precision on this parameter value is likely limited by the uncertainty on the phase curve induced by stellar activity.
                
\begin{figure}
                        \centering
                        \includegraphics[width=0.9\hsize,trim={0.3cm 0.3cm 0.2cm 0.2cm},clip]{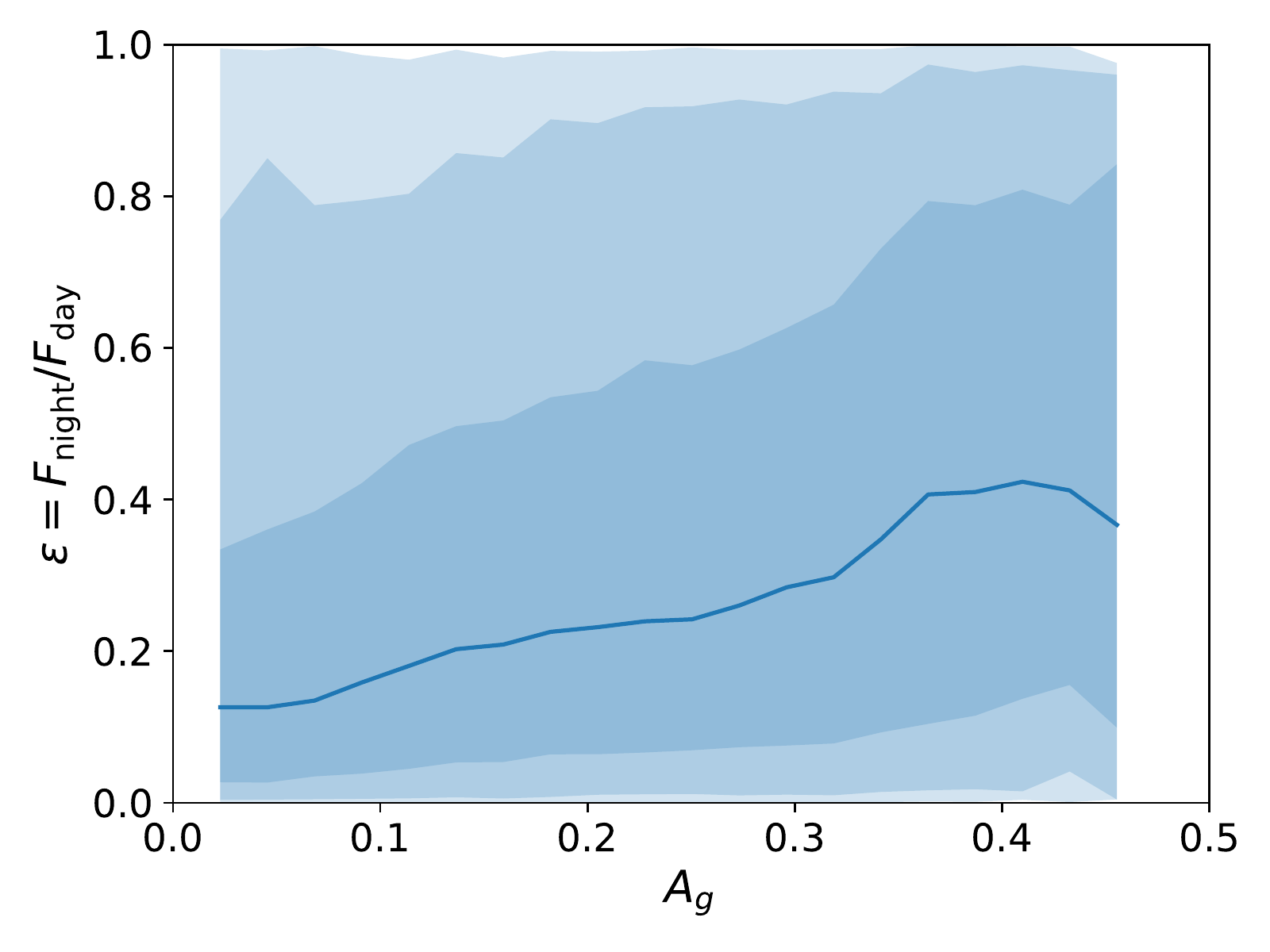}
                        \caption{Heat redistribution efficiency $\varepsilon$ as a function of the geometric albedo $A_g$. The median value of $\varepsilon$ is represented by the blue curve and the shaded areas highlight the $1\,\sigma$, $2\,\sigma,$ and $3\,\sigma$ uncertainty range. The value of $\varepsilon$ is computed as the nightside-to-dayside flux ratio measured by \cheops{}. Low values of $A_g$ indicates low reflectivity and a phase-curve amplitude mostly due to thermal emission. The thermal contribution gets smaller for higher $A_g$ where reflected light starts to be important.}
                        \label{fig:eps_vs_Ag}
                \end{figure}

                As reported in Section~\ref{sssec:priors}, we fitted the data with the nightside flux as a free parameter and found a value consistent with zero. However, we also obtained a large upper uncertainty on $F_\text{night}$ with a $3\,\sigma$ upper limit matching a uniform temperature distribution on the planet surface ($F_\text{night}\approx F_\text{day}$), which means a heat redistribution efficiency of $\varepsilon=F_\text{night}/F_\text{day}\approx1$. This results from the use of Gaussian processes to model the stellar activity: its flexibility was necessary to properly correct the associated signal, but with the downside of adding considerable degeneracies in the shape of the phase-curve model. Using the nightside flux we obtain when this parameter is free, we can nevertheless compute the heat redistribution efficiency in the \cheops{} passband $\varepsilon=0.23_{-0.17}^{+0.35}$. Assuming a fully thermal phase curve ($A_g\to0$), we derive a smaller limit of $\varepsilon\to0.13_{-0.10}^{+0.21}$ (see Fig.~\ref{fig:eps_vs_Ag}). Both values are less than $1.5\,\sigma$ away from 0 and have a $3\,\sigma$ upper limit consistent with 1.
                
                Due to the degeneracy between reflective and thermal components, we could not determine the contribution of each part in the absence of a significant hotspot offset. Therefore, we followed an approach assuming two extreme cases for the phase-curve amplitude.
                
                We first considered the case where the amplitude of the eclipse is only due to reflected light without any thermal contribution from the planet. The conversion of the eclipse depth $\delta_\text{ecl}$ into reflected light is done with Eq.~\ref{eq:pc_lamb_ref} for a circular orbit ($e=0$) and at superior conjunction ($\alpha=0$). It provides a maximum geometric albedo $A_{g, \text{max}}=0.42\pm0.02$, from which we compute a 3\,$\sigma$ upper limit (99.93\% confidence) and determine that the geometric albedo of \planet{} is $A_g < 0.48$. This value is not inconsistent with other measured geometric albedos of ultra-hot Jupiters, even though these objects usually feature dark atmospheres with low reflectivity \citep{wong_2021_tess_pc}.
                
\begin{figure}
                        \centering
                        \includegraphics[width=.9\hsize,trim={2cm 0cm 1cm 0.7cm},clip]{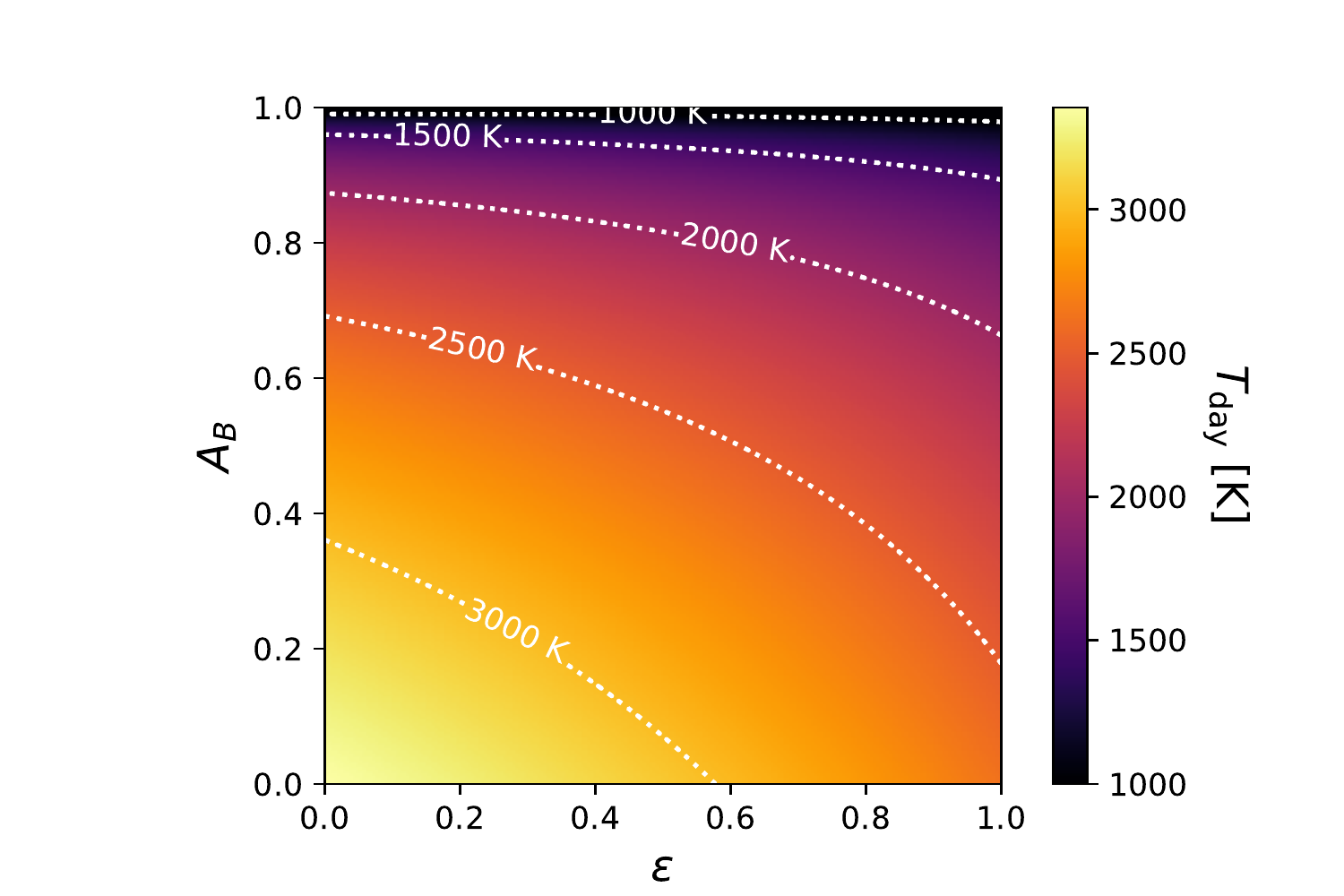}
                        \includegraphics[width=.9\hsize,trim={2cm 0cm 1cm 0.7cm},clip]{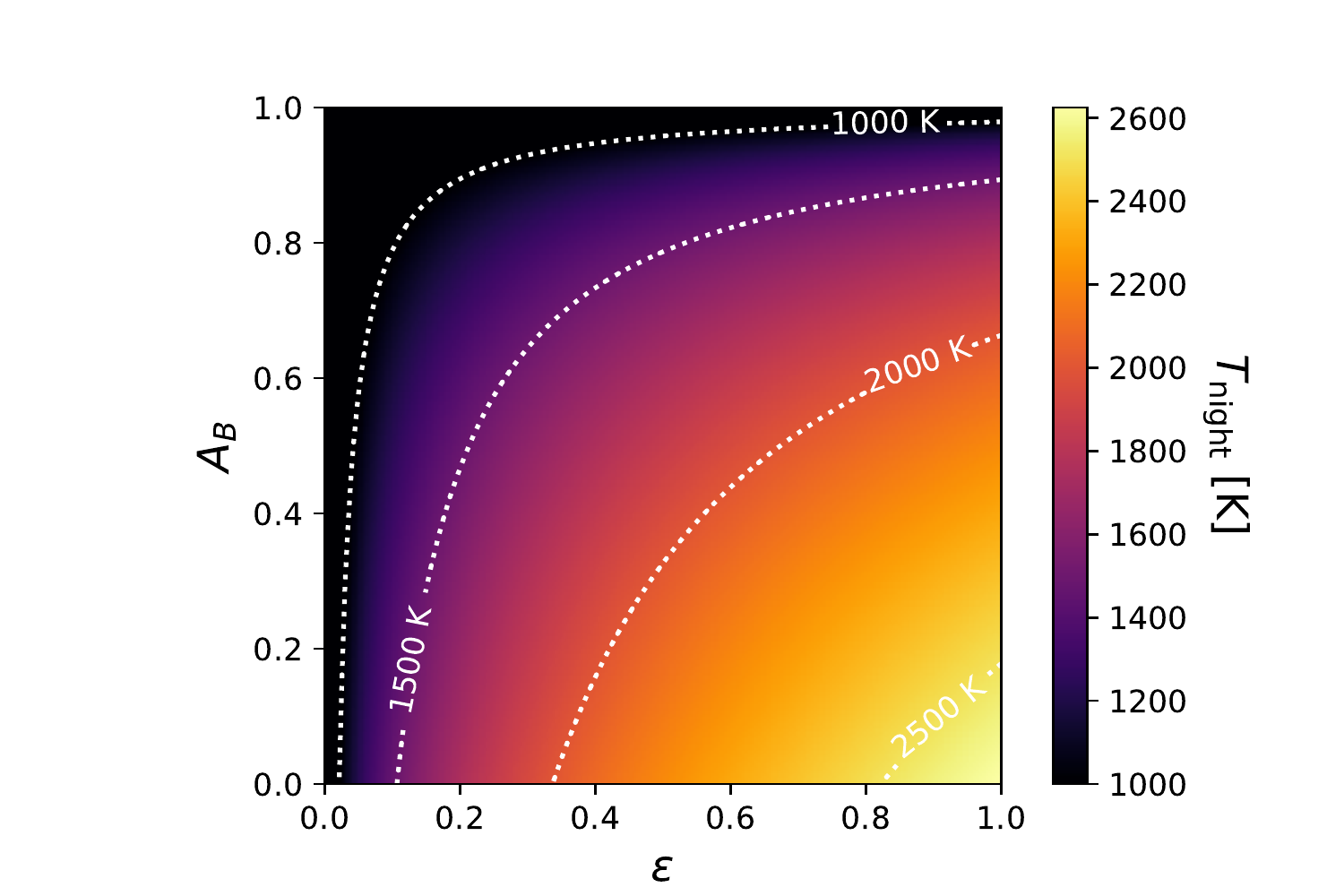}
                        \caption{Dayside (top) and nightside (bottom) effective temperatures of the atmosphere of \planet{} as a function of the Bond albedo $A_B$ and the heat redistribution efficiency $\varepsilon$. The lower limit of the colour scale is set to $1000\,\text{K}$ for a better visualisation. White dotted lines show isothermal profile across the ($A_B,\varepsilon$) plane. The dayside and nightside temperatures are equal when $\varepsilon=1$ and both go down to $0\,\text{K}$ in the limit $A_B\to1$. If the dayside temperature is $3000\,\text{K}$, then the Bond albedo values range from 0 to $\sim\!0.38$, with corresponding values of the heat redistribution efficiency from $\sim\!0.59$ to 0, respectively. In such a case, the maximum nightside temperature would be $\sim\!2250\,\text{K}$.}
                        \label{fig:planet_temp}
                \end{figure}

                We then explored how the thermal emission contributes to the flux as seen through the \cheops{} passband. The power irradiating the planet is given by the following expression:
                \begin{equation}
                        P_\text{irr} = \pi R_p^2\,\left(\frac{R_\star}{a}\right)^2\int_{\lambda=0}^{+\infty}\!\mathcal{S}\!\left(\lambda, T_\star\right) d\lambda
                ,\end{equation}
                where $R_p$ is the planet radius, $R_\star$ is the stellar radius, $a$ is the semi-major axis, and $\mathcal{S}\!\left(\lambda, T_\star\right)$ is the flux emission spectrum of the star as a function of its temperature $T_\star$ and the wavelength $ \lambda$. We can compute the power absorbed by the planetary atmosphere $P_\text{abs}=\left(1-A_B\right)P_\text{irr}$, with $A_B$ being the Bond albedo. The power radiated away by the planet must equal the one absorbed if it is to be at equilibrium. Assuming the planet is a black-body radiator, we can derive a relationship between the average effective temperature of the planet $\tilde{T}_p$ and the irradiating power $P_\text{irr}$ and obtain the following:
                \begin{equation}
                        \tilde{T}_p^4 = \frac{1-A_B}{4\,\sigma_\text{SB}}\left(\frac{R_\star}{a}\right)^2\int_{\lambda=0}^{+\infty}\!\mathcal{S}\!\left(\lambda, T_\star\right) d\lambda
                ,\end{equation}
                where $A_B$ is the Bond albedo and $\sigma_\text{SB}$ is the Stefan-Boltzmann constant. The average effective temperature of the planet  is given by $\tilde{T}_p^4 = 1/{4\pi}\,\iint_\text{planet}T_p^4\,d\Omega$, where $T_p$ is the local effective temperature in the planet atmosphere and $d\Omega$ describes a surface element of the atmosphere.
                Following the parametrisation of \cite{cowan_2011_albedo_epsilon}, we define the dayside and nightside effective temperatures as functions of the heat redistribution efficiency $\varepsilon$ and $\tilde{T}_p$:
                \begin{align}
                        &T_\text{day} = \left(\frac{8-5\varepsilon}{3}\right)^\frac{1}{4}\tilde{T}_p\,,\\
                        &T_\text{night} = \varepsilon^\frac{1}{4}\,\tilde{T}_p\,.
                \end{align}
                We computed the ranges of values of both temperatures when the Bond albedo and the heat redistribution efficiency are allowed to vary within $\left[0, 1\right]$ (see Fig.~\ref{fig:planet_temp}). We used the spectral energy distribution represented in Fig.~\ref{fig:star_sed} for the star and the median value reported in Table~\ref{tab:result_parameters_lamb} for $a/R_\star$. In the limit where no energy is absorbed by the planet ($A_B\to1$), we retrieve a temperature of $0\,\text{K}$ everywhere on the planet. When all the irradiating power is absorbed ($A_B=0$) and the heat redistribution is perfect ($\varepsilon=1$), we obtain a uniform atmospheric temperature $T_\text{uni}=\tilde{T}_p=2625\pm11\,\text{K}$. The maximum dayside temperature is reached when $A_B=\varepsilon=0,$ and one obtains $T_\text{day, max}=3355\pm14\,\text{K}$ for \planet{}.
                
\begin{figure}
                        \centering
                        \includegraphics[width=.9\hsize,trim={2cm 0cm 1.4cm 0.7cm},clip]{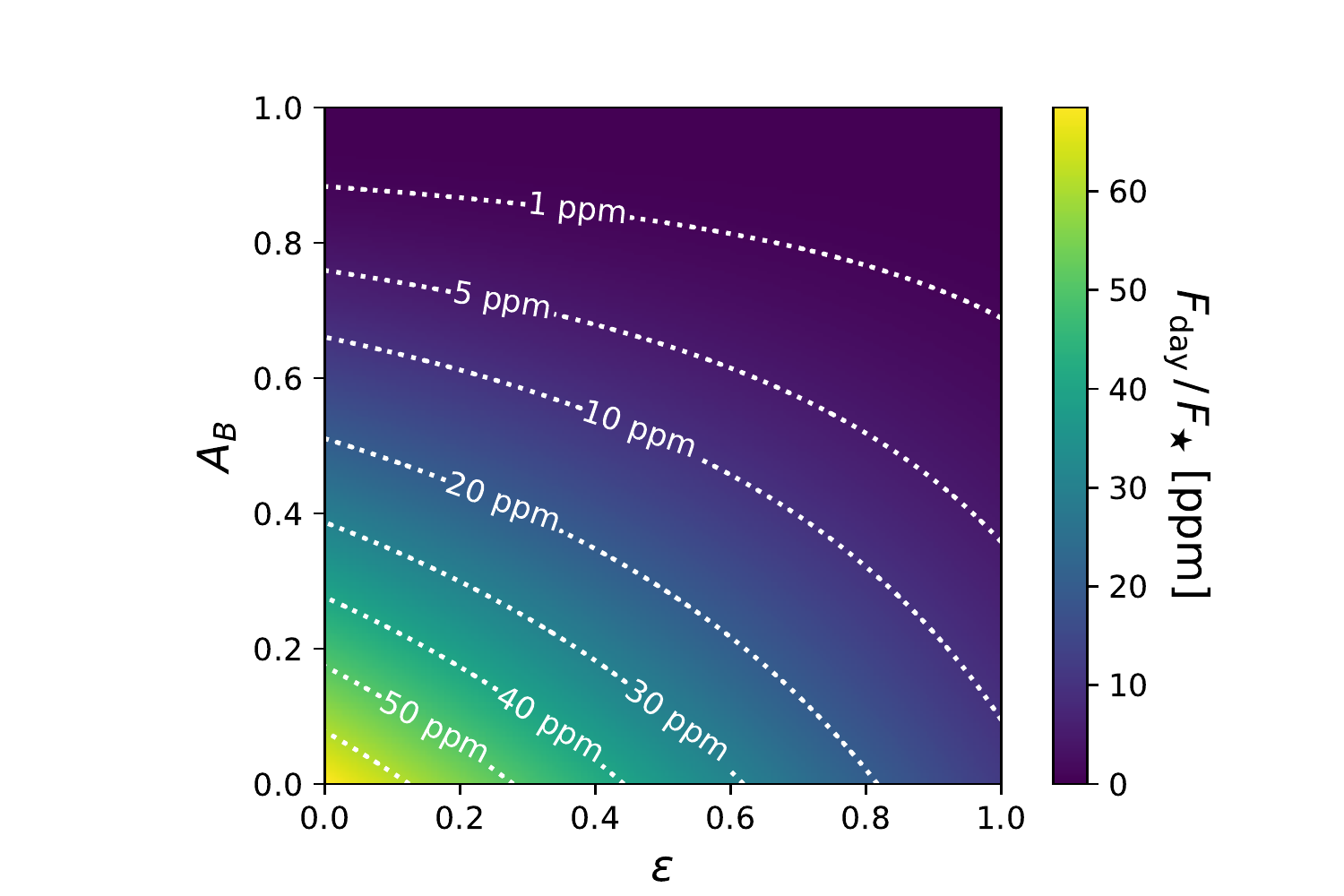}
                        \caption{Dayside thermal flux from \planet{} as a function of the Bond albedo $A_B$ and the heat redistribution efficiency $\varepsilon$. The flux $F_\text{day}$ is normalised by the stellar flux $F_\star$. Both fluxes are computed as seen through \cheops{} passband. White dotted lines show constant flux ratios across the ($A_B,\varepsilon$) plane. If the measured eclipse depth contains a thermal emission component with an amplitude 40\,ppm, the Bond albedo may be as large as $\sim\!0.29$ and as small as zero, and the corresponding heat redistribution efficiency would be zero and $\sim\!0.45,$ respectively.}
                        \label{fig:planet_flux}
                \end{figure}

                From the dayside temperatures, we were able to calculate the corresponding eclipse depth in the \cheops{} passband with the following relationship:
                \begin{equation}
                        \label{eq:dayside_flux}
                        \frac{F_\text{day}}{F_\star}=\left(\frac{R_p}{R_\star}\right)^2\frac{\int_{\lambda=0}^{+\infty}\!\mathcal{S}\!\left(\lambda, T_\text{day}\right) \mathcal{T}_{\rm inst}\!\left(\lambda\right) d\lambda}{\int_{\lambda=0}^{+\infty}\!\mathcal{S}\!\left(\lambda, T_\star\right) \mathcal{T}_{\rm inst}\!\left(\lambda\right) d\lambda}
                ,\end{equation}
                where $R_p$ and $R_\star$ are the planetary and stellar radii, respectively, $\mathcal{S}\!\left(\lambda, T_\text{day}\right)$ and $\mathcal{S}\!\left(\lambda, T_\star\right)$ are the flux emission spectra of the planet and the star, respectively, and $\mathcal{T}_{\rm inst}\!\left(\lambda\right)$ is the \cheops{} passband. Similarly to the computation of the effective temperatures, we used a black-body approximation for the planet, the spectral energy distribution of Fig.~\ref{fig:star_sed} for the star, and the median value of $k=R_p/R_\star$ from Table~\ref{tab:result_parameters_lamb}. The values of $F_\text{day}/F_\star$ are displayed in Fig.~\ref{fig:planet_flux}. We retrieve expected trends such as a low flux for high Bond albedo values and maximal dayside thermal flux when $\varepsilon\to0$.
                
                It is noteworthy that the maximum dayside temperature $T_\text{day, max}$ computed in the case of black-body emission only produces an eclipse depth of $68.5\pm1.8\,\text{ppm}$ in the \cheops{} passband, which is not enough to explain the measured $\delta_\text{ecl}=96.5^{+4.5}_{-5.0}\,\text{ppm}$. The dayside temperature necessary to generate such a thermal flux is $3542\pm14\,\text{K}$ and corresponds to a negative Bond albedo ($A_B\approx-0.24$ for $\varepsilon=0$). From these values, it seems that a fully thermal emission with no reflected light cannot explain the eclipse depth measured with \cheops{} and this in turn implies a minimum geometric albedo $A_g>0.12\pm0.02$.
                
\begin{figure}
                        \centering
                        \includegraphics[width=.95\hsize,trim={0.2cm 0.3cm 0.2cm 0.3cm},clip]{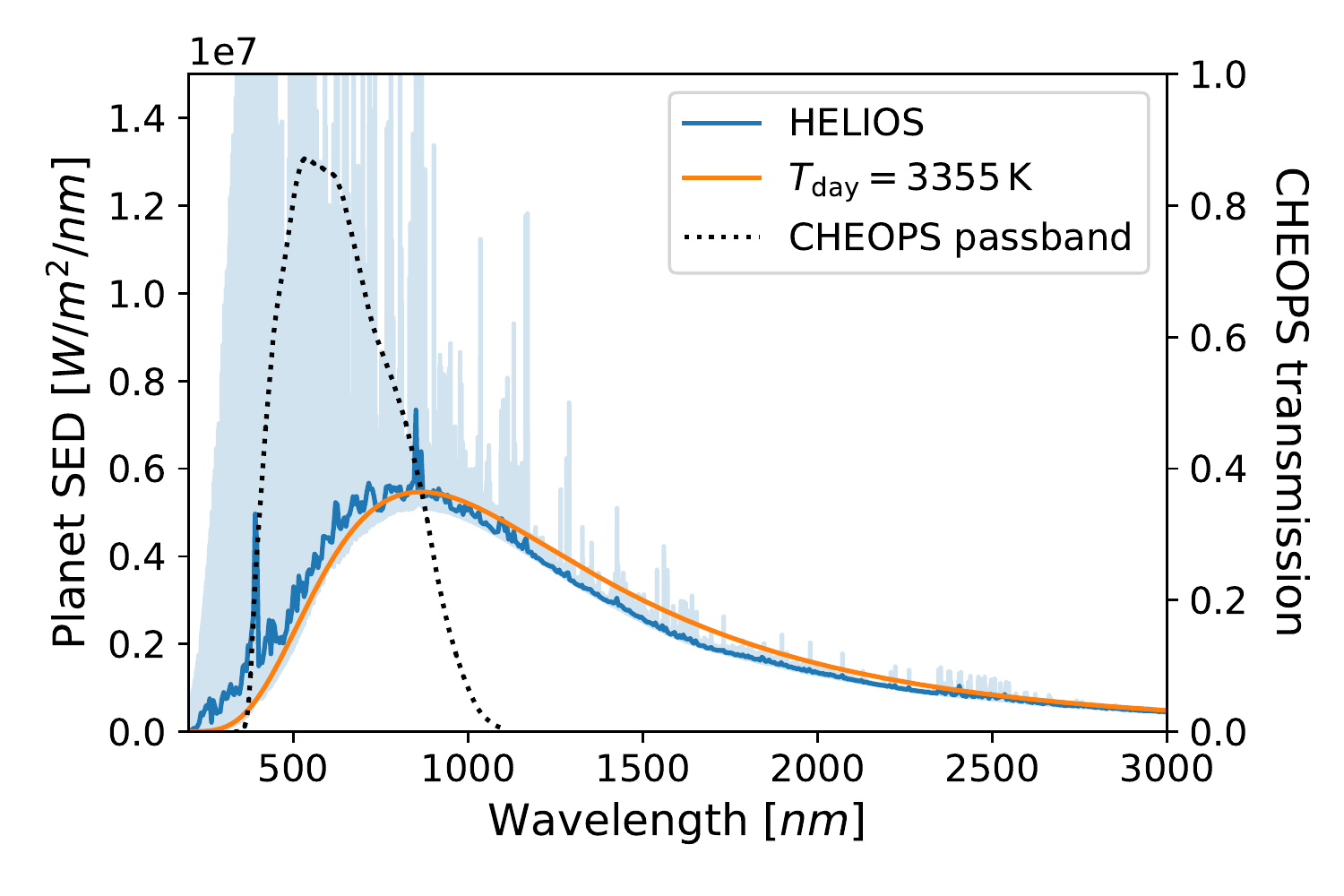}
                        \caption{Spectral energy distribution of the planet \planet{}. The high-resolution synthetic SED computed with HELIOS \citep{malik_2017_helios, malik_2019_helios} is represented in light blue and assumes the irradiating power is fully absorbed and not redistributed ($A_B=\varepsilon=0$). The dark blue line is the HELIOS data averaged over bins of 5\,nm. The black-body profile of the maximum dayside temperature $T_\text{day, max}$ is shown in orange. We note that accounting for emission lines in the planetary spectrum leads to a different SED with respect to the black-body approximation, with more energy in the blue end. This affects the flux measured in the \cheops{} passband (black dotted line), with the dayside of \planet{} appearing slightly brighter.}
                        \label{fig:planet_sed}
                \end{figure}

                To further explore the limits on the dayside thermal flux, we computed a synthetic planetary emission spectrum using the radiative transfer code HELIOS \citep{malik_2017_helios, malik_2019_helios}. The model assumes a cloud-free atmosphere in chemical equilibrium with several sources of opacity as detailed in \cite{lendl_2020_wasp-189b}. The inputs used for the computation are a stellar emission spectrum from the PHOENIX library \citep{husser_phoenix_2013} as shown in Fig.~\ref{fig:star_sed} and the median values of the planetary parameters listed in Table~\ref{tab:result_parameters_lamb}. We also assumed that all the irradiating stellar power is absorbed by the planet ($A_B=0$) and that there is no heat redistribution ($\varepsilon=0$). We obtain the SED represented in Fig.~\ref{fig:planet_sed} with many emission lines in the blue end of the spectrum leading, in turn, to an increase of flux in the \cheops{} passband with respect to the black-body approximation.
                
                We computed the dayside flux in the \cheops{} passband using the HELIOS emission spectrum for the planet in Eq.~\ref{eq:dayside_flux}. The resulting eclipse depth is slightly larger than the black-body case with $77.0\pm2.5\,\text{ppm}$. This value remains inconsistent with the measured eclipse depth by $3.5\,\sigma$ and implies $A_g>0.08\pm0.02$.
                
                In an attempt to maximise the modelled eclipse depth, we performed a final computation using the same HELIOS model for the planet but accounting for a cooler and oblate star. We used a stellar emission spectrum from the PHOENIX library \citep{husser_phoenix_2013} for a stellar temperature of 7800\,K (minimum temperature reached at the equator) and reduced the size of the stellar disc by a factor $1-f_\star$ in Eq.~\ref{eq:dayside_flux} to account for its oblateness. These settings represent a lower limit of the received stellar flux (smaller cooler star) and maximise the detected eclipse depth up to $87.1\pm3.1\,\text{ppm}$, which is marginally consistent ($1.6\,\sigma$) with a fully thermal phase curve and no reflected light. However, this latest result corresponds to an extreme limit where $A_B=\varepsilon=0$ and the stellar flux is expected to be underestimated as we assumed a uniform photospheric temperature equal to the equatorial minimum. As a consequence, the numbers we obtain seem to point toward a non-zero geometric albedo, with a lower limit of $A_g>0.041\pm0.026$ in the most extreme case. These results are consistent with the temperatures reported by \cite{lendl_2020_wasp-189b}, where the dayside temperature computed with $A_g=0$ exceeds the limit of stellar irradiating power.
                
                In light of this study, it is difficult to explain the observed eclipse depth by pure thermal emission from the atmosphere of \planet{}. The conditions we considered to reach marginal consistency are extreme with the smallest and coolest possible star, and a full absorption of the stellar irradiation by the planet. Moreover, the heat redistribution inefficiency maximising the dayside temperature also implies a nightside temperature of nearly $0\,\text{K}$. Therefore, it seems very likely that the planetary signal is not purely thermal and that the atmosphere has a non-negligible reflective component in the \cheops{} passband. As we have shown in this study, deriving a precise lower limit on the geometric albedo strongly depends on the assumptions made on the star, the system architecture and the atmospheric properties of the planet. Nevertheless, it is noteworthy that the limit on the geometric albedo derived using the HELIOS emission spectrum is consistent with $A_g\sim0.1$. Similar geometric albedos have been reported for comparable planets \citep{wong_2021_tess_pc}, even if the current atmospheric models struggle to explain the presence of reflective material at temperatures larger than 3000\,K.
                
                \tess{} \citep{ricker_tess} will observe the \host{} system in April-May 2022 when covering sector 51. With its redder passband, \tess{} is more sensitive to planetary thermal flux than \cheops,{} and the planet-to-star flux ratio is expected to be even larger in the case of hot blue stars such as \host{}. Thus, if the flux from \planet{} is mostly of thermal origin, the eclipse observed with \tess{} should be much deeper than the one reported in this work. Following the method described in this section, we computed the dayside thermal flux in the \tess{} passband as a function of $A_B$ and $\varepsilon$, similarly to Fig.~\ref{fig:planet_flux}. We estimate that \tess{} could measure an eclipse depth due to thermal emission up to $165\pm4\,\text{ppm}$. Such a signal should be detectable with a precision of more than $3\,\sigma$ when combining the eight-to-ten eclipses observed during the two \tess{} orbits ($\sim27\,\text{days}$). An additional significant advantage is the long baseline of the observations, which will allow a more robust correction of the stellar variability.
                A precise determination of the eclipse depth in the \tess{} passband should provide an upper limit on the dayside temperature corresponding to a fully thermal phase-curve amplitude. If one assumes that both \cheops{} and \tess{} probe the same atmospheric layer, and hence the same temperature distribution, one could use this result to constrain the geometric albedo in the \cheops{} passband.

        \subsection{Stellar activity} \label{ssec:results_star}
                
\begin{figure*}
                        \centering
                        \includegraphics[width=\hsize,trim={0.8cm 0cm 2.5cm 1cm},clip]{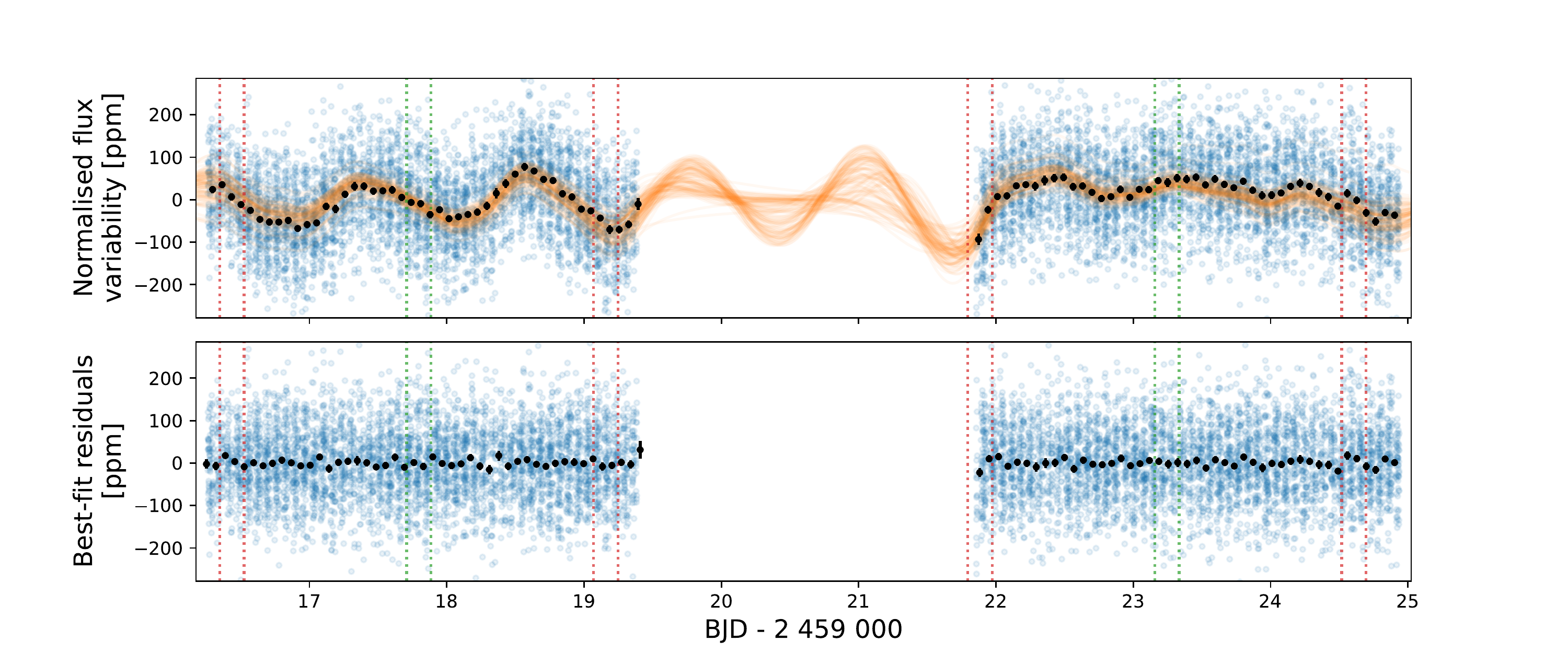}
                        \caption{Photometric variability attributed to stellar rotation as fitted by the Gaussian process. The upper panel shows in blue the time series corrected for instrumental systematics and the planet model, including transit, eclipse, and phase-curve signal. The residuals of the best-fit are represented in the lower panel (blue points). The data points binned once per \cheops{} orbit are shown in black. The faded orange lines are GP models sampled from the posterior distribution highlighting the uncertainty of the fit. The vertical dotted lines mark the time interval of the planetary transits (red) and eclipses (green).}
                        \label{fig:results_stellar_gp}
                \end{figure*}

                We fitted the photometric variability observed in the phase curves with a Gaussian process and attributed to stellar rotation. The extracted variability, corrected from systematics and after subtraction of the planet model, is shown in Fig.~\ref{fig:results_stellar_gp}, where several GP models drawn from the posterior distribution are shown. The signal is well fitted, especially in the first phase curve, and the flexibility of the Gaussian processes is used to adapt to the change of variability in the second part.
                
                We combine the stellar inclination obtained from our analysis with the stellar radius and the projected rotation speed to compute the rotation period of \host. The estimated period $P_\star=1.198^{+0.026}_{-0.025}$\,days matches the peak maximum measured from the Lomb-Scargle periodogram precisely (see Fig.~\ref{fig:pc_raw_variability}) and supports the notion that stellar rotation is indeed the cause of the observed photometric variability.
                
                The damped period of the GP is computed from the undamped period $P_0$ and the quality factor $Q$ and its value $P_\text{damped}=1.32^{+0.22}_{-0.11}$\,days is consistent at $1\sigma$ with the stellar rotation period. The large upper error bar seems to indicate that longer periods might be favoured, which can be caused by the absence of oscillations in the second phase curve.
                
                Our results seem to indicate that the stellar rotation and the photometric variability fitted by the GP are related. If the stellar outer envelope is convective, this likely indicates the presence of stellar spots on \host{}. However, the nature of the outer envelope cannot be clearly assessed and could be radiative. In that case, the mechanism causing this phenomenon is not understood, either due to inhomogeneities on the stellar surface \citep{trust_A-star_rotation_2020} or excited pulsation modes \citep{lee_2020_rotating_cores_pulsations}, as discussed in Section~\ref{ssec:model_star}.

        \subsection{Photometric precision} \label{ssec:results_perf}
        
                The photometric performance of \cheops{} can be evaluated from the residual light curve after correcting for all systematic effects, the stellar activity and removing the planetary model (bottom panel of Fig.~\ref{fig:results_pc}). The noise level in the residuals is evaluated over a given duration by computing the error on the mean of all data points in a time window of the considered duration. The window is moved across the whole time series to estimate the noise at different moments of the observation. The median of the noise levels computed for a given duration gives us a precise estimate of the photometric performances of \cheops. We compute the noise levels using five different methods, which will not be detailed here as it goes beyond the scope of this work; however, all give consistent results.
                Based on this approach, we evaluate the photometric precision of the \cheops{} observations of \host{} (A4-star of magnitude $V=6.6$) to be 10\,ppm over one-hour windows, and it goes down to 5\,ppm over a six-hour window.

\section{Conclusion} \label{sec:conc}

        In this work, we perform the analysis of the \cheops{} observations of the \host{} system that cover four eclipse events and two full phase curves of the ultra-hot Jupiter.
We fit the asymmetric transits with a model accounting for GD and oblateness of the fast rotating star. We measure a planetary radius of $1.600^{+0.017}_{-0.016}\,R_J$, which consistent at $0.7\,\sigma$ with the value $1.619\pm0.021\,R_J$ from \cite{lendl_2020_wasp-189b}, and constrain the stellar oblateness with a polar radius $2.88^{+0.15}_{-0.12}\,\%$ smaller than the equatorial radius. We also robustly determine the system orientation and find that \planet{} is on a polar orbit with a true orbital obliquity $\Psi_p=89.6\pm1.2\deg$. The value is more precise and consistent at $0.9\,\sigma$ with $85.4\pm4.3\deg$ found by \cite{lendl_2020_wasp-189b}. Combined with Doppler tomography, we narrow down the architecture of the system to two possible configurations where the planet transits in front of the receding hemisphere of the star: one with the north stellar pole visible, the other with the south pole visible.
        
        We measure an eclipse depth of $96.5^{+4.5}_{-5.0}\,\text{ppm}$ and find no hints of strong atmospheric winds with a hotspot offset consistent with zero ($-7\pm17\,\deg$). Our value of the eclipse depth is consistent at $1.3\,\sigma$ with the value $87.9\pm4.3\,\text{ppm}$ derived by \cite{lendl_2020_wasp-189b}. Our precision is similar despite two additional eclipses, and we attribute this to the presence of photometric variability. The stellar activity also prevents us from deriving precise constraints on the planet nightside flux or on the heat redistribution efficiency. We could not disentangle the contribution of reflected light and thermal emission to the phase curve amplitude, but we compute an upper limit on the geometric albedo in the \cheops{} passband ($A_g < 0.48$). Using synthetic emission spectra for both the star and the planet, we show that the brightness of \planet{} in the \cheops{} passband is only marginally explained by thermal emission alone and that it would require a Bond albedo of zero and an extremely inefficient energy redistribution. This conclusion is consistent with the work of \cite{lendl_2020_wasp-189b}, where the reported dayside temperature is $3435\pm27\,\text{K}$ and implies a negative Bond albedo (emitted planetary flux larger than stellar irradiating power).
        
        From the transit fit, we infer the inclination of the host star and compute its rotation period ($1.198^{+0.026}_{-0.025}\,\text{days}$). This period matches the photometric variability detected in the data and attributed to stellar activity. As the properties of \host{} do not allow to assess the nature of the outer envelope, we propose several possible explanations. If the outer envelope is convective, then the presence of spots is the most probable cause for the photometric variability. In the case of radiative envelope in A-type stars, we mention two mechanisms that could explain the observations: inhomogeneities on the stellar surface or pulsation modes excited by resonance couplings between the convective core and the radiative envelope.
        
        The \cheops{} time series are obtained with exquisite photometric precision reaching a RMS noise level on the detrended data of 10\,ppm over window length of one\,hour.
Given the nearly polar orientation of the system and the oblateness of \host, one might expect the right ascension of the ascending node of the orbit to precess due to the perturbations induced by the stellar quadrupole moment $J_2$. This effect is used in the case of the Earth for Sun-synchronous spacecraft orbits such as \cheops. In the case of \planet{}, this would be directly measurable by a change of inclination $i_p$, or, equivalently, a change of impact parameter $b$. As mentioned previously, the accurate measurement of the inclination requires to account for the oblateness of the star in the determination of the transit and eclipse durations. Further observations of transits of \planet{} with \cheops{} are being acquired for this purpose. If such an effect is detected, it would allow us to constrain the $J_2$ term of \host{} similarly to how it was done for Kepler-13Ab \citep{masuda_2015_gd_kepler13Ab}.

\begin{acknowledgements}\\
        %Lead author's acknowledgements
        This project has received funding from the European Research Council (ERC) under the European Union's Horizon 2020 research and innovation programme (project {\sc Four Aces}, grant agreement No. 724427).  It has also been carried out in the frame of the National Centre for Competence in Research ``PlanetS'' supported by the Swiss National Science Foundation (SNSF). A.De. acknowledges the financial support of the SNSF.
        %Co-author's acknowledgements
        M.J.Ho. and Y.Al. acknowledge the support of the Swiss National Fund under grant 200020\_172746.\\
        M.Le. acknowledges support from the Swiss National Science Foundation under grant No. PCEFP2\_194576.\\
        S.Sa. has received funding from the European Research Council (ERC) under the European Unions Horizon 2020 research and innovation programme (project {\sc STAREX}, grant agreement No. 833925).\\
        D.Eh. acknowledges support from the European Research Council (ERC) under the European Union's Horizon 2020 research and innovation programme (project {\sc Four Aces}, grant agreement No. 724427).\\
        A.Br. was supported by the SNSA.\\
        S.Ho. gratefully acknowledges CNES funding through the grant 837319.\\
        V.V.Gr. is an F.R.S-FNRS Research Associate.\\
        V.Bo. acknowledges support from the European Research Council (ERC) under the European Union's Horizon 2020 research and innovation programme (project {\sc Four Aces}, grant agreement No. 724427; project {\sc Spice Dune}, grant agreement No. 947634, project {\sc SCORE}, grant agreement No. 851555).\\
        O.De. acknowledges support by FCT - Funda\c c\~ao para a Ci\^encia e a Tecnologia through national funds and by FEDER through COMPETE2020 - Programa Operacional Competitividade e Internacionalizac\~ao by these grants: UID/FIS/04434/2019, UIDB/04434/2020, UIDP/04434/2020, PTDC/FIS-AST/32113/2017 \& POCI-01-0145-FEDER- 032113, PTDC/FIS-AST/28953/2017 \& POCI-01-0145-FEDER-028953, PTDC/FIS-AST/28987/2017 \& POCI-01-0145-FEDER-028987, O.De. is supported in the form of work contract (DL 57/2016/CP1364/CT0004) funded by national funds through FCT.\\
        B.-O.De. acknowledges support from the Swiss National Science Foundation (PP00P2-190080).\\
        L.M.Se. gratefully acknowledges financial support from the CRT foundation under Grant No. 2018.2323 ``Gaseous or rocky? Unveiling the nature of small worlds''.\\
        S.G.So. acknowledges support from FCT through FCT contract No. CEECIND/00826/2018 and POPH/FSE (EC).\\
        T.G.Wi. and A.C.Ca. acknowledge support from STFC consolidated grant numbers ST/R000824/1 and ST/V000861/1, and UKSA grant ST/R003203/1.\\
        R.Al. acknowledges support from the Spanish Ministry of Science and Innovation and the European Regional Development Fund through grants ESP2016-80435-C2-1-R, ESP2016-80435-C2-2-R, PGC2018-098153-B-C33, PGC2018-098153-B-C31, ESP2017-87676-C5-1-R, MDM-2017-0737 Unidad de Excelencia Maria de Maeztu-Centro de Astrobiolog\'ia (INTA-CSIC), as well as the support of the Generalitat de Catalunya/CERCA programme. The MOC activities have been supported by the ESA contract No. 4000124370.\\
        S.C.C.Ba. acknowledges support from FCT through FCT contracts No. IF/01312/2014/CP1215/CT0004.\\
        X.Bo. acknowledges his role as ESA-appointed CHEOPS science team member.\\
        M.De. acknowledges support by the CNES.\\
        The Belgian participation to CHEOPS has been supported by the Belgian Federal Science Policy Office (BELSPO) in the framework of the PRODEX Program, and by the University of Li\`ege through an ARC grant for Concerted Research Actions financed by the Wallonia-Brussels Federation. L.De. is an F.R.S.-FNRS Postdoctoral Researcher.\\
        M.Fr. gratefully acknowledges the support of the Swedish National Space Agency (DNR 65/19, 174/18).\\
        D.Ga. gratefully acknowledges financial support from the CRT foundation under Grant No. 2018.2323 ``Gaseous or rocky? Unveiling the nature of small worlds''.\\
        M.Gi. is an F.R.S.-FNRS Senior Research Associate.\\
        K.G.Is. is the ESA CHEOPS Project Scientist and is responsible for the ESA CHEOPS Guest Observers Programme. She does not participate in, or contribute to, the definition of the Guaranteed Time Programme of the CHEOPS mission through which observations described in this paper have been taken, nor to any aspect of target selection for the programme.\\
        J.La. acknowledges granted access to the HPC resources of MesoPSL financed by the Region Ile de France and the project Equip@Meso (reference ANR-10-EQPX-29-01) of the programme Investissements d'Avenir supervised by the Agence Nationale pour la Recherche.\\
        P.Ma. acknowledges support from STFC research grant number ST/M001040/1.\\
        D.Qu. acknowledges partial support by a grant from the Simons Foundation (PI Queloz, grant number 327127).\\
        I.Ri. acknowledges support from the Spanish Ministry of Science and Innovation and the European Regional Development Fund through grant PGC2018-098153-B- C33, as well as the support of the Generalitat de Catalunya/CERCA programme.\\
        %Other acknowledgements
        This work has made use of data from the European Space Agency~(ESA) mission {\it Gaia} (\url{https://www.cosmos.esa.int/gaia}), processed by the {\it Gaia} Data Processing and Analysis Consortium~(DPAC, \url{https://www.cosmos.esa.int/web/gaia/dpac/consortium}). Funding for the DPAC has been provided by national institutions, in particular the institutions participating in the {\it Gaia} Multilateral Agreement.\\
        %Thanks to the referee
        We thank the referee for the insightful comments that helped improve the quality of this work.
\end{acknowledgements}

\bibliographystyle{aa}
\bibliography{references}

\begin{appendix}
        
        \section{Background thresholds used to clip out data points}
                \begin{table}[ht]
                        \caption{Thresholds above which a correlation between the measured flux and the background level have been detected in the eclipse observations for each of the photometric extraction method.}
                        \label{tab:bkg_thrsh}
                        \centering
                        \begin{tabular}{lc}
                                \hline\hline
                                Photometric extraction method & Threshold value [$\text{e}^-$] \\
                                \hline
                                DRP -- aperture of 22.5\,pixels & $4\,10^6$ \\
                                DRP -- aperture of 25\,pixels & $5\,10^6$ \\
                                DRP -- aperture of 30\,pixels & $4\,10^6$ \\
                                DRP -- aperture of 40\,pixels & $7\,10^6$ \\
                                \pipe{} -- imagettes & -- \\
                                \pipe{} -- sub-arrays & $1\,10^3$ \\
                                \hline\hline
                        \end{tabular}
                        \tablefoot{These values were used to identify and discard photometric data points before the analysis of the light curves. The \pipe{} imagette photometry has no value because the flux-background correlation could not be detected in the data.}
                \end{table}

        \newpage
        \section{Angular parametrisation of the gravity-darkening model}
                \begin{figure}[ht]
                        \centering
                        \includegraphics[width=0.9\hsize,trim={3.5cm 5cm 5cm 4cm},clip]{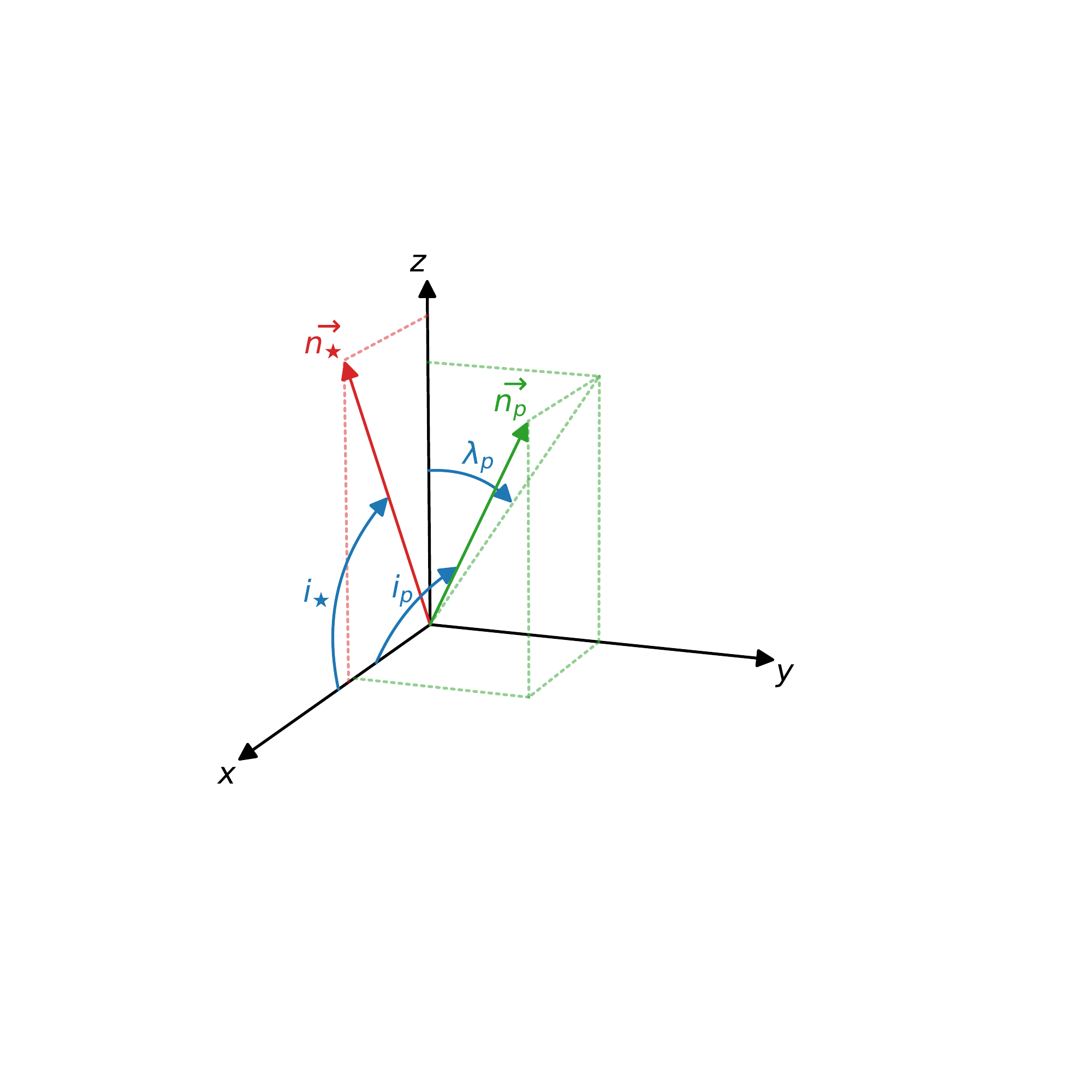}
                        \caption{Geometry used to define the angular parameters of the GD model. The $x$-axis points towards the observer. The $z$-axis is defined so that the stellar spin axis $\protect\overrightarrow{n_\star}$ is contained in the ($x$,~$z$)-plane and its projection onto the plane of the sky is always along $z\geq0$. The $y$-axis completes the orthonormal basis. The orientation of the planetary orbit is defined by the orbital angular momentum unit vector $\protect\overrightarrow{n_p}$ that points perpendicularly to the orbital plane. The stellar inclination $i_\star$ is the angle between the $x$-axis and the stellar spin axis $\protect\overrightarrow{n_\star}$. The inclination of the planetary orbit $i_p$ is the angle between the $x$-axis and the orbital angular momentum unit vector $\protect\overrightarrow{n_p}$. The projected orbital obliquity $\lambda_p$ is the angle between the $z$-axis and the projection of $\protect\overrightarrow{n_p}$ onto the ($y$,~$z$)-plane. Following the convention used in this work, the vector $\protect\overrightarrow{n_p}$ is always located in the half space where $z\geq0$, and we have $\lambda_p>0$ when $\protect\overrightarrow{n_p}$ is in the half space where $y<0$. Thus, the example displayed in the figure has $i_\star>0$, $i_p>0$ and $\lambda_p<0$. The expression of the true orbital obliquity $\Psi_p$ is obtained from the dot product of the two unit vectors: $\cos\!\left(\Psi_p\right)=\protect\overrightarrow{n_\star}\cdot\protect\overrightarrow{n_p}$.}
                        \label{fig:gd_geometry}
                \end{figure}

        \clearpage
        \section{De-trending parameters}
                
                \begin{@twocolumnfalse}
                \begin{table*}[ht]
                        \parbox{\textwidth}{
                        \caption{Model parameters for the normalisation and the systematic noise. The values reported are the one obtained for the phase-curve model with a Lambertian reflector and a sinusoid approximation of the thermal emission. All parameters had uniform priors with the sole purpose of improving convergence speed of the MCMC run.}
                        \label{tab:result_other_parameters}
                        \centering
                        \begin{tabular}{llcr}
                                \hline\hline
                                Parameter & Symbol & Value & Units \\
                                \hline
                                \multirow{6}{*}{Stellar flux used for normalisation} & $F_{\star,\ \text{ecl}_1}$ & ${147277.2}\pm{3.5}$ & $10^3\,\text{e}^-$ \\
                                & $F_{\star,\ \text{ecl}_2}$ & ${146965.4}_{-3.6}^{+3.5}$ & $10^3\,\text{e}^-$ \\
                                & $F_{\star,\ \text{ecl}_3}$ & ${146984.1}_{-3.6}^{+3.7}$ & $10^3\,\text{e}^-$ \\
                                & $F_{\star,\ \text{ecl}_4}$ & ${146207.4}_{-3.6}^{+3.7}$  & $10^3\,\text{e}^-$ \\
                                & $F_{\star,\ \text{PC}_1}$ & ${146500}\pm{20}$ & $10^3\,\text{e}^-$ \\
                                & $F_{\star,\ \text{PC}_2}$ & ${146468}\pm{20}$ & $10^3\,\text{e}^-$ \\
                                \hline
                                \multirow{10}{*}{Quadratic and linear trends} & $c_{2,\ \text{ecl}_1}$ & ${2124}_{-181}^{+188}$ & $\text{ppm}/\text{day}^2$ \\
                                & $c_{1,\ \text{ecl}_1}$ & ${-631}\pm{39}$ & $\text{ppm}/\text{day}$ \\
                                & $c_{2,\ \text{ecl}_2}$ & ${2005}\pm{173}$ & $\text{ppm}/\text{day}^2$ \\
                                & $c_{1,\ \text{ecl}_2}$ & ${-215}_{-29}^{+30}$ & $\text{ppm}/\text{day}$ \\
                                & $c_{2,\ \text{ecl}_3}$ & ${932}_{-160}^{+158}$ & $\text{ppm}/\text{day}^2$ \\
                                & $c_{1,\ \text{ecl}_3}$ & ${-95}\pm{30}$ & $\text{ppm}/\text{day}$ \\
                                & $c_{2,\ \text{ecl}_4}$ & ${-456}_{-154}^{+149}$ & $\text{ppm}/\text{day}^2$ \\
                                & $c_{1,\ \text{ecl}_4}$ & ${381}_{-33}^{+34}$ & $\text{ppm}/\text{day}$ \\
                                & $c_{1,\ \text{PC}_1}$ & ${-34}_{-15}^{+14}$ & $\text{ppm}/\text{day}$ \\
                                & $c_{1,\ \text{PC}_2}$ & ${-14}_{-14}^{+16}$ & $\text{ppm}/\text{day}$ \\
                                \hline
                                \multirow{10}{*}{Roll angle coefficients for the eclipses} & $a_{1,\ \text{ecl}}$ & ${28}\pm{10}$ & ppm \\
                                & $b_{1,\ \text{ecl}}$ & ${92}\pm{43}$ & ppm \\
                                & $a_{2,\ \text{ecl}}$ & ${108}_{-33}^{+34}$ & ppm \\
                                & $b_{2,\ \text{ecl}}$ & ${-24}\pm{16}$ & ppm \\
                                & $a_{3,\ \text{ecl}}$ & ${-16}\pm{16}$ & ppm \\
                                & $b_{3,\ \text{ecl}}$ & ${-30}\pm{21}$ & ppm \\
                                & $a_{4,\ \text{ecl}}$ & ${-13}\pm{11}$ & ppm \\
                                & $b_{4,\ \text{ecl}}$ & ${4}\pm{11}$ & ppm \\
                                & $a_{5,\ \text{ecl}}$ & ${7.4}_{-6.0}^{+6.2}$ & ppm \\
                                & $b_{5,\ \text{ecl}}$ & ${5.6}_{-4.8}^{+4.7}$ & ppm \\
                                \hline
                                \multirow{10}{*}{Roll angle coefficients for the phase curves} & $a_{1,\ \text{PC}}$ & ${-31}_{-3.3}^{+3.2}$ & ppm \\
                                & $b_{1,\ \text{PC}}$ & ${-80}_{-250}^{+240}$ & ppm \\
                                & $a_{2,\ \text{PC}}$ & ${82}_{-177}^{+179}$ & ppm \\
                                & $b_{2,\ \text{PC}}$ & ${-5}_{-48}^{+47}$ & ppm \\
                                & $a_{3,\ \text{PC}}$ & ${16}_{-41}^{+42}$ & ppm \\
                                & $b_{3,\ \text{PC}}$ & ${-21}\pm{101}$ & ppm \\
                                & $a_{4,\ \text{PC}}$ & ${30}\pm{42}$ & ppm \\
                                & $b_{4,\ \text{PC}}$ & ${-5}_{-23}^{+24}$ & ppm \\
                                & $a_{5,\ \text{PC}}$ & ${6.8}\pm{8.2}$ & ppm \\
                                & $b_{5,\ \text{PC}}$ & ${17}\pm{11}$ & ppm \\
                                \hline\hline
                        \end{tabular}
                        }
                \end{table*}
                \end{@twocolumnfalse}

        \clearpage
        \section{Correlation plots of the model parameters}
                
                \begin{@twocolumnfalse}
                \begin{figure*}[ht]
                        \parbox{\textwidth}{
                        \centering
                        \includegraphics[height=.837\vsize]{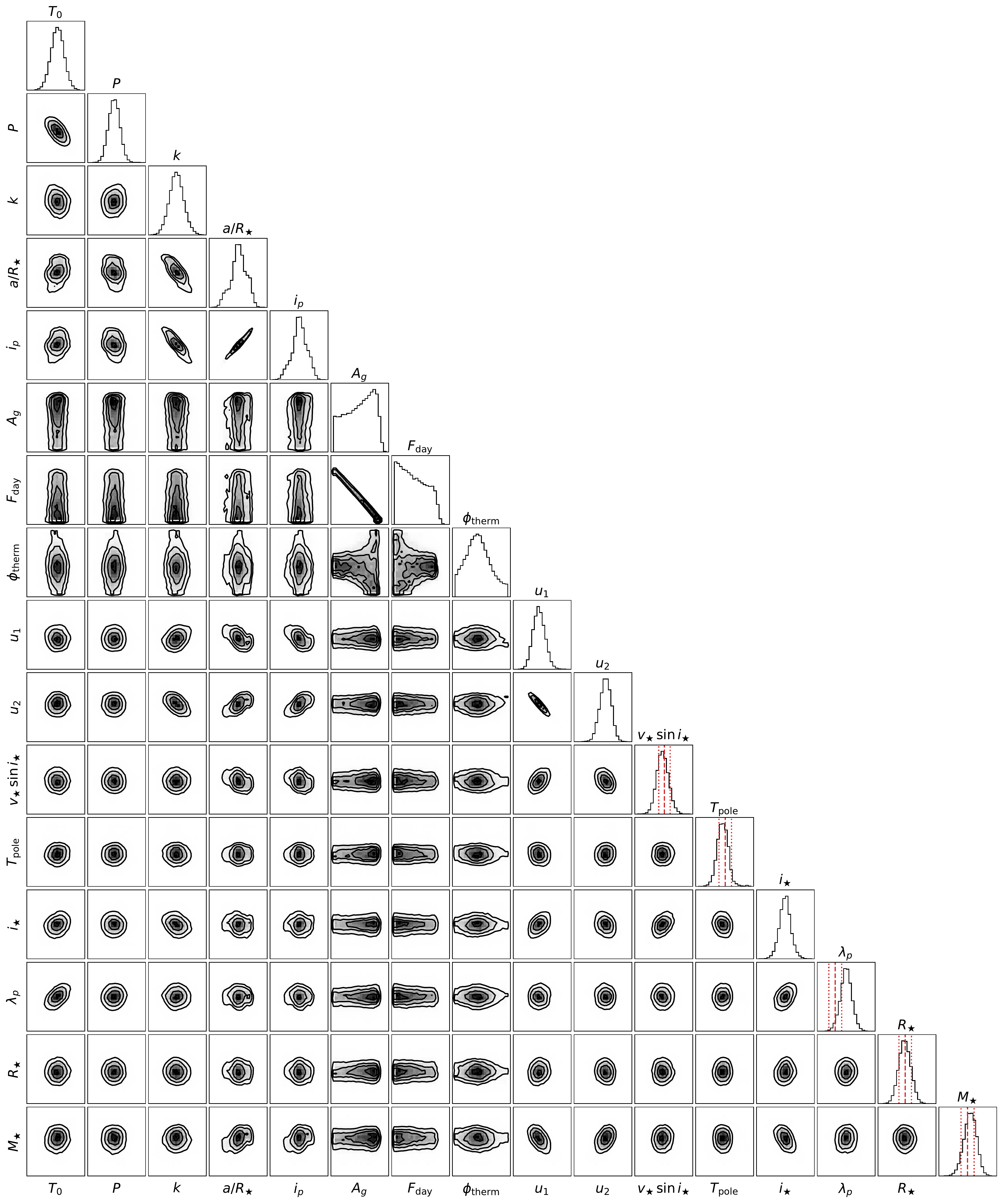}
                        \caption{Correlation plot of the parameters of the planet model made with the code \texttt{corner} \citep{foreman_corner}. The marginalised posterior distribution of each parameter is represented in the diagonal and the 2D histogram contours are shown below. The red vertical lines highlight the normal prior (if any), with the dashed line being the mean value and the dotted line being the $\pm1\sigma$ interval. The values of the axes are not displayed as this plot aims only at revealing parameter correlations (values are reported in Table~\ref{tab:result_parameters_lamb}). The strong linear correlation between the geometric albedo $A_g$ and the dayside thermal flux $F_\text{day}$ is a result of the degeneracy between reflected light and thermal emission. The plot also shows that the hotspot offset $\phi_\text{them}$ is well defined for high $F_\text{day}$ values, as explained in Section~\ref{ssec:results_params}.}
                        \label{fig:corner_plot}       
                        }                 
                \end{figure*}
                \end{@twocolumnfalse}
        
                \clearpage
                \begin{figure}
                        \centering
                        \includegraphics[width=\hsize]{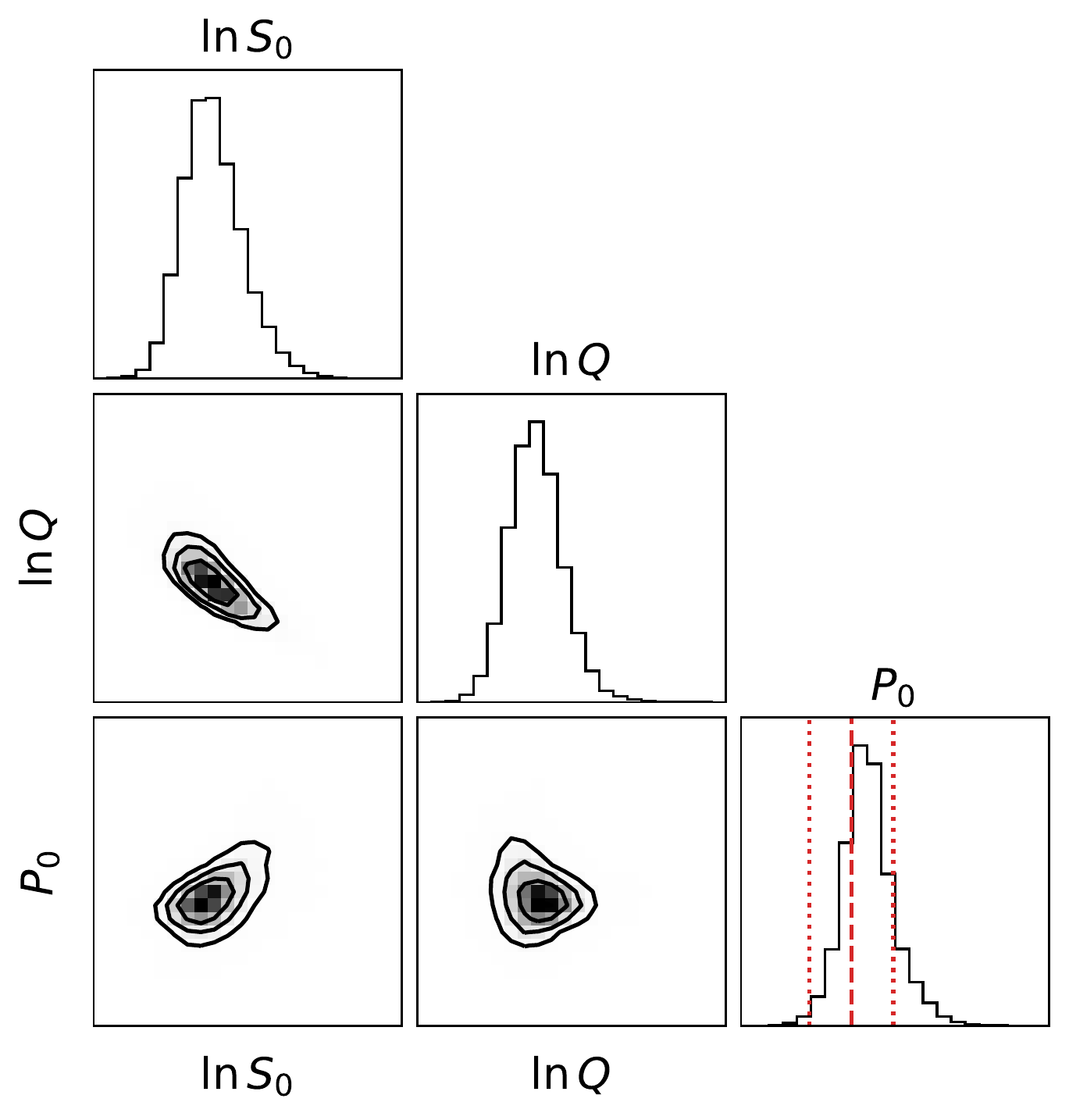}
                        \caption{Same as Fig.~\ref{fig:corner_plot}, but for the hyper-parameters of the Gaussian process used to fit the stellar activity signal.}
                        \label{fig:corner_plot_gp}
                \end{figure}
\end{appendix}

\end{document}